\newif\iffigs\figstrue
\documentclass[a4paper,11pt]{article}
\usepackage{amscd}
\usepackage{latexsym,amsfonts,amssymb,amsmath,lscape,graphics,setspace}
\usepackage{graphicx}
\usepackage{longtable}
\usepackage{graphicx,tikz}
\usepackage{youngtab}
\usepackage{multirow}
\usepackage{color}
\usepackage{mathptmx}
\usepackage{mathtools}
\usepackage{mathrsfs}
\usepackage{commath}
\usepackage{xcolor}
\usepackage{hyperref}
\hypersetup{
    colorlinks = true,
    linkcolor = {darkgray},
    citecolor = black
} \allowdisplaybreaks
\usepackage[all,cmtip]{xy}
\usepackage{slashed,epsfig}
\usepackage{amsmath,amssymb}
\usepackage{mathrsfs}
\usepackage{commath}

\newcommand{\ft}[2]{{\textstyle\frac{#1}{#2}}}

\def\1bar{1\hskip -.275cm -}
\def\2bar{2\hskip -.275cm -}
\def\3bar{3\hskip -.275cm -}

\newsavebox{\uuunit}
\sbox{\uuunit}
                 {\setlength{\unitlength}{0.825em}
                      \begin{picture}(0.6,0.7)
                                      \thinlines
                                      \put(0,0){\line(1,0){0.5}}
                                      \put(0.15,0){\line(0,1){0.7}}
                                      \put(0.35,0){\line(0,1){0.8}}
                                     \multiput(0.3,0.8)(-0.04,-0.02){10}{\rule{0.5pt}{0.5pt}}
                      \end {picture}}

\makeatletter \@addtoreset{equation}{section} \makeatother

\def\bfone{\relax{\rm 1\kern-.35em 1}}

\def\bfone{\relax{\rm 1\kern-.35em 1}}
\font\cmss=cmss10 \font\cmsss=cmss10 at 7pt

\newcommand{\so}{\mathfrak{so}}
\newcommand{\su}{\mathfrak{su}}

\newcommand{\uu}{\mathfrak{u}}

\def\bfone{\relax{\rm 1\kern-.35em 1}}
\def\inbar{\vrule height1.5ex width.4pt depth0pt}
\def\IC{\relax\,\hbox{$\inbar\kern-.3em{\rm C}$}}
\def\ID{\relax{\rm I\kern-.18em D}}
\def\IF{\relax{\rm I\kern-.18em F}}
\def\IH{\relax{\rm I\kern-.18em H}}
\def\II{\relax{\rm I\kern-.17em I}}
\def\IN{\relax{\rm I\kern-.18em N}}
\def\IP{\relax{\rm I\kern-.18em P}}
\def\IQ{\relax\,\hbox{$\inbar\kern-.3em{\rm Q}$}}
\def\IR{\relax{\rm I\kern-.18em R}}
\font\cmss=cmss10 \font\cmsss=cmss10 at 7pt
\def\ZZ{\relax\ifmmode\mathchoice
{\hbox{\cmss Z\kern-.4em Z}}{\hbox{\cmss Z\kern-.4em Z}} {\lower.9pt\hbox{\cmsss Z\kern-.4em Z}}
{\lower1.2pt\hbox{\cmsss Z\kern-.4em Z}}\else{\cmss Z\kern-.4em Z}\fi}

\def\bar{\overline}

\let\shat=\hat
\def\hat{\widehat}

\def\Coe#1.#2.{{#1\over #2}}

\def\coe#1.#2.{\relax{\textstyle {#1 \over #2}}\displaystyle}

\def\to{\rightarrow}
\def\notin{\hbox{{$\in$}\kern-.51em\hbox{/}}}

\def\IE{\relax{{\rm I\kern-.18em E}}}

\def\IGam{\relax{{\rm I}\kern-.18em \Gamma}}

\def\IA{\relax{\hbox{{\rm A}\kern-.82em {\rm A}}}}

\newcommand{\SU}{\mathop{\rm SU}}
\newcommand{\SO}{\mathop{\rm SO}}

\def\n010{\mathrm{N^{0,1,0}}}

\begin{document}
\baselineskip=18pt
\baselineskip 0.6cm
\begin{titlepage}
\setcounter{page}{0}
\renewcommand{\thefootnote}{\fnsymbol{footnote}}
\begin{flushright}
ARC-19-07
\\
\today
\end{flushright}

\vskip 2cm
\begin{center}
{\Large \bf The $\mathcal{N}_3=3\to \mathcal{N}_3=4$ enhancement of
Super Chern-Simons theories in $D=3$, \\
Calabi HyperK\"ahler metrics and
M2-branes on the $\mathcal{C}(\mathrm{N^{0,1,0}})$ conifold} \vskip
1cm
 {\large P.
Fr\'e\,$^{a,b,c}$\footnote{pfre@unito.it}}, {\large A.
Giambrone\,$^{a}$\footnote{alfredo.giambrone@edu.unito.it}}, {\large
P. A. Grassi\,$^{b,c,d} $\footnote{pietro.grassi@uniupo.it}} and
{\large P.
 Vasko\,$^{a,b,c} $\footnote{petr.vasko@to.infn.it}}
 \vskip 1cm
{ \small \centerline{ $^{(a)}$ \it Dipartimento di Fisica,
Universit\`a di Torino,} \centerline{\it via P. Giuria 1, 10125
Torino, Italy.}
\medskip
\centerline{$^{(b)}$
\it INFN, Sezione di Torino,} \centerline{\it
via P. Giuria 1, 10125 Torino, Italy,}
\medskip
\centerline{$^{(c)}$
\it Arnold-Regge Center,}
\centerline{\it via P. Giuria 1,  10125 Torino, Italy, }
\medskip
\centerline{$^{(d)}$
\it Dipartimento di Scienze e Innovazione
Tecnologica,} \centerline{\it Universit\`a del Piemonte Orientale,}
\centerline{\it viale T. Michel, 11, 15121 Alessandria, Italy. }
\medskip

}
\end{center}
\vskip 0.2cm
\centerline{{\bf Abstract}}
\medskip
Considering matter coupled supersymmetric  Chern-Simons theories in
three dimensions we extend the Gaiotto-Witten mechanism of
supersymmetry enhancement $\mathcal{N}_3=3\to \mathcal{N}_3=4$ from
the case where the hypermultiplets span a flat HyperK\"ahler
manifold to that where they live on a curved one. We derive the
precise conditions of this enhancement in terms of generalized
Gaiotto-Witten identities to be satisfied by the tri-holomorphic
moment maps. An infinite class of HyperK\"ahler metrics compatible
with the enhancement condition is provided by the Calabi metrics on
$T^\star \mathbb{P}^{n}$. In this list we find, for $n=2$ the
resolution of the metric cone on $\n010$ which is the unique
homogeneous Sasaki Einstein 7-manifold leading to an
$\mathcal{N}_4=3$ compactification of M-theory. This leads to
challenging perspectives for the discovery of new relations between
the enhancement mechanism in $D=3$, the geometry of M2-brane
solutions and also for the dual description of super Chern Simons
theories on curved HyperK\"ahler manifolds in terms of gauged fixed
supergroup Chern Simons theories. The relevant supergroup is in this
case $\mathrm{SU(3|N)}$ where $\mathrm{SU(3)}$ is the flavor group
and $\mathrm{U(N)}$ is the color group.

\end{titlepage}

\tableofcontents \noindent {}
\setcounter{footnote}{0} \newpage\setcounter{footnote}{0}
\section{Introduction}
Matter coupled Chern-Simons gauge theories are of interest both as
challenging paradigms in quantum  field theory and as theoretical
models for the description of certain condensed matter systems.
\par
At the dawn of the new millennium Chern Simons matter coupled
theories raised to prominence in association with the
$\mathrm{AdS_4/CFT_3}$ gauge-gravity correspondence. Indeed, after
the discovery of the $\mathrm{AdS_5/CFT_4}$ correspondence
\cite{Maldacena:1997re,Kallosh:1998ph,Ferrara:1998ej,Ferrara:1998jm,Ferrara:1997dh},
the programme of the $\mathrm{AdS_4/CFT_3}$ was an obvious
development where all the results of Kaluza-Klein supergravity,
accumulated at the beginning of the eighties could be recycled. In
the years 1998-2000 a rush  started to complete the derivation of
the Kaluza-Klein spectra for all the compactifications of type
$\mathrm{AdS_5\times (G/H)_5}$ and $\mathrm{AdS_4\times (G/H)_7}$.
The idea was to compare such spectra with the towers of primary
conformal fields in the dual gauge theory either in $D=4$, for type
IIB D3-branes, or in $D=3$ for M2-branes. The case of the coset $
\mathrm{T^{(1,1)}} \, \equiv \, \frac{\mathrm{SU(2)_I\times
SU(2)_{II}}}{\mathrm{U(1)}}$, where the denominator group is the
diagonal of the standard $\mathrm{U_{I,II}(1)\subset SU_{I,II}(2)}$
was studied and the results were published in the month of May 1999
\cite{sergiotorino}. The case of the Sasakian homogeneous seven
manifolds listed in table \ref{sasakiani} was actively studied and
the results were published in
\cite{Fabbri:1999mk,Fabbri:1999ay,Fre1999xp,Merlatti:2000ed}.
\par
This was one side of the correspondence: that of supergravity. The
other side, that of the gauge theory, required the determination of
suitable candidates. The first mile-stone in this direction came in
1998 with the paper by Klebanov and Witten \cite{witkleb} where the
geometrical description as a K\"ahler quotient of the metric cone
$\mathcal{C}\left(\mathrm{T^{(1,1)}}\right)$ on the coset
$\mathrm{T^{(1,1)}}$, denominated by them the \textit{conifold}, was
discussed. Indeed, the main point of \cite{witkleb} was the
identification of the pivot role of the K\"ahler quotient in
singling out the field content and the interactions of the dual
gauge theory on the brane world-volume. In the case of $
\mathrm{T^{(1,1)}}$, the metric cone
$\mathcal{C}\left(\mathrm{T^{(1,1)}}\right)$ can be described as the
K\"ahler quotient of $\mathbb{C}^2\times \mathbb{C}^{2\star}$ with
respect to a single $\mathrm{U(1)}$. So Klebanov and Witten outlined
a pattern that, about one year later and in presence of all the
accomplished Kaluza Klein spectra for the relevant Sasakian
manifolds, was generalized to the case $\mathrm{AdS_4/CFT_3}$ in
\cite{Fabbri:1999hw}.
\par
In all cases the (Hyper)-K\"ahler quotient description of the metric
cone on a (tri)-Sasakian manifold is the starting point for the
construction of the dual gauge theory on the brane world-volume. The
coordinates of the linear space of which we perform the quotient are
the (hyper/Wess-Zumino)-multiplets and the \textit{color gauge
group} is accordingly singled out by the quotient. Having singled
out the principles for the second side of the correspondence, the
explicit construction of the dual gauge theories became possible
together with the definition of all the towers of conformal
primaries to be compared with the Kaluza-Klein spectra. Both tasks
were accomplished for the seven-dimensional Sasakians in the already
quoted papers \cite{Fabbri:1999mk,Fre1999xp,ringoni,Billo:2000zs}.
In particular the case of the $\mathcal{N}_3=3,D=3$ gauge theory,
corresponding to the conifold of $\n010$, was derived in
\cite{ringoni}, leading to the mechanism of gaussian integration of
the gauge multiplet degrees of freedom, leaving a quartic
superpotential remnant, that anticipated of about nine years the scheme
used in \cite{Aharony:2008ug} to obtain the ABJM model.
Indeed in \cite{ringoni} it was shown that, generalizing to
non abelian gauge groups  a mechanism already discovered in \cite{Kapustin:1999ha} for abelian ones,
the addition of Chern-Simons interactions to an $\mathcal{N}_3=4,D=3$
Yang-Mills gauge theory breaks supersymmetry to $\mathcal{N}_3=3$.
\par
The $\mathcal{N}_3=4,3$ gauge theories can be identified as special
subclasses of $\mathcal{N}_3=2,D=3$ gauge theories, whose general
form was described in \cite{Fabbri:1999ay} for linear
representations, and was generalized to arbitrary K\"ahler and
HyperK\"ahler manifolds in \cite{pappo1}, which introduced also a
more compact and geometrical notation for the lagrangian. Utilizing
the off-shell formulation of $\mathcal{N}_3=2,D=3$ gauge theories of
\cite{Fabbri:1999ay,pappo1}, in \cite{ringoni} it was advocated that
in the infrared strong coupling limit the gauge coupling constant
$g^2$  goes to infinity while the dimensionless Chern-Simons
coupling constant $\alpha$ states finite. In this way all kinetic
terms of the fields belonging to the gauge multiplets are suppressed
and the latter fields can be integrated out leaving an
$\mathcal{N}_3=3$ matter coupled Chern-Simons gauge theory whose
superpotential has the following very special form:
\begin{equation}
    \mathcal{W}\, = \, - \frac{1}{8\alpha} \, \mathcal{P}^+_\Lambda \,
\mathcal{P}^+_\Sigma \, \mathfrak{m}_{\Lambda\Sigma}
\label{quartaPot}
\end{equation}
where $\mathcal{P}_+^\Lambda$ denote the holomorphic part of the
moment-maps for the triholomorphic action of the gauge group
generators $T^\Lambda$  on the HyperK\"ahler manifold ${HK}_{2n}$
spanned by the hypermultiplets. The gauge group is generically
denoted $\mathcal{G}$, its Lie algebra is denoted $\mathbb{G}$ and
$\mathfrak{m}_{\Lambda\Sigma}$ is an invariant non-degenerated
quadratic form on $\mathbb{G}$. As we stress later on,
$\mathfrak{m}_{\Lambda\Sigma}$ is not necessarily the Cartan Killing
form and it is not necessarily positive definite. The full scalar
potential for these theories takes the form:
\begin{eqnarray}
    V_{scalar} & = & \frac{1}{6} \,
\left( \partial_i \mathcal{W} \, \partial_{j^\star}
\overline{\mathcal{W}} \,g^{ij^\star} \, + \,
\mathbf{m}^{\Lambda\Sigma} \,
\mathcal{P}^3_\Lambda \, \mathcal{P}^3_\Sigma \, \right) \nonumber\\
\mathbf{m}^{\Lambda\Sigma}(u,v) & \equiv &
\frac{1}{4\alpha^2}\,\mathfrak{m}^{\Lambda\Gamma} \,
\mathfrak{m}^{\Sigma\Delta} \, k_\Gamma^i \, k_\Delta^{j^\star} \,
g_{ij^\star} \label{potentissimo}
\end{eqnarray}
where $\mathcal{P}^3_\Sigma$ are the real components of the
tri-holomorphic moment maps for the action of $\mathcal{G}$ on the
HyperK\"ahler manifold ${HK}_{2n}$, while $g_{ij^\star}$ denotes the
components of its HyperK\"ahler metric $\mathbf{g}$ and
$k_\Gamma^i,k_\Gamma^{j^\star}$ are the components of the Killing
vectors generating $\mathcal{G}$. Indeed the metric $\mathbf{g}$
must admit the gauge group $\mathcal{G}$ as isometry group.
\begin{table}
  \centering
  {\small \begin{tabular}{|c||c|c|c|l|}
\hline
 $\mathcal{N}$ & Name & Coset &$\begin{array}{c}
   \mbox{Holon.} \\
   \so(8) \mbox{ bundle } \
 \end{array}$ & Fibration \\
  \hline
  8 & $\mathbb{S}^7$ & $\frac{\mathrm{SO(8)}}{\mathrm{SO(7)}}$ & 1 &
  $ \left \{ \begin{array}{l}
    \mathbb{S}^7 \, \stackrel{\pi}{\Longrightarrow} \, \mathbb{P}^3 \\
    \forall \, p \, \in \, \mathbb{P}^3 \, ; \, \pi^{-1}(p) \, \sim \, \mathbb{S}^1\\
  \end{array}  \right. $ \\
  \hline 2 & $\mathrm{M^{1,1,1}}$ & $\frac{\mathrm{SU(3)\times SU(2)\times U(1)}}{\mathrm{SU(2) \times U(1) \times U(1)
  }}$ & $\mathrm{SU(3)}$ & $ \left \{ \begin{array}{l}
    M^{1,1,1} \, \stackrel{\pi}{\Longrightarrow} \, \mathbb{P}^2  \, \times \, \mathbb{P}^1\\
    \forall \, p \, \in \, \mathbb{P}^2  \, \times \, \mathbb{P}^1\, ; \, \pi^{-1}(p) \, \sim \, \mathbb{S}^1\\
  \end{array}  \right. $ \\
  \hline
   2 & $\mathrm{Q^{1,1,1}}$ & $\frac{\mathrm{SU(2)\times SU(2)\times SU(2) \times U(1)}}{\mathrm{U(1) \times U(1) \times
  U(1) }}$ & $\mathrm{SU(3)}$ & $ \left \{ \begin{array}{l}
    Q^{1,1,1} \, \stackrel{\pi}{\Longrightarrow} \, \mathbb{P}^1  \, \times \, \mathbb{P}^1\, \times \,\mathbb{P}^1  \\
    \forall \, p \, \in \, \mathbb{P}^1  \, \times \, \mathbb{P}^1\, \times \,\mathbb{P}^1  \, ; \,
    \pi^{-1}(p) \, \sim \, \mathbb{S}^1\\
  \end{array}  \right. $ \\
  \hline
   2 & $V^{5,2}$ & $\frac{\mathrm{SO(5)}}{\mathrm{SO(2)}}$ & $\mathrm{SU(3)}$ &
  $ \left \{ \begin{array}{l}
    \mathrm{V^{5,2}} \, \stackrel{\pi}{\Longrightarrow} \, M_a \, \sim \, \mbox{quadric in } \mathbb{P}^4  \\
    \forall \, p \, \in \, \, M_a \, \, ; \, \pi^{-1}(p) \, \sim \, \mathbb{S}^1\\
  \end{array}  \right. $ \\
  \hline
   3 & $\mathrm{N^{0,1,0}}$ & $\frac{\mathrm{SU(3)\times SU(2)}}{\mathrm{SU(2)\times U(1)}}$ &
   $\mathrm{SU(2)}$ &
  $ \left \{ \begin{array}{l}
    \mathrm{N^{0,1,0}} \, \stackrel{\pi}{\Longrightarrow} \, \mathbb{P}^2  \\
    \forall \, p \, \in \, \, \mathbb{P}^2 \, \, ; \, \pi^{-1}(p) \, \sim \, \mathbb{S}^3\\
    \hline
    \mathrm{N^{0,1,0}} \, \stackrel{\pi}{\Longrightarrow} \, \frac{\mathrm{SU(3)}}{\mathrm{U(1)}\times
    \mathrm{U(1)}}  \\
    \forall \, p \, \in \, \, \frac{\mathrm{SU(3)}}{\mathrm{U(1)}\times \mathrm{U(1)}} \, \, ; \, \pi^{-1}(p)
    \, \sim \, \mathbb{S}^1
  \end{array}  \right. $ \\
  \hline
  \end{tabular}}
  \caption{The homogeneous $7$-manifolds that admit at least $2$ Killing spinors are all sasakian or
  tri-sasakian. This is evident from the fibration structure of the $7$-manifold, which is
  either a fibration
  in circles $\mathbb{S}^1$ for the $\mathcal{N}=2$ cases or a fibration in $\mathbb{S}^3$ for the unique
  $\mathcal{N}=3$ case corresponding to the $\mathrm{N}^{0,1,0}$ manifold.
  Since this latter is also an $\mathcal{N}=2$
  manifold, there is in addition the $\mathbb{S}^1$ fibration.}\label{sasakiani}
\end{table}
\par
In 2007 Bagger and Lambert presented their version of the
$\mathcal{N}_3=8$ Chern-Simons theory \cite{Bagger:2007vi, Anton1, Anton2}. Their
work allowed us to understand how $\mathcal{N}>3$ enhancements might
arise  starting from an $\mathcal{N}_3=3$ model. Few months after
this discovery, all the formulations with $4 \leq \mathcal{N} \leq
8$ were constructed, utilizing the mechanism of gaussian integration
of the physical fields of the vector multiplets, originally
introduced for the case of the compactification on the tri-Sasakian
$\n010$ manifold in \cite{ringoni}. Supersymmetric Chern-Simons
theories were completely classified in the case when the scalar
sector parameterizes a flat manifold. The key point was to
understand how to specialize the $\mathcal{N}_3=3$ theory in order
to enhance the R-symmetry.
\par
An interesting construction is that presented by Gaiotto and Witten
in \cite{gaiottowitten}. Their starting point is an
$\mathcal{N}_3=1$ theory with the field content of an
$\mathcal{N}_3=3$ one. Adding a suitable superpotential the theory
becomes $\mathcal{N}_3=3$ supersymmetric. By means of a restriction
imposed on the superpotential one obtains an $\mathcal{N}_3=4$
supersymmetric theory. Further restrictions lead to higher
$\mathcal{N}$-extended supersymmetric theories. An important feature
is that these restrictions are equivalent to suitable choices of the
gauge group and of the matter representation.
\par
The setup of \cite{ringoni} shows that for general groups and
general couplings the Chern Simons interactions  break R-symmetry
from $\SO(4)$ to $\SO(3)$ and consequently also supersymmetry from
$\mathcal{N}_3=4$ to $\mathcal{N}_3=3$, as we already explained.
\par
Yet one can try to specialize the theory in order to recover
$\SO(4)$ R-symmetry and this is the main issue of the present paper.
\par
Another important discovery was made by Gaiotto and Witten, always
dealing with the case when the scalar multiplets span a flat target
manifold of K\"ahler, HyperK\"ahler or even more restricted
holonomy.  They found that the enhancement to $\mathcal{N}\geq 4$
supersymmetry implies also the existence of a Lie super-algebra
$\mathfrak{G}$ whose bosonic part is the Lie algebra $\mathbb{G}$ of
the gauge group $\mathcal{G}$. This issue was thoroughly
investigated in \cite{farillo}. The authors of this paper worked
directly with the formulation of the super Chern Simons
matter coupled theories obtained after the elimination of the non-dynamical fields and with the final superpotential written in terms of dynamical fields. They showed
that the crucial issues for the supersymmetry enhancement are the
following:
\begin{enumerate}
  \item suitable choices of the gauge group $\mathcal{G}$ with its
  related Lie algebra $\mathbb{G}$ which is not necessarily
  semisimple, rather it typically also involves  abelian $\uu(1)$
  factors,
  \item suitable choices of complex or symplectic linear representations
  $\mathcal{D}(\mathbb{G})$ to which the
  scalar multiplets are assigned,
  \item a suitable choice of a non-degenerate, yet not positive
  definite $\mathcal{G}$-invariant metric $\mathfrak{m}_{\Lambda\Sigma}$
  on the Lie algebra $\mathbb{G}$.
\end{enumerate}
\par
In all instances classified in \cite{farillo}, the above
enumerated choices correspond to the embedding
$\mathbb{G}\hookrightarrow \mathfrak{G}$ of the bosonic Lie algebra
into a super-Lie algebra, the representations of the scalar
multiplets being the same of the fermionic generators of
$\mathfrak{G}$. In certain cases the metric $\mathfrak{m}$ is the restriction
to the bosonic generators of the super Cartan-Killing metric of
$\mathfrak{G}$.
\par
 This provides a challenging Occam's
razor in the classification of supersymmetric Chern-Simons theories.
Indeed, this brings us to another interesting feature discovered by
Kapustin and Saulina \cite{Kapustin:2009cd}. These authors showed
that the same Lie super-algebra $\mathfrak{G}$ can be used to
construct a Chern-Simons supergauge theory, namely a pure Chern-Simons
theory whose gauge group is the supergroup $\mathfrak{G}$.
Quantizing such a topological theory \`{a} la BRST and introducing
the ghosts for the fermionic part of the supergauge symmetry, after
topological twist, these latter can be identified with the matter
multiplets of the standard supersymmetric Chern Simons theory of the
bosonic subalgebra $\mathbb{G}\subset \mathfrak{G}$ \footnote{In the work of \cite{Anton4} the reversed path, from
supergroup theory, in the case Achucarro-Tonwnsed supergravity  to supersymmetric Chern-Simons theory
has been used} . This relation
between the $\mathcal{N}_3=4$ supersymmetric Chern-Simons theory and
the supergroup Chern Simons one, described by Kapustin and Saulina,
is somehow reminiscent of the relation between the Neveu-Schwarz and
the Green-Schwarz formulations of superstrings, where one trades
world volume supersymmetry for supersymmetry in the target space.
Kapustin and Saulina advocated that the supergroup formulation is
helpful to build supersymmetric Wilson-loops \cite{Anton3}.
\par
The supergroup Chern Simons formulation is well established in the
flat scalar manifold case. Instead, what might be the relevant
supergroup $\mathfrak{G}$ and what might be its role in
$\mathcal{N}_3=4$ enhanced super Chern Simons theories on curved
HyperK\"ahler manifolds is  not  clear yet. This issue will be
addressed in future publications \cite{futurocernoso}.
\par
Indeed, as already noticed, supersymmetric Chern-Simons theories
were mostly constructed assuming that the scalar sector
parameterizes a flat K\"ahler manifold which, in the
$\mathcal{N}\geq 3$ has to be HyperK\"ahler. More general cases with
curved HyperK\"ahler manifolds were only sketched. In the
formulation of \cite{pappo1} the scalar fields parameterize a
generic K\"ahler or HyperK\"ahler manifold and the gauge group is
the isometry group of such a manifold. In addition, one has suitable
superpotential functions.
\par
The goal of the present paper is to show that, within the more
general setup of \cite{pappo1}, where the hypermultiplets span
generic HyperK\"ahler manifolds $HK_{2n}$, Chern-Simons
$\mathcal{N}_3=3$ gauge theories are enhanced to $\mathcal{N}_3=4$
supersymmetry, if and only if the tri-holomorphic moment-maps
$\mathcal{P}^{\pm,3}_\Lambda$ of the $HK_{2n}$ isometry group
$\mathcal{G}$ (the gauge group)\footnote{Here we refer only to those isometries acting tri-holomorphically on the HyperK\"ahler space.}, satisfy the following
differential-algebraic constraints:
\begin{alignat}{3}\label{MMcon_triv}
&\partial_i\left(\mathcal{P}^+\cdot\mathcal{P}^+\right)&&=
\partial_{\bar{\ell}}\left(\mathcal{P}^-\cdot\mathcal{P}^-\right)=&&0 \nonumber\\
&\partial_i\left(\mathcal{P}^+\cdot\mathcal{P}^3\right)
&&=\partial_{\bar{\ell}}\left(\mathcal{P}^-\cdot\mathcal{P}^3\right)=&&0 \nonumber\\
&\partial_{\bar{\ell}}\left(\mathcal{P}^+\cdot\mathcal{P}^3\right)
&&=\partial_i\left(\mathcal{P}^-\cdot\mathcal{P}^3\right)=&&0
\end{alignat}
together with
\begin{equation} \label{MMcon_int}
\partial_i\partial_{\bar{\ell}}\left(2\mathcal{P}^3\cdot \mathcal{P}^3-
\mathcal{P}^+\cdot\mathcal{P}^-\right)=0.
\end{equation}
In the above formulae the scalar product is taken with respect to
the previously mentioned non-degenerate invariant metric
$\mathfrak{m}_{\Lambda\Sigma}$, whose signature is not necessarily
positive (or negative) definite.
\par
Those above are a weaker formulation of the constraints introduced
by Gaiotto and Witten that have the same appearance without
derivatives.
\par
Once the constraints (\ref{MMcon_triv}-\ref{MMcon_int}) have been
established the obvious question is \textit{which examples do we
know of non-trivial HyperK\"ahler manifolds endowed with continuous
isometries whose moment maps  satisfy these constraints?}. The first
example was noted by Kapustin and Saulina and it is provided by the
time honored Eguchi Hanson space $EH$. This HyperK\"ahler manifold
is $T^\star \mathbb{P}^1$, namely the total space of the cotangent
bundle to the one-dimensional complex projective space:
$\mathbb{P}^1\sim \mathbb{S}^2$. The isometry group acting tri-holomorphically on the
corresponding Ricci flat HyperK\"ahler metric is $\mathrm{SU(2)}$
and in appendix \ref{capponefinale} we review the appropriate
calculation of its moment maps, showing that they satisfy the
necessary constraints for enhancement.
\par
Actually the Eguchi-Hanson manifold is the first in an infinite
series of HyperK\"ahler manifolds, i.e. the $T^\star \mathbb{P}^n$
manifolds, endowed with the Calabi HyperK\"ahler metrics that were
explicitly constructed in \cite{Cvetic:2001zb}, using a Maurer
Cartan differential form approach. Such a construction is reviewed
and applied to the case of interest to us in section
\ref{conifoldus}. Indeed we make the conjecture that the enhancement
constraints (\ref{MMcon_triv}-\ref{MMcon_int}) hold true for the
$\mathrm{SU(n+1)}$ isometry of the Calabi HyperK\"ahler metric on
$T^\star \mathbb{P}^n$ for all values of $n\in \mathbb{N}$ and  in
appendix \ref{su3momenti} we explicitly prove our conjecture for the
case $n=2$.
\par
The Calabi metric on $T^\star \mathbb{P}^2$ is not a randomly chosen
case rather it has a profound physical relevance. Indeed it
corresponds to the resolution of the conic singularity at the tip of
the metric cone $\mathcal{C}\left(\mathrm{N^{0,1,0}}\right)$. As
displayed in table \ref{sasakiani}, the coset manifold
$\mathrm{N^{0,1,0}}$ is the unique tri-holomorphic, homogeneous
Sasaki-Einstein manifold that exists in $7$-dimensions.  Somehow, as
we already remarked above, $\mathrm{N^{0,1,0}}$ is the
$7$-dimensional analogue of the Sasaki-Einstein homogeneous manifold
$\mathrm{T^{1,1}}$ in $5$-dimensions. It leads to a compactification
of M-theory on
\begin{equation}\label{adiesson010}
  \mathcal{M}_{11} \, = \, \mathrm{ AdS_4} \times \mathrm{N^{0,1,0}}
\end{equation}
which preserves $\mathcal{N}_4=3$ supersymmetry and whose Kaluza
Klein spectrum was explicitly calculated and organized into
$\mathrm{Osp(3|4)}\times \mathrm{SU(3)}$ supermultiplets in
\cite{Fre1999xp,ringoni}. In particular in \cite{ringoni} the Kaluza
Klein spectrum was compared with the spectrum of conformal operators
of a dual conformal field theory whose structure follows from the
description of the metric cone
$\mathcal{C}\left(\mathrm{N^{0,1,0}}\right)$ as a HyperK\"ahler
quotient of $\mathbb{C}^3 \times \mathbb{C}^{3^\star}$ with respect
to the tri-holomorphic action of a $\mathrm{U(1)}$ group. All this
is just synoptic with the Klebanov--Witten  construction of the
conformal gauge theory dual to the $\mathrm{AdS_5\times T^{1,1}}$
compactification of type IIB supergravity \cite{witkleb}. There the
metric cone $\mathcal{C}\left(\mathrm{T^{1,1}}\right)$ is described
as the K\"ahler quotient of $\mathbb{C}^2 \times
\mathbb{C}^{2^\star}$ with respect to the holomorphic action of a
$\mathrm{U(1)}$ group. In this synopsis the smooth Calabi
HyperK\"ahler metric on $T^\star \mathbb{P}^2$ is the analogue of
the Ricci flat K\"ahler metric on the conifold resolution
constructed and discussed in
\cite{PandoZayas:2000ctr,dueparpandoro,Benvenuti:2005qb}.
\par
In the HyperK\"ahler quotient the level $\kappa$ of the moment map
plays the role of resolution parameter. For $\kappa\, = \, 0$ we
have the singular metric cone, while for $\kappa \neq 0$ we obtain
the Calabi metric on the smooth manifold $T^\star \mathbb{P}^2$.
From the M2-brane gauge-theory viewpoint the $\mathrm{U(1)}$ gauge
group (which becomes $\mathrm{U(N)}$ for $N$ M2-branes) is the
\textit{color group}, while $\mathrm{SU(3)}$ is the global
\textit{flavor group}.
\par
It is interesting to remark that if we do not gauge the flavor
symmetry $\mathrm{SU(3)}$, the Chern Simons conformal gauge theory
of the color group $\mathrm{U( N)}$ on the boundary
of $\mathrm{AdS_4}$ is, as discussed in \cite{ringoni}, an
$\mathcal{N}_3=3$ superconformal theory with $\mathrm{Osp(3|4)}$
symmetry. On the other hand if we gauge also the flavor group
$\mathrm{SU(3)}$, we obtain an $\mathcal{N}_3=4$ superconformal
Chern Simons theory with $\mathrm{Osp(4|4)}$ symmetry. Indeed, as we
prove in appendix \ref{piattume}, with a suitable choice of the
metric $m^{\Lambda\Sigma}$, that we specify there, the case of the
Lie algebra:
\begin{equation}\label{carbillo}
  \mathbb{G} \, = \, \su(3) \oplus \su(N) \oplus \uu(1) \,\,\, \,\,\, \,\,\, N \neq 3
\end{equation}
falls into the classification of \cite{farillo} and the
corresponding moment maps satisfy the enhancement constraints
(\ref{MMcon_triv}-\ref{MMcon_int}). Such a flavor-color conformal
theory is of the flat HyperK\"ahler type and hence, according to
Kapustin and Saulina, it is equivalent to a gauged--fixed supergroup
Chern-Simons theory\footnote{See also the recent paper on Supergravity Chern-Simons theory \cite{Anton4}}, the supergroup being:
\begin{equation}\label{granellini}
    \mathfrak{G} \, = \, \mathrm{\su(3|}N\mathrm{)}
\end{equation}
Integrating out the color degrees of freedom in the
\textit{supersymmetric bosonic Chern Simons formulation} one obtains
an $\mathcal{N}_3\, = \, 4$ theory with gauge group the
$\mathrm{SU(3)}$ flavor group and target space the Calabi
HyperK\"ahler manifold $T^\star \mathbb{P}^2$. What happens after an
analogous integration in the equivalent \textit{supergroup
Chern-Simons formulation} is what we plan to explore in
\cite{futurocernoso}.
\par
Our paper is organized as follows. Section \ref{cappodue} summarizes
the general structure of $\mathcal{N}_3=3$ Chern-Simons gauge theory
on curved scalar manifolds as geometrically formulated in
\cite{pappo1}. Section \ref{corella} is the main core of the present
article. Utilizing the appropriate quaternionic vielbein formalism
for HyperK\"ahler and Quaternionic K\"ahler manifolds introduced in
\cite{Andrianopoli:1996cm} and systematically reviewed in
\cite{advancio} we show that we can rewrite the $\mathcal{N}_3=3$
Chern Simons theory in a manifestly $\mathcal{N}_3 =4$ form \`{a} la
Gaiotto--Witten whenever the weak constraints
(\ref{MMcon_triv}-\ref{MMcon_int}) are satisfied. Section
\ref{conifoldus} deals with the case of the HyperK\"ahler Calabi
metric on $T^\star \mathbb{P}^2$ and its relation with the
$N^{0,1,0}$-compactification of M-theory. In subsections
\ref{zaffone},\ref{theoryofN010} we recall the HyperK\"ahler
quotient construction of the metric cone
$\mathcal{C}\left(\n010\right)$ and how it was used in
\cite{ringoni} to determine the structure of the superconformal
theory dual to the (\ref{adiesson010}) compactification. In
subsection \ref{n010pope1} we discuss the resolution of the conifold
singularity which we do in two different but equivalent ways: in
subsection \ref{dividendus} we resolve the singularity uplifting the
moment map to a non vanishing level $\kappa$ in the HyperK\"ahler
quotient, while in subsection \ref{mauricartisu3}, following the
approach of \cite{Cvetic:2001zb}, we perform the direct construction
of the Calabi HyperK\"ahler metric utilizing the Maurer Cartan forms
of $\mathrm{SU(3)}$ on the coset $\n010$. In particular in eq.s
(\ref{curvatura}) and (\ref{baldus}) we present the intrinsic
components of the Riemann tensor and of the $\mathrm{Usp(4)}$
curvature 2-form $\mathbb{R}^{\alpha\beta}$ that, up to our
knowledge, were not yet explicitly available in the literature.
Section \ref{omma22} presents in the utilized notations the explicit
form of the \textit{square integrable self-dual closed $(2,2)$-form}
existing on $T^\star \mathbb{P}^2$ equipped with the Calabi metric
\cite{Cvetic:2001zb}. This item is very important in order to
construct M2-brane solutions of D=11 supergravity with an internal
self-dual $4$-form flux on the transverse space, which preserves
half of the supersymmetries preserved by the fluxless solution
\cite{Cerchiai:2018shs}. Finally section \ref{concluddo} contains
our conclusions.
\par
The several appendices contain the details of lengthy calculations,
in particular those of the moment maps on curved and flat spaces.
\section{${\cal N}_3=3$ supersymmetric Chern-Simons theories}
\label{cappodue}
$ \mathcal{N}_3=3,D=3$ Chern Simons gauge theories are just a
particular subclass of $\mathcal{N}_3=2,D=3$ Chern Simons field
theory. Hence we start from the general form of the latter that was
systematized in \cite{pappo1}.
\subsection{The Lagrangian of the $\mathcal{N}_3=2$ Chern Simons gauge
theory}
\par The  lagrangian of $\mathcal{N}_3=2$ Chern-Simons Gauge Theory, as systematized in \cite{pappo1}, takes the following form:
\begin{eqnarray}
\mathcal{L}_{CSoff}&=& - \, \alpha \, \mbox{Tr} \,\left(\mathfrak{F}
\, \wedge \, \mathcal{A} \, + \, \frac{2}{3}\, \mathcal{A}\wedge \,
\mathcal{A}\wedge \, \mathcal{A}\right)\,+\,\left( \frac{1}{2} \,
g_{ij^\star} \, \Pi^{m\mid i} \, \nabla \bar{z}^{j^\star} \, + \,
\bar{\Pi}^{m\mid j^\star} \, \nabla z^{i}\right)\, \wedge \, e^n \,
\wedge \,  e^p \, \epsilon_{mnp}\nonumber\\ \null &&  - \,
\frac{1}{6} \, \, g_{ij^\star} \, \Pi^{m\mid i} \, \bar{\Pi}^{m\mid
j^\star} \, e^r \,  \wedge \, e^s \, \wedge \,  e^t \,
\epsilon_{rst}\nonumber\\ && +\,{\rm i}\frac{1}{2} \, g_{ij^\star}
\, \left(\bar{\chi}^{j^\star} \, \gamma^m \,\nabla \chi^{i} \, + \,
\bar{\chi}^{i}_c \, \, \gamma^m \, \nabla \chi^{i^\star}_c\right) \,
\wedge \, e^n \, \wedge \,  e^p \, \epsilon_{mnp} \nonumber\\ &&
+\big( \, - \, \frac{1}{3} \, M^\Lambda \,\left(
\partial_i k^j_\Lambda \, g_{j\ell^\star} \, \bar{\chi}^{\ell^\star} \,\chi^i \, - \,  \partial_{i^\star}
k^{j^\star}_\Lambda \, g_{j\ell^\star} \bar{\chi}^\ell_c \,
\chi^{i^\star}_c \right) \, + \, \frac{\alpha}{3} \, \left(
\bar{\lambda}^\Lambda \, \lambda^\Sigma \, + \,
\bar{\lambda}^\Lambda_c \, \lambda^\Sigma_c\right) \,
\mathbf{\mathfrak{m}}_{\Lambda\Sigma}\nonumber\\ && \, +\,
{\rm i}\, \frac{1}{3} \, \left( \bar{\chi}^{j^\star}_c \,
\lambda^\Lambda \, k^i_\Lambda \, - \, \bar{\chi}^{i}_c \,
\lambda^\Lambda \, k^{j^\star}_\Lambda\right) \, g_{ij^\star} \,
 + \, \frac{1}{6} \, \left(
\partial_i\partial_j \mathcal{W}\, \bar{\chi}^i_c \, \chi^j \, + \, \partial_{i^\star}\partial_{j^\star}
{\overline{\mathcal{W}}}\, \bar{\chi}^{i^\star} \,
\chi^{j^\star}_c\right) \nonumber\\
&& -V(M,D,\mathcal{H},z,\bar{z}) + \mathfrak{a} \, \mathscr{R}_{i j^\star k l^\star}\bar{\chi}^{j^\star} \chi^i \bar{\chi}_c^{k} \chi_c^{j^\star}  \, \big) \epsilon_{mnp} \, e^m \, \wedge \, e^n \, \wedge \, e^p
\label{pastiglialeone}
\end{eqnarray}
where:
\begin{enumerate}
  \item The complex scalar fields $z^i$ span a K\"ahler manifold $\mathcal{M}_{K}$, $g_{ij^\star}$ denoting its
  K\"ahler metric, $\mathscr{R}_{ij^\star k l^\star}$ denoting its curvature 2-form. The coefficient $\mathfrak{a}$ is fixed by supersymmetry.
  \item $\Pi^{m|i}$ are auxiliary fields that are identified with the world volume
  derivatives of the scalar
  $z^i$ by their own equation of motion.
  \item The one--forms $e^m$ denote the dreibein of the world volume.
  \item $\mathcal{A}^\Lambda$ is the gauge-one form of the gauge group $\mathcal{G}$.
  \item $\lambda^\Lambda$ are the gauginos, namely the spin $\ft 12$ partners of
  the gauge bosons $\mathcal{A}^\Lambda$
  \item $\chi^i$ are the chiralinos, namely the spin $\ft 12$ partners of the Wess-Zumino scalars $z^i$.
  \item $M^\Lambda$ are the real scalar fields in the adjoint of the gauge group that
  complete the gauge
  multiplet together with the gauginos and the gauge bosons.
  \item $\mathcal{W}(z)$ is the superpotential.
  \item $k^i_\Lambda$ are the Killing vectors of the K\"ahler metric of $\mathcal{M}_{K}$,
  associated with the generators
  of the gauge group.
  \item $\mathfrak{m}^{\Lambda\Sigma}=\mathfrak{m}^{\Sigma\Lambda}$
  denotes a non degenerate, $\mathcal{G}$-invariant metric on the Lie Algebra
  $\mathbb{G}$, which is not necessarily semisimple. The metric
  $\mathfrak{m}^{\Lambda\Sigma}$ is not necessarily
  positive-definite and as a consequence the scalar potential is not
  necessarily positive definite.
  \item The coeffcient $\mathfrak{a}$, which we do not calculate since we do not need it, is fixed by supersymmetry invariance of $\mathscr{L}_{CSoff}$.
\end{enumerate}
The scalar potential in terms of physical and auxiliary fields is
the following one:
\begin{eqnarray} V\left(M,D,\mathcal{H},z,\bar{z}\right)&=&
\left(\frac{\alpha}{3} \,  M^\Lambda
\,\mathbf{\mathfrak{m}}_{\Lambda\Sigma}\,  - \, \frac{1}{6}\,
\mathcal{P}_\Sigma(z,\bar{z})\, + \, \frac{1}{6} \, \zeta_{I} \,
\mathfrak{C}_\Sigma^I \right) \, D^\Sigma \, + \, \frac{1}{6}
M^\Lambda \, M^\Sigma \, k^i_\Lambda \, k_\Sigma^{j^\star} \,
g_{ij^\star} \nonumber\\ && +\,\frac{1}{6} \left(\mathcal{H}^i \,
\partial_i \mathcal{W} \, + \, \mathcal{H}^{\ell^\star} \,
\partial_{\ell^\star} {\overline{\mathcal{W}}} \right) \, - \, \frac{1}{6}
\, g_{i\ell^\star} \,\mathcal{H}^i \,
\mathcal{H}^{\ell^\star} \label{karamella}
\end{eqnarray}
where $\mathcal{P}_\Sigma(z,\bar{z})$ are the moment maps associated
with each generator of the gauge-group, $\zeta_{I}$ are the
Fayet-Iliopoulos parameters associated with each generator of the
center of the gauge Lie algebra, $\mathcal{H}^i$ are the complex
auxiliary fields of the Wess-Zumino multiplets and $D^\Lambda$ are
the auxiliary scalars of the vector multiplets. By
$\mathfrak{C}^I_\Sigma$ we denote the projector onto a basis of
generators of the Lie Algebra center
$\mathfrak{z}\left[\mathbb{G}\right]$.
\par
 In these theories the gauge multiplet does not propagate and it is essentially made of lagrangian
multipliers for certain constraints. Indeed, the auxiliary fields,
the gauginos and the vector multiplet scalars have algebraic field
equations so that they can be eliminated by solving such equations
of motion. The vector multiplet auxiliary scalars $D^\Lambda$ appear
only as lagrangian multipliers of the constraint:
\begin{equation}\label{Msolvo} M^\Lambda \, = \,
\frac{1}{2\alpha} \, {\mathbf{\mathfrak{m}}}^{\Lambda \Sigma}\,\left(
\mathcal{P}_\Sigma \, -\, \zeta_I \, \mathfrak{C}_\Sigma^I \right)
\end{equation}
while the variation of the auxiliary fields $\mathcal{H}^{j^\star}$
of the Wess Zumino multiplets yields:
\begin{equation}\label{eliminoH}
\mathcal{H}^{i} \, = \, g^{ij^\star} \, \partial_{j^\star} \,
\overline{\mathcal{W}} \quad ; \quad
\overline{\mathcal{H}}^{j^\star} \, = \, g^{ij^\star} \,
\partial_{i} \, {\mathcal{W}}
\end{equation}
On the other hand, the equation of motion of the field $M^\Lambda$
implies:
\begin{equation}\label{gospadi} D^\Lambda \, =
\, - \, \frac{1}{\alpha} \, \mathfrak{m}^{\Lambda \Gamma}
g_{ij^\star} \, k_\Gamma^{i} \, k^{j^\star}_\Sigma\, M^\Sigma \, =
\, - \, \frac{1}{2\,\alpha^2} \,g_{ij^\star}\,
\mathfrak{m}^{\Lambda \Gamma}\, k_\Gamma^{i} \,
k^{j^\star}_\Sigma\,\mathfrak{m}^{\Sigma \Delta}\,\left(
\mathcal{P}_\Delta \, -\, \zeta_I \,  \mathfrak{C}_\Delta^I \right)
\end{equation}
which finally resolves all the auxiliary fields in terms of
functions of the physical scalars.
\par Upon use of both constraints (\ref{Msolvo}) and
(\ref{eliminoH}) the scalar potential takes the following positive
definite form:
\begin{eqnarray}
V(z,\bar{z}) & = & \frac{1}{6} \, \left( \partial_i \mathcal{W} \,
\partial_{j^\star} \overline{\mathcal{W}} \,g^{ij^\star} \, + \,
\mathbf{m}^{\Lambda\Sigma} \, \left(\mathcal{P}_\Lambda \, - \,
\zeta_I \, \mathfrak{C}_\Lambda^I\right) \, \left(\mathcal{P}_\Sigma
\, - \, \zeta_J \, \mathfrak{C}_\Sigma^J \right) \right)\nonumber\\
\mathbf{m}^{\Lambda\Sigma}(z,\bar{z}) & \equiv &
\frac{1}{4\alpha^2}\,\mathfrak{m}^{\Lambda\Gamma} \,
\mathfrak{m}^{\Sigma\Delta} \, k_\Gamma^i \, k_\Delta^{j^\star}
\, g_{ij^\star} \label{quadraPot}
\end{eqnarray}
In a similar way the gauginos can be resolved in terms of the
chiralinos:
\begin{equation}\label{eliminogaugino} \lambda^\Lambda \, = \, -\,
\frac{1}{2\alpha} \, \mathfrak{m}^{\Lambda\Sigma} \, g_{ij^\star}
\chi^i \, k^{j^\star}_\Sigma \quad ; \quad \lambda^\Lambda_c \, = \,
-\, \frac{1}{2\alpha} \, \mathfrak{m}^{\Lambda\Sigma} \,
g_{ij^\star} \chi^{j^\star} \, k^{i}_\Sigma
\end{equation}
In this way if we were able to eliminate also the gauge one form
$\mathcal{A}$, the Chern-Simons gauge theory would reduce to a
theory of Wess-Zumino multiplets with additional interactions. The
elimination of $\mathcal{A}$, however, is not possible in the
nonabelian case and it is possible in the abelian case only through
duality nonlocal transformations. This is the corner where
interesting nonperturbative dynamics is hidden.
\subsection{The structure of ${\cal N}_3=3$ Chern Simons
gauge theories}\label{N=3gaugetheory} The ${\cal N}_3=3$ case is
just a particular case in the class of theories described in the
previous section since a theory with ${\cal N}_3=3$ SUSY, must {\it
a fortiori} be an ${\cal N}_3=2$ theory. In \cite{Fabbri:1999ay} ,
the case of ${\cal N}_3=4$ theories was also considered, within the
${\cal N}_3=2$ class. These latter are obtained through dimensional
reduction of an ${\cal N}_4=2$ theory in four--dimensions. The main
issue in such a dimensional reduction is the enhancement of the
$D=4$ $R$-symmetry, which is $\mathrm{USp(2)}$ to $\mathrm{SO(4)}$
in $D=3$. Indeed, since each $D=4$ Majorana spinor splits, under
dimensional reduction on a circle $\mathbb{S}^1$, into two $D=3$
Majorana spinors, the number of three--dimensional supercharges is
just twice the number of $D=4$ supercharges:
\begin{equation}
  {\cal N}_3 = 2 \, \times \, {\cal N}_4
\label{n34}
\end{equation}
The mechanism of such enhancement of $R$-symmetry is analyzed in
detail in Appendix \ref{rsimmaconvo}. Such analysis is quite
relevant to the main issue of the present paper which is the
retrieval of the $\so(4)$ R-symmetry algebra naturally produced by
the dimensional reduction when special conditions are satisfied by
the hypermultiplet interactions. In the absence of such conditions
the $D=3$ $R$-symmetry being instead reduced to $\so(3)$ by Chern
Simons interaction.
\par
Indeed the ${\cal N}_3=3$ case corresponds to an intermediate
situation. It is an ${\cal N}_3=2$ theory with the field content of
an ${\cal N}_3=4$ one, but with additional ${\cal N}_3=2$
interactions that respect  three out of the four supercharges
obtained through dimensional reduction. Using an ${\cal N}_3=2$
superfield formalism and the notion of twisted chiral multiplets it
was shown in \cite{kapustin} that for abelian gauge theories these
additional ${\cal N}_3=3$ interactions are
\begin{enumerate}
  \item A Chern Simons term, with coefficient $\alpha$
  \item A mass-term  with coefficient $\mu=\alpha$ for the
  chiral field $Y^\Lambda$ in the adjoint of the color gauge group.
  By this latter we denote the complex field belonging,
  in four dimensions, to the ${\cal N}_4=2$ gauge
  vector multiplet.
\end{enumerate}
In \cite{ringoni} the authors  retrieved for non-abelian gauge
theories the same result as that found by the authors of
\cite{kapustin} for abelian theories. In \cite{ringoni} the
construction was presented in the component formalism which is
better suited to discus the relation between the world--volume gauge
theory and the geometry of the transverse cone
$\mathcal{C}(\mathcal{M}_7)$. Let us also remark that the arguments
used in \cite{Aharony:2008ug} are the same which were spelled out
ten years earlier in \cite{ringoni}. In this section we summarize in
the more general notations of \cite{pappo1}, based on HyperK\"ahler
metrics and the tri-holomorphic moment maps, the general form of a
non abelian $ {\cal N}_3=3$ Chern Simons gauge theory in three
dimensions as it was obtained in \cite{ringoni}.
\subsubsection{The field content and the interactions} The strategy
of \cite{ringoni} was that of writing the ${\cal N}_3=3$ gauge
theory as a special case of an ${\cal N}_3=2$ theory, whose general
form was discussed in the previous section. For this latter the
field content is given by:
\begin{equation}
   \begin{array}{|c|c|c|c|}
   \hline
     \mbox{multipl. type $\, /\,\mathrm{SO(1,2)}$ spin}  &  1 & \ft 1 2 & 0 \\
     \hline
     \hline
     \null & \null &\null & \null \\
     \mbox{vector multipl.} &  \underbrace{A^\Lambda_\mu}_{\mbox{gauge field}} &
     \underbrace{\left( \lambda^{+\Lambda},\lambda^{-\Lambda}
     \right)}_{\mbox{gauginos}}  &\underbrace{ M^\Lambda}_{\mbox{real scalar}} \\
     \hline
     \null &  \null &\null & \null \\
     \mbox{chiral multip.} &  \null &\underbrace{\left( \chi^{+i},\chi^{-i^*}
     \right)}_{\mbox{chiralinos}}  & \underbrace{ z^i,\ \bar z^{i^*}}_
     {\mbox{complex scalars}}
     \\
     \hline
   \end{array}
\label{fieldcont}
\end{equation}
and  the complete Lagrangian was given in the previous sections. In
particular the complete Chern Simons Lagrangian  before the
elimination of the auxiliary fields was displayed in
eq.(\ref{pastiglialeone}).
\par
The Chern-Simons ${\cal N}_3=3$ case is obtained when the  following
conditions are fulfilled:
\begin{itemize}
\item The spectrum of chiral multiplets is made of $\mbox{dim}
\mathcal{G} \, + \, 2n$ complex fields arranged in the following way
 \begin{equation}\label{cagullo}
   z^i \, = \, \left\{ \begin{array}{rccl}
   Y^\Lambda & = & \mbox{complex fields} & \mbox{in the \textbf{adjoint rep. of the color group}} \\
   q^j & = & \left( \begin{array}{c} u^a \\ v_b \\\end{array}\right) & \left \{
   \begin{array}{l}
   \mbox{$2n$ complex fields spanning a \textbf{HyperK\"ahler manifold} ${HK}_{2n}$}\\
   \mbox{which is invariant under a }\\
   \mbox{\textbf{triholomorphic action} of the gauge group $\mathcal{G}$}\\
   \end{array} \right.\\
   \end{array} \right.
.\end{equation}
\item the K\"ahler potential has the following form:
\begin{equation}\label{kallettus}
  \mathcal{K}(Y,u,v) \, = \, \hat{\mathcal{K}}(u,v)
\end{equation}
where $\hat{\mathcal{K}}(u,v)$ is the K\"ahler potential of the
Ricci-flat HyperK\"ahler metric of the HyperK\"ahler manifold
${HK}_{2n}$. The assumption that $ \mathcal{K}(Y,u,v)$ does not
depend on $Y^\Lambda$ implies that the kinetic term of  these
scalars vanishes turning them into auxiliary fields that can be
integrated away.
  \item The superpotential $\mathcal{W}(z)$ has the following form:
  \begin{equation}
  \mathcal{W}(Y,u,v)=\mathfrak{m}^{\Lambda\Sigma}\,\left(Y_\Lambda\, \mathcal{P}^+_\Sigma(u,v)\,
  +\,2\,\alpha\,\,Y_\Lambda\,Y_\Sigma\right)
  \label{suppotn3}
  \end{equation}
where $\mathcal{P}_\Sigma^+(u,v)$ denotes the holomorphic part of
the triholomorphic moment map induced by the triholomorphic action
of the color  group on ${HK}_{2n}$.
\end{itemize}
The reason why these two choices make the theory ${\cal N}_3=3$
invariant is simple: the first choice corresponds to assuming the
field content of an ${\cal N}_3=4$ theory which is necessary since
${\cal N}_3=3$ and ${\cal N}_3=4$ supermultiplets are identical. The
second choice takes into account that the metric of the
hypermultiplets must be HyperK\"ahler and that the gauge coupling
constant was sent to infinity. The third choice introduces an
interaction that preserves ${\cal N}_3=3$ supersymmetry but breaks
(when $\alpha \ne 0$) ${\cal N}_3=4$ supersymmetry.
\par
Going back to the off-shell Chern Simons lagrangian given in
eq.(\ref{pastiglialeone}) one can perform the elimination of the
auxiliary fields that now include $Y^\Lambda,D^\Lambda,M^\Lambda,
\mathcal{H}^i$ at the bosonic level and the gauginos
$\lambda^\Lambda, \lambda^\Lambda_c,\chi^\Lambda,\chi^\Lambda_c$ at
the fermionic level (note that there are two more non propagating
gauginos coming from the chiral multiplet in the adjoint
representation of the gauge group). We do not enter the details of
the integration over the non propagating fermions and we just
consider the bosonic lagrangian emerging from the integration over
the auxiliary bosonic fields. The first integration to perform is
that over the auxiliary field $\mathcal{H}^\Lambda$. This is simply
the lagrangian multiplier of the constraint:
\begin{equation}\label{cartacallus}
  \partial_\Lambda \mathcal{W} \, = \, 0 \quad \Rightarrow \quad Y_\Lambda \, = \,
  \frac{1}{4\alpha} \, \mathcal{P}_\Lambda^+(u,v)
\end{equation}
Substituting this back into the lagrangian  yields a potential with
the same structure as that in eq.(\ref{quadraPot}) but with a
modified superpotential which becomes quadratic in the holomorphic
momentum maps:
\begin{eqnarray}
    V(u,v) & = & \frac{1}{6} \,
\left( \partial_i \mathfrak{W} \, \partial_{j^\star}
\overline{\mathfrak{W}} \,g^{ij^\star} \, + \,
\mathbf{m}^{\Lambda\Sigma} \,
\mathcal{P}^3_\Lambda \, \mathcal{P}^3_\Sigma \, \right) \nonumber\\
\mathbf{m}^{\Lambda\Sigma}(u,v) & \equiv &
\frac{1}{4\alpha^2}\,\mathfrak{m}^{\Lambda\Gamma} \,
\mathfrak{m}^{\Sigma\Delta} \, k_\Gamma^i \, k_\Delta^{j^\star} \, g_{ij^\star} \\
\mathfrak{W}& = & - \frac{1}{8\alpha} \, \mathcal{P}^+_\Lambda \,
\mathcal{P}^+_\Sigma \, \mathfrak{m}^{\Lambda\Sigma} \label{quartaPot}
\end{eqnarray}
here by $\mathfrak{W}$ we mean the on-shell superpotential.
Altogether the supersymmetric lagrangian of the $\mathcal{N}_3\, =\,
3$ Chern Simons theory, after gaussian integration of the non
propagating fields, takes the following form\footnote{Here and in the sequel, we do not take care of the quartic fermionic interaction which is not relevant in our discussion.}:
\begin{eqnarray}
\mathcal{L}_{CSon} &=& - \,\alpha \, \mbox{Tr} \,\left( \mathfrak{F}
\, \wedge \, \mathcal{A} \, + \, \frac{1}{3}\, \mathcal{A}\wedge \,
\mathcal{A}\wedge \, \mathcal{A}\right)\, +\, \frac{1}{6} \,
g_{ij^\star} \,   \nabla ^{m} q^{i} \, \nabla_{m} \bar{q}^{j^\star}
\, e^r \, \wedge \, e^s \, \wedge \,  e^t \, \epsilon_{rst} \nonumber\\
  \null && +\,{\rm i}\frac{1}{2} \, g_{ij^\star} \, \left(\bar{\chi}^{j^\star} \, \gamma^m \,\nabla \chi^{i}
  \, + \,  \bar{\chi}^{i}_c \, \, \gamma^m \,
           \nabla \chi^{j^\star}_c\right) \, \wedge \, e^n \, \wedge \,  e^p \, \epsilon_{mnp} \nonumber\\
&& + \, \left(  \frac{{\rm i}}{12\alpha} \, \left( \partial_\ell
k_{j^\star}^\Lambda \, \bar{\chi}^{j^\star} \chi^\ell \,
 - \, \partial_{\ell^\star}k_j^\Lambda \, \bar{\chi}_c^j \chi_c^{\ell^\star}  \right)\,
 \mathcal{P}^3_\Lambda \, - \, \frac{1}{6\alpha} \, \mathfrak{m}_{\Lambda \Sigma}
 \left( \bar{\chi}^{\ell^\star} \, k_{\ell^\star}^{\Lambda}  \chi^{j} \, k_{j}^{\Sigma} \,
 + \,  \bar{\chi}^{j}_c \, k_{j}^{\Lambda} \chi^{\ell^\star}_c \, k_{\ell^\star}^{\Sigma}  \right)  \right. \nonumber\\
&& \left. + \, \frac{1}{6} \, \left(
\partial_{i}\partial_{j}\mathfrak{W} \, \bar{\chi}_c^i
\, \chi^j \, + \,
\partial_{i^\star}\partial_{j^\star}\overline{\mathfrak{W}}
\, \bar{\chi}^{i^\star} \, \chi_c^{j^\star}\right) \, + \,
\frac{1}{6} \,
   \partial_i \mathfrak{W} \, \partial_{j^\star} \overline{\mathfrak{W}} \,g^{ij^\star} \right. \nonumber\\
                   && \left. + \, \frac{1}{24\alpha^2} \, \mathfrak{m}^{\Lambda \Gamma}
                   \, \mathfrak{m}^{\Sigma \Delta} \, k_{\Gamma}^i \, k_{\Delta}^{j^\star}
                   \, g_{i j^\star} \, \mathcal{P}^3_\Lambda\, \mathcal{P}^3_\Sigma \, \right)\, \wedge
                   \, e^m \,\wedge \, e^n \, \wedge \,  e^p \, \epsilon_{mnp} \label{alpenliebe}
\end{eqnarray}
\section{HyperK\"ahler manifolds in the hypermultiplet sector and the supersymmetry enhancement}
\label{corella} Given the above result we take the following two
steps:
\begin{description}
  \item[a)] Still maintaining full generality we try to rearrange the items contained in the
  lagrangian (\ref{alpenliebe})
in such a way  as to bring into evidence the HyperK\"ahler structure
of the scalar manifold and its holonomy group.
  \item[b)] Next we introduce the constraints (\ref{MMcon_triv}-\ref{MMcon_int}) on the
  moment maps and we show that when they hold true the lagrangian
  (\ref{alpenliebe}) can be further elaborated in such a way as to
  become structurally similar to the Lagrangian of the
  Gaiotto-Witten theory \cite{gaiottowitten}. In this way the
  $R$-symmetry and the supersymmetry enhancements are revealed for
  general HyperK\"ahler manifolds (curved ones included), whose
  tri-holomorphic isometries satisfy the constraint (\ref{MMcon_triv}-\ref{MMcon_int})
  at the level of their moment maps. Note also,  that, as we already stressed, equations
  (\ref{MMcon_triv}-\ref{MMcon_int}) encode weaker constraints with respect to those so
  far discussed in the literature.
\end{description}
 Let us start with our programme.
\subsection{Quaternionic vielbein for HyperK\"ahler manifolds and
moment maps} Following the notations of \cite{Andrianopoli:1996cm}
we recall that a HyperK\"ahler manifold ${HK_{2n}}$ is a $4
n$-dimensional real manifold endowed with a metric $h$:
\begin{equation}
d s^2 = h_{u v} (q) d q^u \otimes d q^v   \quad ; \quad u,v=1,\dots,
4  m \label{qmetrica}
\end{equation}
and three complex structures
\begin{equation}
(\mathbf{J}^x) \,:~~ T({HK_{2n}}) \, \longrightarrow \, T({HK_{2n}}) \qquad
\quad (x=1,2,3)
\end{equation}
that satisfy the quaternionic algebra
\begin{equation}
\mathbf{J}^x \mathbf{J}^y = - \delta^{xy} \, \mathbf{1} \,  +  \,
\epsilon^{xyz} \mathbf{J}^z \label{quatalgebra}
\end{equation}
and respect to which the metric is hermitian:
\begin{equation}
\forall   \mbox{\bf X} ,\mbox{\bf Y}  \in   T{HK_{2n}}   \,: \quad h
\left( \mathbf{J}^x \mbox{\bf X}, \mathbf{J}^x \mbox{\bf Y} \right )   = h \left(
\mbox{\bf X}, \mbox{\bf Y} \right ) \quad \quad
  (x=1,2,3)
\label{hermit}
\end{equation}
From eq.(\ref{hermit}) it follows that one can introduce a triplet
of 2-forms
\begin{equation}
\begin{array}{ccccccc}
\mathbf{K}^x& = &\mathbf{K}^x_{u v} d q^u \wedge d q^v & ; & \mathbf{K}^x_{uv} &=&   h_{uw}
(\mathbf{J}^x)^w_v \cr
\end{array}
\label{iperforme}
\end{equation}
that provides the generalization of the concept of K\"ahler form
occurring in the complex case. The triplet $\mathbf{K}^x$ is named the {\it
HyperK\"ahler} form. It is an $\mathrm{SU(2)}$ Lie--algebra valued
2--form in the same way as the K\"ahler form is a $\mathrm{U(1)}$
Lie--algebra valued 2--form. The space is HyperK\"ahler if the
2-forms in this triplet are all closed:
\begin{equation}\label{chiusuralampo}
    \mathrm{d}\mathbf{K}^x \, = \, 0
\end{equation}
As a consequence of the above structure the manifold ${HK_{2n}}$ has
a holonomy group of the following type:
\begin{eqnarray}
{\rm Hol}({HK_{2n}})&=& \mathbf{1} \otimes {\cal H} \quad
(\mbox{HyperK\"ahler})
\nonumber \\
{\cal H} & \subset & \mathrm{Usp(2n)} \label{olonomia}
\end{eqnarray}
Hence introducing flat indices $\{A,B,C= 1,2\},
\{\alpha,\beta,\gamma = 1,.., 2n\}$  that run, respectively, in the
fundamental representations of $\mathrm{SU(2)}$ and
$\mathrm{\mathrm{USp(2n)}}$, we can find a vielbein 1-form
\begin{equation}
{\cal U}^{A\alpha} = {\cal U}^{A\alpha}_u (q) d q^u
\label{quatvielbein}
\end{equation}
such that
\begin{equation}
h_{uv} = {\cal U}^{A\alpha}_u {\cal U}^{B\beta}_v
\mathbb{C}_{\alpha\beta}\epsilon_{AB} \label{quatmet}
\end{equation}
where $\mathbb{C}_{\alpha \beta} = - \mathbb{C}_{\beta \alpha}$ and
$\epsilon_{AB} = - \epsilon_{BA}$ are, respectively, the flat
$\mathrm{USp(2n)}$ and $\mathrm{USp(2)} \sim \mathrm{SU(2)}$
invariant metrics. The vielbein ${\cal U}^{A\alpha}$ is covariantly
closed with respect to a flat $\mathrm{SU(2)}$-connection
$\omega^z$:
\begin{equation}\label{flattaconna}
    d\omega^x \, + \, \ft 12 \, \epsilon^{xyz} \, \omega^{y} \wedge
    \omega^{z} \, = \, 0
\end{equation}
and to some $\mathrm{USp(2n)}$-Lie Algebra valued connection
$\Delta^{\alpha\beta} = \Delta^{\beta \alpha}$:
\begin{eqnarray}
\nabla {\cal U}^{A\alpha}& \equiv & d{\cal U}^{A\alpha} +{i\over 2}
\omega^x (\epsilon \sigma_x\epsilon^{-1})^A_{\phantom{A}B}
\wedge{\cal U}^{B\alpha} \nonumber\\
&+& \Delta^{\alpha\beta} \wedge {\cal U}^{A\gamma}
\mathbb{C}_{\beta\gamma} =0 \label{quattorsion}
\end{eqnarray}
where $(\sigma^x)_A^{\phantom{A}B}$ are the standard Pauli matrices.
For them we utilize the conventions shown in formula (\ref{3dgamma})
and we set $\epsilon_{AB} \, = \, i \, \sigma_2$. Furthermore ${
\cal U}^{A\alpha}$ satisfies  the reality condition:
\begin{equation}
{\cal U}_{A\alpha} \equiv ({\cal U}^{A\alpha})^* = \epsilon_{AB}
\mathbb{C}_{\alpha\beta} {\cal U}^{B\beta} \label{quatreality}
\end{equation}
Eq.(\ref{quatreality})  defines  the  rule to lower the symplectic
indices by means   of  the  flat  symplectic   metrics
$\epsilon_{AB}$   and $\mathbb{C}_{\alpha \beta}$. More specifically
we can write a stronger version of eq.(\ref{quatmet}):
\begin{eqnarray}
({\cal U}^{A\alpha}_u {\cal U}^{B\beta}_v + {\cal U}^{A\alpha}_v
{\cal
 U}^{B\beta}_u)\mathbb{C}_{\alpha\beta}&=& h_{uv} \epsilon^{AB}\nonumber\\
({\cal U}^{A\alpha}_u {\cal U}^{B\beta}_v + {\cal U}^{A\alpha}_v
{\cal U}^{B\beta}_u) \epsilon_{AB} &=& h_{uv} {1\over n}
\mathbb{C}^{\alpha \beta} \label{piuforte}
\end{eqnarray}
\noindent We have also the inverse vielbein ${\cal U}^u_{A\alpha}$
defined by the equation
\begin{equation}
{\cal U}^u_{A\alpha} {\cal U}^{A\alpha}_v = \delta^u_v \label{2.64}
\end{equation}
Flattening a pair of indices of the Riemann tensor ${\cal
R}^{uv}_{\phantom{uv}{ts}}$ we obtain
\begin{equation}
{\cal R}^{uv}_{\phantom{uv}{ts}} {\cal U}^{\alpha A}_u {\cal
U}^{\beta B}_v =
 \mathbb{R}^{\alpha\beta}_{ts}\epsilon^{AB}
\label{2.65}
\end{equation}
\noindent where $\mathbb{R}^{\alpha\beta}_{ts}$ is the field
strength of the $\mathrm{USp(2n) }$ connection $\Delta^{\alpha\beta}
\, = \,\Delta^{\beta\alpha}$ :
\begin{equation}
d \Delta^{\alpha\beta} + \Delta^{\alpha \gamma} \wedge
\Delta^{\delta \beta} \mathbb{C}_{\gamma \delta} \equiv
\mathbb{R}^{\alpha\beta} = \mathbb{R}^{\alpha \beta}_{ts} dq^t
\wedge dq^s \label{2.66}
\end{equation}
Eq. (\ref{2.65}) is the explicit statement that the Levi Civita
connection associated with the metric $h$ has a holonomy group
contained in $\mathbf{1} \otimes \mathrm{USp(2n) }$. Consider now
eq.s~(\ref{quatalgebra},\ref{iperforme}). We easily deduce the
following relation:
\begin{equation}
h^{st} \mathbf{K}^x_{us} \mathbf{K}^y_{tw} = -   \delta^{xy} h_{uw} +
  \epsilon^{xyz} \mathbf{K}^z_{uw}
\label{universala}
\end{equation}
Eq.(\ref{universala}) implies that the intrinsic components of the
HyperK\"ahler 2-forms $\mathbf{K}^x$ yield a representation of the quaternion
algebra. Hence we can write:
\begin{equation}
\mathbf{K}^x =\,\frac{\rm i}{2}\, \mathbb{C}_{\alpha\beta} (\sigma ^x)_{AB}
\, {\cal U}^{\alpha A} \wedge {\cal U}^{\beta B} \label{2.69}
\end{equation}
where the second index of the Pauli matrix has been lowered with
$\epsilon_{BC}$.
\par
Recalling now that a HyperK\"ahler manifold is also a complex
K\"ahler manifold we can introduce complex coordinates and vielbein
with respect to a reference complex structure that we choose to be
that associated with $\mathbf{K}^z$. Than $\mathbf{K}^z$ is the
K\"ahler 2-form of $HK_{2n}$ and we have:
\begin{eqnarray}
  \mathbf{K}^z &=& i \, g_{ij^\star} \, dz^i \wedge d\bar{z}^{j^\star} \, =
  \,i \,\mathbf{e}^{\alpha} \wedge \bar{\mathbf{e}}_{\alpha}\nonumber\\
   &=& i \, {\cal U}^{1\alpha}\wedge {\cal
   U}^{2\beta}\mathbb{C}_{\alpha\beta}\label{bartolo1}
\end{eqnarray}
where $\mathbf{e}^{\alpha}$ is a set of complex vielbein one-forms
such that:
\begin{equation}
    ds^2 \, = \, g_{ij^\star} \, dz^i \otimes d\bar{z}^{j^\star} \,
    = \, \sum_{\alpha=1}^n \, \mathbf{e}^{\alpha} \otimes
    \bar{\mathbf{e}}_{\alpha} \quad ; \quad
    \bar{\mathbf{e}}_{\alpha} \, \equiv \,\left(
    \mathbf{e}^{\alpha}\right)^\star \label{bartolo2}
\end{equation}
Utilizing our basis of Pauli matrices we also find:
\begin{equation}
    \mathbf{K}^\pm \, = \, \mathbf{K}^x\pm i \mathbf{K}^y \quad ; \quad \mathbf{K}^+ \,
    = \,  - \, i \,{\cal U}^{1\alpha}\wedge {\cal
   U}^{1\beta}\mathbb{C}_{\alpha\beta} \quad ; \quad \mathbf{K}^- \,
   = \, \, i \,  {\cal U}^{2\alpha}\wedge {\cal
   U}^{2\beta}\mathbb{C}_{\alpha\beta}\label{bartolo3}
\end{equation}
Once the complex vielbein $\mathbf{e}^\alpha$ are found there is a
universal way of writing the quaternionic vielbein ${\cal
U}^{A\alpha}$ so that eq.s(\ref{bartolo1}-\ref{bartolo3}) are
satisfied, namely:
\begin{equation}\label{parascopio}
    {\cal U}^{1\alpha} \, = \,\mathbf{e}^{\alpha} \quad ; \quad {\cal U}^{2\beta} \,
    = \,\mathbb{C}^{\alpha\beta} \, \bar{\mathbf{e}}^{\beta}
\end{equation}
In this way we get:
\begin{equation}\label{fulmivelluto}
    \mathbf{K}^+ \, = \,- i \,\mathbf{e}^{\alpha} \wedge \mathbf{e}^{\beta} \,
    \mathbb{C}_{\alpha\beta} \quad ; \quad \mathbf{K}^- \,
     = \, i \,\bar{\mathbf{e}}_{\alpha} \wedge \bar{\mathbf{e}}_{\beta} \,
    \mathbb{C}^{\alpha\beta}
\end{equation}
The above structure is very useful for the calculation of the
relation between Killing vectors of an isometry group $\mathcal{G}$
of the HyperK\"ahler metric and their associated moment maps. Let us
denote $\mathbf{k}_\Lambda$ such Killing vectors closing the Lie
algebra $\mathbb{G}$, whose structure constants we denote
$f^\Lambda_{\phantom{\Lambda}\Gamma\Delta}$ as usual:
\begin{eqnarray}
  \mathbf{k}_\Lambda &=& {k}_\Lambda^i\, \partial_i + {k}_\Lambda^{i^\star}\,
  \partial_{i^\star}  \nonumber\\
  \left[\mathbf{k}_\Gamma\, , \, \mathbf{k}_\Delta\right ]
  &=& f^\Lambda_{\phantom{\Lambda}\Gamma\Delta} \, \mathbf{k}_\Lambda
  \label{killoni}
\end{eqnarray}
Utilizing the complex vielbein:
\begin{equation}\label{groningen}
   \mathbf{e}^\alpha \, = \, e^\alpha_i \,  dz^i +  e^\alpha_{i^\star} \,
   d\bar{z}^{i^\star} \quad ; \quad \bar{\mathbf{e}}_\alpha \,
    = \, \bar{e}_{\alpha\,i} \,  dz^i +  \bar{e}_{\alpha \,
   i^\star} \, dz^{i^\star}
\end{equation}
it is convenient to introduce the flat components of the Killing
vectors:
\begin{equation}\label{risotto}
    k^\alpha_\Lambda \, = \, e^\alpha_i \, {k}_\Lambda^i +
    e^\alpha_{i^\star}\, {k}_\Lambda^{i^\star} \quad ; \quad
    k_{\alpha,\Lambda} \, = \, \bar{e}_{\alpha\,i}  {k}_\Lambda^i \,
   + \, \bar{e}_{\alpha \,i^\star} \,{k}_\Lambda^{i^\star}
\end{equation}
and from the definition of the tri-holomorphic moment maps:
\begin{equation}\label{sacchetto}
    \mathbf{i}_\Lambda \mathbf{K}^x \, = \, - \,\mathrm{d}\mathcal{P}^x_\Lambda
\end{equation}
we obtain\footnote{As usual we denote anholonomic derivatives
$\partial_\alpha = \mathbf{e}^i_\alpha \partial_i +
\mathbf{e}^{j^\star}_\alpha \partial_{j^\star}$  where
$\mathbf{e}^i_\alpha$ is the inverse vielbein. $\partial^\alpha$ is
the complex conjugate of $\partial_\alpha$.}:
\begin{eqnarray}
  k_\Lambda^\alpha  &=& i \partial^\alpha \, \mathcal{P}^3_\Lambda \,
  = \, - \,\frac{i}{2} \, \mathbb{C}^{\alpha\beta} \, \partial_\beta \,
  \mathcal{P}^+_\Lambda \nonumber\\
   k_{\Lambda\,\alpha}  &=& -i \partial_\alpha \, \mathcal{P}^3_\Lambda \, = \, \frac{i}{2} \,
   \mathbb{C}_{\alpha\beta} \, \partial^\beta \,
  \mathcal{P}^-_\Lambda \label{anholokill}
\end{eqnarray}
\subsection{Making the $R$-symmetry exiplicit and the enhancement}
In this new frame we can make  $R$-symmetry explicit and we can
conveniently study its enhancement.\\ The scalar kinetic term will
involves the gauged quaternionic vielbein
\begin{equation}
  \frac{1}{24}\mathbb{C}_{\alpha \beta} \epsilon_{AB} \langle \shat{\mathscr{U}}^{A \alpha}
   \, , \,  \shat{\mathscr{U}}^{B \beta} \rangle  e^r \wedge e^s \wedge e^t \epsilon_{rst}
\end{equation}
where
\begin{eqnarray}
  \langle \shat{\mathscr{U}}^{A \alpha} ,  \shat{\mathscr{U}}^{B \beta} \rangle
  &=& \eta^{mn}  \shat{\mathscr{U}}^{A \alpha}_m \shat{\mathscr{U}}^{A \alpha}_n \\
  \shat{\mathscr{U}}^{A \alpha} &=& \mathscr{U}^{A \alpha}_i \nabla q^i
  + \mathscr{U}^{A \alpha}_{j^\star} \nabla \overline{q}^{j^\star} \\
  \nabla q^i &=& dq^i + k^i_\Lambda \mathscr{A}^\Lambda = \nabla_m q^i dx^m
\end{eqnarray}
The fermionic kinetic term can be rewritten in an $\rm{SU}(2)$ invariant form:
\begin{equation}
 \frac{\rm{i}}{2} \, \chi_{\dot{A} \alpha} \gamma^m \nabla \chi^{\dot{A} \alpha}
 \wedge e^n \wedge e^p \epsilon_{mnp}
\end{equation}
where
\begin{equation}
  g_{ij^\star} \, = \, \mathbb{C}_{\alpha \beta} \mathscr{U}^{1\alpha}_i
  \mathscr{U}^{2\beta}_{j^\star} \, = \,  \mathbb{C}_{\alpha \beta}
  \mathscr{U}^{1\alpha}_{j^\star} \mathscr{U}^{2\beta}_{i}
\end{equation}
\begin{eqnarray}
  \left\{  \chi^{\dot{A} \alpha} \right\} &=& \left\{ \chi^i
  \mathscr{U}^{1\alpha}_i \, , \, \chi_c^{j^\star} \mathscr{U}^{2\alpha}_{j^\star} \right\} \nonumber\\
 \left\{  \chi_{\dot{A} \alpha} \right\} &=& \mathbb{C}_{\alpha \beta}\left\{
 \overline{\chi}^{j^\star} \mathscr{U}^{2\beta}_{j^\star} \, ,
  \, -\overline{\chi}_c^i \mathscr{U}^{1\beta}_{i} \right\}
\end{eqnarray}
We can also think of this latter as the reduction from four to three
dimensions of the kinetic term for a Majorana spinor,
$\chi^{\dot{1}}$ and $\chi^{\dot{2}}$ being its opposite chirality
projections. In three dimensions the $\dot{A}$ index plays the role
of the $\rm{SU}(2)_L$ R-symmetry while the $\rm{SU}(2)$ rotating the complex structures plays the role of the $\rm{SU}(2)_R$
R-symmetry. To recover $\mathscr{N}_3=4$  supersymmetry we should be
able to write the interactions in an $\rm{SU}(2)_L \times
\rm{SU}(2)_R$ invariant form. We can do this when the constraints
(\ref{MMcon_triv}-\ref{MMcon_int}) hold true. They can be rewritten
with anholonomic derivatives as follows
\begin{eqnarray}
  \partial_{\alpha} \left( \mathscr{P}^+ \cdot \mathscr{P}^+ \right) & = & 0 \\ \label{mom++}
  \partial^\alpha \left( \mathscr{P}^- \cdot \mathscr{P}^- \right) & = & 0 \\
  \partial_{\alpha} \left( \mathscr{P}^3 \cdot \mathscr{P}^+ \right) & = & 0 \\
  \partial^{\alpha} \left( \mathscr{P}^3 \cdot \mathscr{P}^- \right) & = & 0 \label{mom3-} \\
  \partial^{\beta} \partial_{\alpha} \left( 2\mathscr{P}^3 \cdot \mathscr{P}^3 - \mathscr{P}^+
  \cdot \mathscr{P}^- \right) & = & 0 \label{mom+-}
\end{eqnarray}
Considering the scalar potential
\begin{equation}
 V_{pot} \, = \, \frac{1}{24\alpha^2} \mathfrak{m}^{\Lambda \Gamma}
  \mathfrak{m}^{\Sigma \Delta} k^i_\Gamma k^{j^\star}_\Delta g_{i j^\star}
  \mathscr{P}^3_\Lambda \mathscr{P}^3_\Sigma = \frac{1}{48\alpha^2}\mathfrak{m}^{\Lambda \Gamma}
  \mathfrak{m}^{\Sigma \Delta} k^\alpha_\Gamma k_{\alpha \Delta} \mathscr{P}^3_\Lambda
  \mathscr{P}^3_\Sigma
\end{equation}
we can use the definition of the tri-holomorphic moment maps and
eq.(\ref{mom3-}) to rewrite it as
\begin{equation}
V_{pot} \, = \,  \frac{1}{192\alpha^2}\mathfrak{m}^{\Lambda \Gamma}
\mathfrak{m}^{\Sigma \Delta}
  \partial^\alpha \mathscr{P}^-_\Gamma \partial_{\alpha}\mathscr{P}^+_{\Delta}
  \mathscr{P}^3_\Lambda \mathscr{P}^3_\Sigma = - \frac{1}{192\alpha^2}\mathfrak{m}^{\Lambda \Gamma}
  \mathfrak{m}^{\Sigma \Delta} \mathscr{P}^-_\Gamma \partial_{\alpha}\mathscr{P}^+_{\Delta}
  \partial^\alpha \mathscr{P}^3_\Lambda \mathscr{P}^3_\Sigma
\end{equation}
thanks to the moment map equivariance we obtain
\begin{equation}
V_{pot} \, = \,    \frac{\rm{i}}{192\alpha^2}f^{\Gamma \Sigma \Pi}
\mathscr{P}^-_\Gamma
    \mathscr{P}^3_{\Sigma} \mathscr{P}^+_\Pi
\end{equation}
where we define
\begin{equation}
  f^{\Lambda \Sigma \Pi} \equiv \mathfrak{m}^{\Lambda \Gamma}
  \mathfrak{m}^{\Sigma \Delta} f^\Pi_{\Gamma \Delta}
\end{equation}
In this definition $\mathfrak{m}^{\Lambda \Sigma}$ is a
non-degenerate invariant quadratic form on the Lie algebra
$\mathbb{G}$, a priori different from the Cartan Killing metric,
which might be degenerate if the Lie algebra is not semisimple. So
$f^{\Lambda \Sigma \Pi}$ is not necessarily completely
antisymmetric. $f^{\Lambda \Sigma \Pi}$ is completely antisymmetric
if $\mathfrak{m}^{\Lambda \Sigma} = \kappa^{\Lambda \Sigma}$ is the
Cartan-Killing metric of a simple Lie algebra. In the case of a Lie
algebra which is the direct sum of some finite number of simple Lie
algebras and abelian ones, $\mathfrak{m}^{\Lambda \Sigma}$ can be
chosen to be block-diagonal. Each block corresponding to a simple
part is proportional to the respective Cartan-Killing metric. Each
block corresponding to an abelian addend is a generic non-degenerate
invariant quadratic form on it. Also in this case $f^{\Lambda \Sigma
\Pi}$ is totally antisymmetric. It turns out that this freedom in
the definition of $\mathfrak{m}^{\Lambda \Sigma}$ is essential in
order to satisfy the moment map constraints in the case of free
hypermultiplets. In any case, assuming that $f^{\Lambda \Sigma
\Pi}$ is completely antisymmetric we obtain
\begin{equation}
 V_{pot} \, = \,  \frac{\rm{i}}{192\alpha^2}f^{\Gamma \Sigma \Pi}
  \mathscr{P}^-_\Gamma \mathscr{P}^3_{\Sigma}
  \mathscr{P}^+_\Pi = \frac{\rm{1}}{576\alpha^2}f^{\Gamma \Sigma \Pi}
  \mathscr{P}^x_\Gamma \mathscr{P}^y_{\Sigma} \mathscr{P}^z_\Pi \epsilon_{xyz}
\end{equation}
Thanks to eq.(\ref{mom++}) the other contributions to the scalar potential vanish.
For the same reason the only surviving interactions from the Yukawa coupling are
\begin{equation}
 Yuk \, = \, - \, \frac{1}{12\alpha} \mathfrak{m}^{\Lambda \Sigma} \mathscr{P}^3_{\Sigma}
  \partial_\alpha \partial^\gamma \mathscr{P}^3_\Lambda \left( \chi_{\dot{1}\gamma}
  \chi^{\dot{1} \alpha} - \mathbb{C}_{\beta \gamma} \mathbb{C}^{\alpha \rho}
  \chi_{\dot{2}\rho} \chi^{\dot{2} \beta}\right) - \, \frac{1}{6\alpha}
  \mathfrak{m}^{\Lambda \Sigma} \left( j_{2\dot{1}\Lambda} j^{2\dot{1}}_\Sigma
  + j_{1\dot{2}\Lambda}j^{1\dot{2}}_\Sigma\right)
\end{equation}
where
\begin{eqnarray}
  j^{A\dot{B}}_\Lambda &=& k^{A\alpha}_\Lambda \chi^{\dot{B}\beta} \mathbb{C}_{\alpha \beta} \nonumber\\
  j_{A\dot{B}\Lambda} &=& k_{A\alpha\Lambda} \chi^{\dot{B}\beta} \mathbb{C}^{\alpha \beta} \\
  k^{A \alpha}_\Lambda &=& \mathscr{U}^{A\alpha}(k_\Lambda) \nonumber\\
  k_{A \alpha\Lambda} &=& \epsilon_{AB} \mathbb{C}_{\alpha \beta}\mathscr{U}^{B\beta}(k_\Lambda)
\end{eqnarray}
Now we use eq.(\ref{mom+-}) to obtain
\begin{eqnarray}
 Yuk & = &  - \, \frac{1}{12\alpha} \mathfrak{m}^{\Lambda \Sigma} \left( \frac{1}{4}
 \partial_{\alpha}\mathscr{P}^+_{\Sigma} \partial^\gamma \mathscr{P}^-_\Lambda
 -  \partial_\alpha \mathscr{P}^3_{\Sigma} \partial^\gamma \mathscr{P}^3_\Lambda\right)
 \left( \chi_{\dot{1}\gamma} \chi^{\dot{1} \alpha}
  - \mathbb{C}_{\beta \gamma} \mathbb{C}^{\alpha \rho}
  \chi_{\dot{2}\rho} \chi^{\dot{2} \beta}\right)\nonumber\\
  &&
   - \, \frac{1}{6\alpha} \mathfrak{m}^{\Lambda \Sigma}
   \left( j_{2\dot{1}\Lambda} j^{2\dot{1}}_\Sigma + j_{1\dot{2}\Lambda}j^{1\dot{2}}_\Sigma\right)
\end{eqnarray}
We can express derivatives of the moment maps in terms of Killing vectors
thanks to eq.(\ref{anholokill}).
We obtain an $\rm{SU}(2)_L \otimes \rm{SU}(2)_R$ invariant interaction
\begin{equation}
   - \, \frac{1}{12\alpha} \mathfrak{m}^{\Lambda \Sigma}
   \left( j_{2\dot{1}\Lambda} j^{2\dot{1}}_\Sigma
   + j_{1\dot{2}\Lambda}j^{1\dot{2}}_\Sigma  + j_{1\dot{1}\Lambda} j^{1\dot{1}}_\Sigma
   + j_{2\dot{2}\Lambda} j^{2\dot{2}}_\Sigma \right) =
    - \, \frac{1}{12\alpha} j_{A\dot{B}} \cdot j^{A\dot{B}}
\end{equation}
This result was obtained utilizing the following relations
\begin{eqnarray}
   \mathbb{C}_{\alpha \beta} \mathscr{U}^{1\alpha}_{i} \mathscr{U}^{2\beta}_{j}
   =  \mathbb{C}_{\alpha \beta} \mathscr{U}^{1\alpha}_{i^\star} \mathscr{U}^{2\beta}_{j^\star}
   =  \mathbb{C}_{\alpha \beta} \mathscr{U}^{2\alpha}_i \mathscr{U}^{2\beta}_{j}
    = \mathbb{C}_{\alpha \beta} \mathscr{U}^{1\alpha}_{i^\star} \mathscr{U}^{1\beta}_{j^\star}
     =  \mathbb{C}_{\alpha \beta} \mathscr{U}^{2\alpha}_i \mathscr{U}^{2\beta}_{j^\star}
     = \mathbb{C}_{\alpha \beta} \mathscr{U}^{1\alpha}_i \mathscr{U}^{1\beta}_{j^\star} = 0 \nonumber
\end{eqnarray}
Summarizing, we have shown that when the moment map constraints
(\ref{MMcon_triv}-\ref{MMcon_int}) are satisfied the
$\mathscr{N}_3=3$ supersymmetric Chern-Simons theory takes the
following $\mathscr{N}_3=4$ form:
\begin{eqnarray}
  \mathcal{L}^{\mathscr{N}=4}_{CS} &=& - \,\alpha \, \mbox{Tr}
  \,\left( \mathfrak{F} \, \wedge \, \mathcal{A} \,
  + \, \frac{1}{3}\, \mathcal{A}\wedge \, \mathcal{A}\wedge \, \mathcal{A}\right)\, \nonumber\\
  \null && +  \left( \frac{1}{4}\mathbb{C}_{\alpha \beta} \epsilon_{AB}
  \langle \shat{\mathscr{U}}^{A \alpha} \, , \,  \shat{\mathscr{U}}^{B \beta} \rangle + \rm{i}
  \, \chi_{\dot{A} \alpha} \gamma^m \nabla_m \chi^{\dot{A} \alpha} \right. \nonumber\\
&& \left.  - \, \frac{1}{2\alpha} j_{A\dot{B}} \cdot j^{A\dot{B}}
+ \frac{\rm{1}}{96\alpha^2}f^{\Gamma \Sigma \Pi} \mathscr{P}^x_\Gamma
\mathscr{P}^y_{\Sigma} \mathscr{P}^z_\Pi \epsilon_{xyz} \right)\, \wedge \rm{Vol}(\mathscr{M}_3)
\end{eqnarray}
\section{The Calabi HyperK\"ahler manifold $T^\star \mathbb{P}^2$
and the resolution of the conifold
$\mathcal{C}(\mathrm{N^{0,1,0}})$}
\label{conifoldus} The manifolds
$\mathrm{N^{p,q,r}}$:
\begin{equation}\label{corifeo}
\mathrm{N^{p,q,r}} \, = \, \frac{\mathrm{SU(3)\times
U_Y(1)}}{\mathrm{U_I(1)\times U_{II}(1)}}
\end{equation}
were introduced by Castellani and Romans in 1984
\cite{Castellani:1983tc} as $7$-dimensional Einstein manifolds with
Killing spinors, useful in the programme of Kaluza-Klein
supergravity, namely for Freund-Rubin compactifications of D=11
supergravity of the type:
\begin{equation}\label{curante}
    \mathcal{M}_{11} \, = \, \mathrm{AdS_4} \times \left(\frac{\mathrm{G}}{\mathrm{H}}\right)_7
\end{equation}
The manifolds $\mathrm{N^{p,q,r}}$ are defined as follows (see
\cite{castdauriafre}, 2nd vol., sect. V.6.2).
\par
Let $\lambda_\Sigma$ ($\Sigma=1,\dots,8$) be the standard Gell--Mann
matrices\footnote{We recall the explicit expression of the Gell-Mann
matrices since we need them in the sequel. In this way we fix
normalizations.}
\begin{equation}\label{gelmanus}
    \begin{array}{ccccccc}
       \lambda_1 & = & \left(
\begin{array}{ccc}
 0 & 1 & 0 \\
 1 & 0 & 0 \\
 0 & 0 & 0 \\
\end{array}
\right) & ; & \lambda_2 & = & \left(
\begin{array}{ccc}
 0 & -i & 0 \\
 i & 0 & 0 \\
 0 & 0 & 0 \\
\end{array}
\right) \\
       \lambda_3 & = & \left(
\begin{array}{ccc}
 1 & 0 & 0 \\
 0 & -1 & 0 \\
 0 & 0 & 0 \\
\end{array}
\right) & ; & \lambda_4 & = & \left(
\begin{array}{ccc}
 0 & 0 & 1 \\
 0 & 0 & 0 \\
 1 & 0 & 0 \\
\end{array}
\right) \\
       \lambda_5 & = & \left(
\begin{array}{ccc}
 0 & 0 & -i \\
 0 & 0 & 0 \\
 i & 0 & 0 \\
\end{array}
\right) & ; & \lambda_6 & = & \left(
\begin{array}{ccc}
 0 & 0 & 0 \\
 0 & 0 & 1 \\
 0 & 1 & 0 \\
\end{array}
\right) \\
       \lambda_7 & = & \left(
\begin{array}{ccc}
 0 & 0 & 0 \\
 0 & 0 & -i \\
 0 & i & 0 \\
\end{array}
\right) & ; & \lambda_8 & = & \left(
\begin{array}{ccc}
 \frac{1}{\sqrt{3}} & 0 & 0 \\
 0 & \frac{1}{\sqrt{3}} & 0 \\
 0 & 0 & -\frac{2}{\sqrt{3}} \\
\end{array}
\right)
     \end{array}
\end{equation}
The eight generators of the $\mathrm{SU(3)}$ group in the
fundamental defining representation can be chosen as the following
eight anti-hermitian matrices:
\begin{equation}\label{strutturicosti}
    \mathbf{t}_\Sigma \, \equiv \, \ft{i}{2} \, \lambda_\Sigma \quad ; \quad \left [ \mathbf{t}_\Lambda \,
    , \, \mathbf{t}_\Sigma \right ] \, = \, \mathfrak{f}_{\Lambda \Sigma}^{\phantom{IJ}\Delta}\mathbf{t}_\Delta
\end{equation}
The tensor $\mathfrak{f}_{\Lambda \Sigma}^{\phantom{IJ}\Delta}$ encodes the
$\mathrm{SU(3)}$ structure constants. Let moreover $i \,\mathbf{Y}$
denote the generator of the extra group $\mathrm{U_Y(1)}$. The coset
manifold (\ref{corifeo}) is completely determined by specifying the
two generators of the $\mathrm{U_I(1)}$ and $\mathrm{U_{II}(1)}$
factors of the subgroup $\mathrm{H}\subset \mathrm{G}\, = \,
\mathrm{SU(3)}\times \mathrm{U(1)}$. One sets:
\begin{eqnarray}
  \mathbf{h}_{\mathrm{I}} &=& -\frac{2}{\sqrt{3p^2+q^2+2r^3}\,\sqrt{3p^2+q^2}}\,
  \left(\sqrt{3}r\,p\, \mathbf{t}_8
  + r\,q \, \mathbf{t}_3 \, - \, \ft{i}{2}
  \, \left(3 \, p^2 + q^2\right) \, \mathbf{Y} \right) \nonumber \\
  \mathbf{h}_{\mathrm{II}} &=& - \, \frac{1}{\sqrt{3p^2+q^2}} \left( -\, q \,
  \mathbf{t}_8 + \sqrt{3} \,p \, \mathbf{t}_3 \right)
\end{eqnarray}
where $p,q,r$ are coprime integers. As shown in the original paper
and in \cite{castromwar}, the local geometry of the manifolds
(\ref{corifeo}) depends only on the ratio $3p/q$ while the integer
$r$ is related with their fundamental group. When we set $p=0,r=0$
the generator $\mathbf{h}_{\mathrm{I}}$ just becomes $i \mathbf{Y}$,
while the generator $\mathbf{h}_{\mathrm{II}}$ becomes
$\mathbf{t}_8$. Hence we find:
\begin{equation}\label{carnevaleaferragosto}
    \mathrm{N^{0,1,0}} \, \sim \, \frac{\mathrm{SU(3)}}{\mathrm{U_8(1)}}
\end{equation}
where $\mathrm{U_8(1)}$ is generated by $\mathbf{t}_8$.
\par
The manifold $\mathrm{N^{0,1,0}}$ figures in the very short list of
homogeneous Sasaki-Einstein $7$ manifolds which is recalled in table
\ref{sasakiani}. Since the subgroup $\mathrm{U_8(1)}$ has an
$\mathrm{SU(2)}$ normalizer in $\mathrm{SU(3)}$, it follows that we
can also write:
\begin{equation}\label{cominato}
    \mathrm{N^{0,1,0}} \, \sim \,
    \frac{\mathrm{SU(3)}}{\mathrm{U_8(1)}} \, \sim \, \frac{\mathrm{SU(3)
    \times SU(2)}}{\mathrm{SU(2)\times U(1)}}
\end{equation}
showing that $\mathrm{N^{0,1,0}}$ is not only sasakian, rather it is
also tri-sasakian, admitting two kind of fibrations.
\par
In the first fibration $\mathrm{N^{0,1,0}}$ is seen as a
circle-bundle over the flag manifold:
\begin{equation}\label{flag}
    \mathfrak{m}^F_{6} \, \equiv \,\frac{\mathrm{SU(3)}}{\mathrm{U(1)\times U(1)}}
\end{equation}
the group $\mathrm{U(1)\times U(1)}$ being the maximal torus. Namely
we have:
\begin{equation}\label{cavernicolo}
  \frac{ \mathrm{SU(3)}}{\mathrm{U_8(1)}}\, \sim \,  \mathrm{N^{0,1,0}}\,
  \stackrel{\pi}{\longrightarrow} \,
\mathfrak{m}^F_{6}
    \quad ; \quad \forall p\in \mathfrak{m}^F_{6} \, : \, \pi^{-1}(p) \,
    \sim \, \mathrm{U(1)}
\end{equation}
\par
In the second fibration $\mathrm{N^{0,1,0}}$ is seen as an
$\mathbb{S}^3$-fibration over $\mathbb{P}^2$
\begin{equation}\label{cominato}
    \mathrm{N^{0,1,0}} \, \sim \,
    \frac{\mathrm{SU(3)\times SU(2)}}{\SU(2)\times \mathrm{U_8(1)}} \,\,
    \stackrel{\pi}{\longrightarrow} \,
\mathbb{P}^2 \quad ; \quad \forall p\in \mathbb{P}^2 \, : \,
\pi^{-1}(p) \,
    \sim \, \mathrm{SU(2)}
\end{equation}
The peculiarity of M-theory compactification on
\begin{equation}\label{catetere}
    \mathcal{M}_{11} \, = \, \mathrm{AdS_4} \times \mathrm{N^{0,1,0}}
\end{equation}
is that it leads to $\mathcal{N}_4=3$ rather than $\mathcal{N}_4=4$
supersymmetry in $D=4$, as one might expect from the
$\mathrm{SU(2)}$-holonomy of the internal seven-manifold. Indeed,
notwithstanding such holonomy, the differential equation for the
Killing spinors can be integrated only for three, rather than for
four of them \cite{Castellani:1983tc}.
\par
Correspondingly in \cite{Termonia:1999cs},\cite{Fre1999xp} the
complete Kaluza Klein spectrum of M-theory on the background
(\ref{catetere}) was derived and organized into supermultiplets of
the supergroup $\mathrm{Osp(3|4)}$ each supermultiplet being
assigned to a tower of irreducible representations of the isometry
group $\mathrm{SU(3)}$ that determines its mass and charges.
\subsection{The $\mathrm{N^{0,1,0}}$ manifold from the $D=3$ gauge
theory viewpoint} \label{zaffone} As discussed in general terms in
\cite{pappo1} and summarized in \cite{Bruzzo:2017fwj}, whenever we
have an $\mathrm{AdS_4} \times \mathcal{M}_7$ solution of $D=11$
supergravity we can construct its associated $M2$-brane solution
that interpolates between a locally Minkowskian $Mink_{1,10}$ flat
manifold at infinity and the $\mathrm{AdS_4} \times \mathcal{M}_7$
manifold at the brane-horizon $r\to 0$. The general structure of the
$\mathcal{M}_{11}$ metric in the M2-brane solution is of the form:
\begin{equation}\label{scodellarotta}
    ds^2_{M2-brane} \, = \, H^{-2/3}(y) \, ds^2_{Mink_{1,2}}\, + \,
    H^{1/3}(y)\, ds^2_{\mathcal{C}(\mathcal{M}_7)}
\end{equation}
where $ds^2_{Mink_{1,2}}$ is the flat Minkowski metric on the
three-dimensional brane world volume and
\begin{equation}\label{crudelia}
    ds^2_{\mathcal{C}(\mathcal{M}_7)} \, = \, dr^2 + r^2 \, ds^2_{\mathcal{M}_7}
\end{equation}
is the metric of the metric cone $\mathcal{C}(\mathcal{M}_7)$ over
the Einstein $7$-manifold $\mathcal{M}_7$.
\par
Whenever $\mathcal{M}_7$ is sasakian, namely it admits at least two
Killing spinors, the metric cone $\mathcal{C}(\mathcal{M}_7)$ is,
according to an equivalent definition of sasakian manifolds, a
Ricci-flat K\"ahler manifold $K_4$. This does not exclude that $K_4$
might be singular. Indeed, for all sasakian homogeneous
$7$-manifolds different from the round $7$-sphere, $K_4$ has a
singularity at the tip of the cone and therefore is a
\textit{conifold}.
\par
One is therefore interested in \textit{crepant resolutions} of this
conifold singularity and we shall address this problem from the
point of view of the gauge theory living on the brane world-volume.
\par
The lagrangian of $\mathcal{N}_3=3$ Chern-Simons Gauge Theory, as
systematized in \cite{pappo1} within the family of $\mathcal{N}_3=2$
Chern Simons gauge theories, takes the form discussed in section
\ref{cappodue} and presented in eq.s
(\ref{pastiglialeone},\ref{kallettus},\ref{suppotn3},\ref{quartaPot}).
\subsection{The ${\cal N}_3=3$ gauge theory corresponding to the
$\n010$ compactification}\label{theoryofN010} Having clarified the
structure of a generic ${\cal N}_3=3$ gauge theory let us consider,
as an illustration, the specific one associated with the $\n010$
seven--manifold following the presentation of \cite{ringoni}. As
explained above the manifold $\n010$ is the circle bundle inside
$\mathcal{O}(1,1)$ over the flag manifold $\mathfrak{m}^F_6$ (see
eq.s(\ref{flag}-\ref{cavernicolo}). Furthermore as also explained in
\cite{Fabbri:1999hw} (see eq.(B.2)), the base manifold
$\mathfrak{m}^F_6$ can be algebraically described as the following
quadric
\begin{equation}
  \sum_{i=1}^{3} \, u^i \, v_i = 0
\label{vanlocus}
\end{equation}
in $\mathbb{P}^2 \times \mathbb{P}^{2*}$, where $u^i$ and $v_i$
($i=1,2,3$) are the homogeneous coordinates of $\mathbb{P}^2$ and
$\mathbb{P}^{2*}$, respectively.
\par
Hence a complete description of the metric cone $\mathcal{C}\left(
\n010\right) $ can be given by writing the following equations in $
\mathbb{C}^3 \times \mathbb{C}^{3*}$:
\begin{equation}
 \mathcal{C}\left(
 \n010\right)= \left \{ \begin{array}{rcll}
    |u^i|^2-|v_i|^2 & = & 0 & \mbox{fixes equal the radii of $\mathbb{P}^2$
    and $\mathbb{P}^{2*}$}  \\
    2 \, u^i \, v_i & = & 0 & \mbox{cuts out the quadric locus} \\
    \left( u^i \, e^{i\theta} , v_i \, e^{-i\theta}\right)
    &\simeq &\left( u^i,v_i\right) &\mbox{identifies points of $\mathrm{U(1)}$
    orbits}\
  \end{array}\right.
\label{metrcon}
\end{equation}
Eq.s (\ref{metrcon}) can be easily interpreted as the statement that
the cone $K_4 \, = \, {\cal C}\left(\n010\right)$ is the
HyperK\"ahler quotient of a flat three-dimensional quaternionic
space with respect to the triholomorphic action of a $\mathrm{U(1)}$
group. Indeed the first two equations in (\ref{metrcon}) can be
rewritten as the vanishing of the triholomorphic moment map of a
$\mathrm{U(1)}$ group. It suffices to identify:
\begin{eqnarray}
{\cal P}_3 & = & -\left( |u^i|^2 -|v_i|^2 \right)  \nonumber\\
{\cal P}_- & =& 2 v_i u^i \label{agniu1}
\end{eqnarray}
Comparing with eq.s (\ref{quartaPot}) we see that the cone
$\mathcal{C} (\n010)$ can be correctly interpreted as the space of
classical vacua in an abelian ${\cal N}_3=3$ gauge theory with $3$
hypermultiplets in the fundamental representation of a flavor group
$\mathrm{SU(3)}$.
\par
If the color group is $\mathrm{U(1)}$ there is only one value for
the index $\Lambda$. The potential is a positive definite quadratic
form in the moment maps with minimum at zero which is attained when
the moment map vanishes.
\par
Relying on this geometrical picture of the transverse space to an
$M2$--brane living on AdS$_4 \times \n010$, in \cite{ringoni} was conjectured that the ${\cal N}_3=3$ non--abelian gauge theory
whose infrared conformal point is dual to $D=11$ supergravity
compactified on AdS$_4 \times \n010$ should have the following
structure:
\begin{equation}
 \begin{array}{crcl}
   \mbox{gauge group} & \mathcal{ G}_{gauge} & = & \mathrm{SU(N)}_1 \times
   \mathrm{SU(N)}_2 \\
   \null & \null & \null & \null \\
   \mbox{flavor group} & \mathcal{G}_{flavor} & = & \mathrm{SU(3)}  \\
   \null & \null & \null & \null \\
   \mbox{color representations of the hypermultiplets} & \left[ \begin{array}{c}
  u  \\
  v
\end{array} \right] & \Rightarrow & \left[ \begin{array}{c}
  \left({\bf N}_1,{\bf \bar N}_2\right)  \\
  \left({\bf \bar N}_1, {\bf N}_2\right)
\end{array} \right]  \\
\null & \null & \null & \null \\
\mbox{flavor representations of the hypermultiplets} & \left[
\begin{array}{c}
  u  \\
  v
\end{array} \right] & \Rightarrow & \left[ \begin{array}{c}
  {\bf 3}  \\
 {\bf \bar 3}
\end{array} \right]  \\
 \end{array}
\label{assegnati}
\end{equation}
More explicitly and using an ${\cal N}_3=2$ notation we can say that
the field content of the theory proposed in \cite{ringoni} is given
by the following chiral fields, that are all written as $N \times N$
matrices:
\begin{equation}
  \begin{array}{cccc}
    Y_1 & = & \left(Y_1\right)^{\Lambda_1}_{\phantom{\Lambda_1}\Sigma_1 } &
    \mbox{adjoint of $\mathrm{SU(N)}_1$} \\
   Y_2 & = & \left(Y_2\right)^{\Lambda_2}_{\phantom{\Lambda_2}\Sigma_2}  &
    \mbox{adjoint of $\mathrm{SU(N)}_2$ }\\
     u^i & = & \left(u^i\right)^{\Lambda_1}_{\phantom{\Lambda_1}\Sigma_2 } &
    \mbox{in the $({\bf 3},{\bf N}_1,{\bf \bar N}_2)$} \\
    v_i & = & \left(v_i\right)^{\phantom{\Sigma_1}\Lambda_2}_{\Sigma_1}  &
    \mbox{in the $({\bf 3},{\bf \bar N}_1,{\bf N}_2)$} \\
  \end{array}
\label{matricicole}
\end{equation}
and the superpotential before integration on the auxiliary fields
$Y$ can be written as follows:
\begin{equation}
  \mathcal{W} = 2 \, \left [\mbox{Tr} \left( Y_1\, u^i \, v_i \right) +
  \mbox{Tr}\left( Y_2\, v_i \, u_i \right) +  \alpha_1 \mbox{Tr}\left(
  Y_1 \, Y_1 \right ) + \alpha_2 \, \mbox{Tr} \left( Y_2 \, Y_2 \right) \right]
\label{suppotmat}
\end{equation}
where $\alpha_{1,2}$ are the Chern Simons coefficients associated
with the $\mathrm{SU(N)}_{1,2}$ simple gauge groups, respectively.
Setting:
\begin{eqnarray}
\alpha_1 & = & \pm \alpha_2 = \alpha \label{galpha}
\end{eqnarray}
and integrating out the two fields $Y_{1,2}$ that have received a
mass by the Chern Simons mechanism, in \cite{ringoni} it was obtained
the following effective quartic superpotential:
\begin{equation}
  \mathfrak{W}^{eff}= -\ft 1 2 \, \ft {1}{\alpha} \left[
  \mbox{Tr} \left( v_i \, u^i \, v_j \, u^j  \right) \pm
  \mbox{Tr} \left( u^i \, v_i \, u^j \, u_j\right)   \right]
\label{effepot}
\end{equation}
The vanishing relations one can derive from the above superpotential
are the following ones:
\begin{equation}
  u^i \, v_j \, u^j = \pm u^j \, v_j \,  u^i \quad ; \quad
   v_i \, u^j \, v_j = \pm v_j \, u^j \,  v_i
\label{vanrel}
\end{equation}
Consider now the chiral conformal superfields one can write in this
theory:
\begin{equation}
  \Phi^{i_1\, i_2 \, \dots \, i_k}_{j_1 \, j_2 \, \dots \, j_k}
  \equiv
  \mbox{Tr} \left( u^{(i_1} \, v_{(j_1} \, u^{i_2} \, v_{j_2} \, \dots
  \, u^{i)_k} \, v_{j_k)} \right)
\label{chiralop}
\end{equation}
where the round brackets denote symmetrization on the indices. The
above operators have $k$ indices in the fundamental representation
of $\mathrm{SU(3)}$ and $k$ indices in the antifundamental one, but
they are not yet assigned to the irreducible representation:
\begin{equation}
  M_1=M_2 = k
\label{irredu}
\end{equation}
as it is predicted both by general geometric arguments and by the
explicit evaluation of the Kaluza Klein spectrum of hypermultiplets
\cite{Fre1999xp}. To be irreducible the operators (\ref{chiralop})
have to be traceless. This is what is implied by the vanishing
relation (\ref{vanrel}) if we choose the minus sign in
eq.(\ref{galpha}).
\par
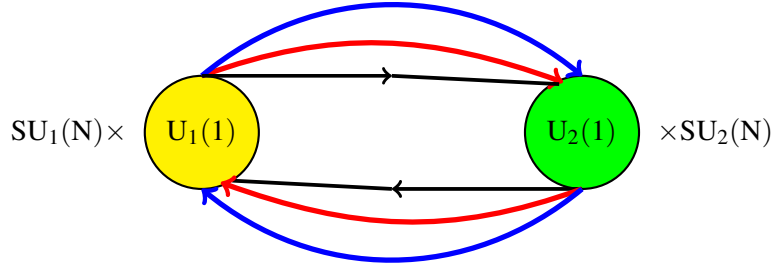
\begin{figure}
\begin{center}
\begin{tikzpicture}[scale=0.50]
\draw [thick] [fill=yellow] (-5,0) circle (1.5cm); \node at (-5,0)
{$\mathrm{U_1(1)}$}; \draw [thick] [fill=green] (5,0) circle
(1.5cm); \node at (5,0) {$\mathrm{U_2(1)}$}; \draw [red, line
width=0.07cm] [->] (-5,1.5) to [out=20,in=160] (4.5,1.33); \draw
[blue,line width=0.07cm] [->] (-5,1.5) to [out=40,in=140] (5,1.5);
\draw [black,line width=0.048cm] [->] (-5,1.5)-- (0,1.5); \draw
[black,line width=0.05cm] (0,1.5)-- (4.4,1.28); \draw [black,line
width=0.05cm] [->] (5,-1.5)-- (0,-1.5); \draw [black,line
width=0.05cm] (0,-1.5)-- (-4.2,-1.28); \draw [red,line width=0.07cm]
[->] (5,-1.5) to [out=-160,in=-20] (-4.5,-1.33); \draw [blue, line
width=0.07cm] [->] (5,-1.5) to [out=-140,in=-40] (-5,-1.5); \node at
(8.5,0) {$\times \mathrm{SU_2(N)}$}; \node at (-8.5,0) {$
\mathrm{SU_1(N)}\times$};
\end{tikzpicture}
\caption{\label{N010quiver} The quiver diagram describing the D=3
gauge theory corresponding to the a stack of M2-branes with
transverse 8-dimensional space provided by the metric cone on the
coset manifold $\mathrm{N^{010}}$ }
\end{center}
\end{figure}

The field content and the structure of this $\mathcal{N}_3=3$ Chern
Simons gauge theory is encoded in the quiver diagram displayed in
fig.\ref{N010quiver}. The similar quiver diagram associated with the
Eguchi Hanson space is pictured in fig.\ref{EHquiver}
\par
In \cite{ringoni} it was noticed that for $\mathrm{N^{0,1,0}}$ the
form of the superpotential, which is dictated by the Chern-Simons
term, is strongly reminiscent of the superpotential considered in
\cite{witkleb}. Indeed, the CFT theory associated with
$\mathrm{N^{0,1,0}}$ has many analogies with the simpler cousin
$\mathrm{T^{1,1}}$ \cite{sergiotorino}. However it was stressed in
\cite{ringoni} that there is also a crucial difference, pertaining
to a general phenomenon that was discussed for the case of
compactifications on $\mathrm{M^{1,1,1}}$ and $\mathrm{Q^{1,1,1}}$
in \cite{Fabbri:1999mk} and \cite{Fabbri:1999hw}. The moduli space
of vacua of the abelian theory is isomorphic to the cone ${\cal
C}\left( \n010\right)$. When the theory is promoted to a non-abelian
one, there are naively conformal operators whose existence is in
contradiction with geometric expectations and with the KK spectrum,
in this case the hypermultiplets that do not satisfy
relation~(\ref{irredu}). Differently from what happens for
$\mathrm{T^{1,1}}$ \cite{witkleb}, the superpotential in eq.
(\ref{effepot}) is not sufficient for eliminating these redundant
non-abelian operators.
\par
Ten years later in a paper by Gaiotto et al \cite{Gaiotto:2009tk},
it was advocated that, maintaining the same flavor-group assignments
and the same color group, the  color representation assignments of
the hypermultiplets that lead to the correct dual CFT are slightly
different from those shown in eq. (\ref{matricicole}) since in
addition to the bi-fundamental representation one needs also the two
fundamental ones.
\par
In any case it is appropriate to stress that, on the basis of the
general form of the $\mathcal{N}_3=3$ gauge theory discussed above
as a particular case of the general $\mathcal{N}_3=2$ theory, it was
just in \cite{ringoni} that the structure of an $\mathcal{N}_3=3,
D=3$ Chern-Simons gauge theory, corner stone of the famous ABJM
model\cite{Aharony:2008ug}, was for the first time derived in the
literature. Indeed in \cite{ringoni} it was just conjectured that
the gauge coupling constant $g$ flows to infinity at the infrared
conformal point, so that the effective lagrangian is obtained from
the general one by letting $\mathbf{e}=\frac{1}{g^2} \to 0$. It was in
\cite{ringoni} that the conversion of the $Y^\Lambda$ field into a
lagrangian multiplier was for the first time observed, leading to
the generation of an effective superpotential of type
(\ref{effepot}).
\par
\subsection{Resolution of the conifold singularity for
$\mathcal{C}(\n010)$ } \label{n010pope1} The shaking news of paper
\cite{Cvetic:2001zb} is that the resolution of the singularity for
the conifold $\mathcal{C}(\n010)$ is provided by a HyperK\"ahler
8-dimensional manifold which is the total space of the cotangent
bundle of $\mathbb{P}^2$, namely:
\begin{equation}\label{caridello}
    \mathcal{M}_8 \, = \, HK^{(2)}_{Calabi} \, \sim \,T^\star \mathbb{P}^2
\end{equation}
The HyperK\"ahler metric on $T^\star \mathbb{P}^2$ is the Calabi
metric which admits, as the authors show, a general representation
for all $T^\star \mathbb{P}^{1+n}$ and this justifies the name
$HK^{(n)}_{Calabi}$ for these HyperK\"ahler manifolds of real
dimensions $4n+4$.
\par
In the present section, following the guide-lines of
\cite{Cvetic:2001zb} we explicitly derive the HyperK\"ahler metric
of $HK^{(2)}_{Calabi}$ as the resolution of the singular conifold
$\mathcal{C}(\n010)$ and we advocate that this resolution just
corresponds, in eq. (\ref{metrcon}), to lifting the real component
of the moment map to a non vanishing level:
\begin{equation}
 HK_{Calabi}^{(1)}\, = \left \{ \begin{array}{rcllll}
   {\cal P}_3 & \equiv & |u^i|^2-|v_i|^2 & = & \kappa \neq 0 &    \\
  {\cal P}_+ & \equiv &  2 \, u^i \, v_i & = & 0 & \\
    \left( u^i \, e^{i\theta} , v_i \, e^{-i\theta}\right)
    &\simeq &\left( u^i,v_i\right) &
  \end{array}\right.
\label{metrcon}
\end{equation}
The Calabi metric  is a generalization of the Eguchi Hanson metric
and it is indeed HyperK\"ahler. What happens is that there are two
routes one can follow to generalize the Eguchi Hanson case:
\begin{equation}\label{stuperspace}
    EH \, = \, \left\{\begin{array}{ccccc}
                        T^\star \mathbb{P}^1 & \stackrel{\mbox{generalization}}{\longrightarrow} & T^\star \mathbb{P}^{n+1} &
                        \mbox{dim}_\mathbb{R}=4n+4 & \text{HyperK\"ahler} \\
                        \null & \null & \null & \null \\
                        O_{\mathbb{P}^1}(-2) & \stackrel{\mbox{generalization}}{\longrightarrow} &
                        O_{\mathbb{P}^{1+n}}(-2-n) &
                        \mbox{dim}_\mathbb{R}=2n+4 & \text{K\"ahler}
                      \end{array}
    \right.
\end{equation}
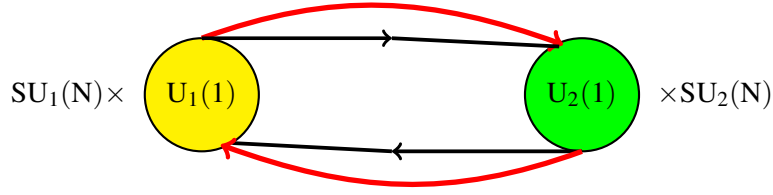
\begin{figure}
\begin{center}
\begin{tikzpicture}[scale=0.50]
\draw [thick] [fill=yellow] (-5,0) circle (1.5cm); \node at (-5,0)
{$\mathrm{U_1(1)}$}; \draw [thick] [fill=green] (5,0) circle
(1.5cm); \node at (5,0) {$\mathrm{U_2(1)}$}; \draw [red, line
width=0.07cm] [->] (-5,1.5) to [out=20,in=160] (4.5,1.33); \draw
[black,line width=0.048cm] [->] (-5,1.5)-- (0,1.5); \draw
[black,line width=0.05cm] (0,1.5)-- (4.4,1.28); \draw [black,line
width=0.05cm] [->] (5,-1.5)-- (0,-1.5); \draw [black,line
width=0.05cm] (0,-1.5)-- (-4.2,-1.28); \draw [red,line width=0.07cm]
[->] (5,-1.5) to [out=-160,in=-20] (-4.5,-1.33);
 \node at
(8.5,0) {$\times \mathrm{SU_2(N)}$}; \node at (-8.5,0) {$
\mathrm{SU_1(N)}\times$};
\end{tikzpicture}
\caption{\label{EHquiver} The quiver diagram associated with the
$\mathbb{C}^2/\mathbb{Z}_2$ Kleinian singularity, whose resolution
is the Eguchi Hanson HyperK\"ahler manifold. In the two nodes, which
correspond to the two irreducible one-dimensional representations of
$\mathbb{Z}_2$ we place the two gauge groups
$\mathrm{SU_{1,2}(N)\times U_{1,2}(1)}$. The scalar multiplets
correspond to the two directed lines going from one to the other
node and are in the bi-fundamental representation of the mentioned
node groups. The HyperK\"ahler quotient is done with respect to the
relative $\mathrm{U(1)}$ group, the overall $\mathrm{U(1)}$ being
the irrelevant barycentric group.}
\end{center}
\end{figure}
The route in the first line of eq.(\ref{stuperspace}) is that
followed by the authors of \cite{Cvetic:2001zb} who indeed
constructed HyperK\"ahler Calabi metrics for all values of $n$
utilizing the tri-Sasaki Einstein manifold
$\frac{\mathrm{SU(n+2)}}{\mathrm{U(n)}}$ as a starting point. The
route in the second line is that followed  in the resolution of
$\mathbb{C}^{n+2}/\mathbb{Z}_{n+2}$ singularities. For
$n=\mathrm{odd}$ the second line exists and it is always K\"ahler
but it can have no comparison with the first line. Instead for
$n=\mathrm{even}$ we might conjecture some relation as in the
Eguchi-Hanson case. This issue will be addressed elsewhere.
\par
Next we derive the mentioned resolution step by step.
\subsection{The resolution via HyperK\"ahler quotient}
\label{dividendus} Here we perform the HyperK\"ahler quotient
$\mathbb{C}^3 \times \mathbb{C}^{\star 3} \slash \, \slash_{HK}
U(1)$. We introduce three complex coordinates $\{ u^i \} \,\,\,
i:1,2,3$ of $\mathbb{C}^3$ and the dual coordinates $\{ v_i \}$ of
$\mathbb{C}^{\star 3}$. We introduce the flat K\"ahler potential
\begin{equation}
  \mathscr{K}_{\mathbb{C}^3 \times \mathbb{C}^{\star 3}} = \overline{u}_i u^i + \overline{v}^i v_i
\end{equation}
We define the tri-holomorphic ${\rm U}(1)$ action, $\{ u^i \, , \,
v_i \} \rightarrow \{ e^{{\rm i} \phi} u^i \, , \, e^{-{\rm i} \phi}
v_i \}$. In order to perform the HyperK\"ahler quotient we identify
points of ${\rm U}(1)$ orbits and we set the ${\rm U}(1)$
tri-holomorphic moment map levels:
\begin{eqnarray}
  \mathscr{P}^3(u,v,\overline{u},\overline{v}) & = & \overline{u}_i u^i \, - \, \overline{v}^i v_i  =  \kappa \nonumber\\
  \mathscr{P}^+(u,v) & = &  u^i v_i  =  0 \nonumber\\
  \{ u^i \, , \, v_i \} &  \simeq  & \{ e^{{\rm i} \phi} u^i \, , \, e^{-{\rm i} \phi} v_i \} \label{algebro}
\end{eqnarray}
This defines the HyperK\"ahler manifold $
\widetilde{\mathscr{C}(N^{0,1,0})}$ which is the resolution of the
conifold $\mathscr{C}(N^{0,1,0})$, $\kappa$ being related to the
resolution parameter.

\subsubsection{Solving the algebraic constraints}
First, we consider the complexification of ${\rm U}(1)$, $e^{i \phi}
\rightarrow e^{-\Phi}$, and we set the following gauge
\begin{eqnarray}
  w^i &=& \frac{u^i}{u^3} = e^{\Phi}u^i \nonumber\\
  z_j &=& u^3 v_j = e^{-\Phi} v_j \quad
\end{eqnarray}
This implies $w^3=1$ while from $\mathscr{P}^+=0$ we obtain
$z^3=-w^a z_a  $ where $ a=1,2$. We can identify $w^a$ with the
inhomogenous coordinates of $\mathbb{P}^2$ and $z_a$ with the fibre
coordinates of $T^\star \mathbb{P}^2$. Now, we find the element
$\Phi$ that lifts the real moment map from $0$ to $\kappa$
\begin{equation}
  \mathscr{P}^3=\kappa \Leftrightarrow \Phi_\pm = -\frac{1}{2}\log{
  \left( \frac{\kappa \pm \sqrt{\kappa^2 + 4R}}{2(1 + \overline{w}_a w^a)} \right)}
\end{equation}
where $R=(1+\overline{w}_a w^a)(\overline{z}^a
z_a + \overline{(w^a z_a)}(w^a z_a))$.\\
This solution defines the immersion $i: T^\star \mathbb{P}^2 \simeq
\widetilde{\mathbb{C}(N^{0,1,0})} \longrightarrow \mathbb{C}^3
\times \mathbb{C}^{\star 3}$.
\subsubsection{The K\"ahler potential and the metric}
The K\"ahler potential of the HyperK\"ahler manifold described by
(\ref{algebro}) has the following form:
\begin{eqnarray}
  \mathscr{K}_{T^\star \mathbb{P}^2} &=&
  i^\star \mathscr{K}_{\mathbb{C}^3 \times \mathbb{C}^{\star 3}} \,- \, \alpha \, \kappa \,
  \Phi_+ \nonumber\\
  &=& \sqrt{4R +\kappa^2} \, + \, \frac{\alpha}{2} \, \kappa \,
  \log{\left( \kappa + \sqrt{\kappa^2 + 4R}\right)} \,
  - \, \frac{\alpha}{2}\,\kappa \,\log{\left( 1 + \overline{w}_a w^a \right)} \nonumber\\
  &\equiv&  F(R) \, - \, \frac{\alpha}{2}\,\kappa \,
  \log{\left( 1 + \overline{w}_a w^a \right)} \label{calogero}
\end{eqnarray}
from which we obtain the K\"ahler metric
$g_{T^\star \mathbb{P}^2} = g_{I J^\star} dq^I \otimes
d\overline{q}^{J^\star} = \partial_I \partial_{J^\star}
\mathscr{K}_{T^\star \mathbb{P}^2} \, dq^I \otimes
d\overline{q}^{J^\star}$, $q^I = \left( w^a , z_a\right)$. \\
The coefficient $\alpha$ has to be fixed. Indeed, the above metric
must be Ricci-flat. First, we compute the determinant of the metric.
Then, we solve $det(g_{I J^\star})=\delta \in \mathbb{R}$
in terms of $\alpha$. \newpage We obtain:
\begin{equation}
  det(g_{I J^\star}) \, = \,
  \frac{1}{4} F'(R) \left(R F''(R)+F'(R)\right) \left(\alpha ^2
  \kappa ^2+8 R^2 F'(R)^2-6 \alpha  \kappa  R F'(R)\right) \, = \, \delta
\end{equation}
From this latter we find
\begin{equation}
 F'(R) \, = \, \frac{\alpha  \kappa +\sqrt{\alpha ^2 \kappa ^2+16 R \sqrt{\delta }}}{4 R}
\end{equation}
while from (\ref{calogero})
\begin{equation}
 F'(R) \, = \, \frac{\alpha  \kappa ^3-\alpha  \kappa ^2 \sqrt{\kappa ^2+4 R}+4
 \alpha  \kappa  R+8 R \sqrt{\kappa ^2+4 R}}{16 R^2+4 \kappa ^2 R}
\end{equation}
So we find the unique solution $\alpha = -2 $ and \, $\delta = 1$.

\subsection{The resolution via Maurer Cartan equations and the Calabi HyperK\"ahler
manifold} \label{mauricartisu3} Let $\mathscr{E}^\Lambda$ be a set of
left-invariant one forms associated with the generators of
$\mathrm{SU(3)}$ normalized as in eq.(\ref{strutturicosti}).
Eventually they will be obtained as in equation (\ref{piretti}) from
the coset representative $\mathbb{L}_{\mathrm{N^{010}}}$ of the
$7$-manifold of interest to us. Yet this is not relevant for the
explicit construction of the Calabi HyperK\"ahler manifolds. What is
relevant is that they satisfy the Maurer Cartan equations of
$\mathrm{SU(3)}$ explicitly written below
\begin{eqnarray}
  0 &=& \mathrm{d}\mathcal{E}^1-\mathcal{E}^2\wedge \mathcal{E}^3-\frac{1}{2} \mathcal{E}^4\wedge
   \mathcal{E}^7+\frac{1}{2} \mathcal{E}^5\wedge \mathcal{E}^6 \nonumber\\
  0 &=& \mathrm{d}\mathcal{E}^2+\mathcal{E}^1\wedge \mathcal{E}^3-\frac{1}{2} \mathcal{E}^4\wedge
   \mathcal{E}^6-\frac{1}{2} \mathcal{E}^5\wedge \mathcal{E}^7 \nonumber\\
  0 &=& \mathrm{d}\mathcal{E}^3-\mathcal{E}^1\wedge \mathcal{E}^2-\frac{1}{2} \mathcal{E}^4\wedge
   \mathcal{E}^5+\frac{1}{2} \mathcal{E}^6\wedge \mathcal{E}^7\nonumber\\
  0 &=& \mathrm{d}\mathcal{E}^4+\frac{1}{2} \mathcal{E}^1\wedge \mathcal{E}^7+\frac{1}{2}
   \mathcal{E}^2\wedge \mathcal{E}^6+\frac{1}{2} \mathcal{E}^3\wedge
   \mathcal{E}^5-\frac{1}{2} \sqrt{3} \mathcal{E}^5\wedge \mathcal{E}^8 \nonumber\\
  0 &=& \mathrm{d}\mathcal{E}^5-\frac{1}{2} \mathcal{E}^1\wedge \mathcal{E}^6+\frac{1}{2}
   \mathcal{E}^2\wedge \mathcal{E}^7-\frac{1}{2} \mathcal{E}^3\wedge
   \mathcal{E}^4+\frac{1}{2} \sqrt{3} \mathcal{E}^4\wedge \mathcal{E}^8 \nonumber\\
  0 &=& \mathrm{d}\mathcal{E}^6+\frac{1}{2} \mathcal{E}^1\wedge \mathcal{E}^5-\frac{1}{2}
   \mathcal{E}^2\wedge \mathcal{E}^4-\frac{1}{2} \mathcal{E}^3\wedge
   \mathcal{E}^7-\frac{1}{2} \sqrt{3} \mathcal{E}^7\wedge \mathcal{E}^8 \nonumber\\
  0 &=& \mathrm{d}\mathcal{E}^7-\frac{1}{2} \mathcal{E}^1\wedge \mathcal{E}^4-\frac{1}{2}
   \mathcal{E}^2\wedge \mathcal{E}^5+\frac{1}{2} \mathcal{E}^3\wedge
   \mathcal{E}^6+\frac{1}{2} \sqrt{3} \mathcal{E}^6\wedge \mathcal{E}^8 \nonumber\\
  0 &=& \mathrm{d}\mathcal{E}^8-\frac{1}{2} \sqrt{3} \mathcal{E}^4\wedge \mathcal{E}^5-\frac{1}{2}
   \sqrt{3} \mathcal{E}^6\wedge \mathcal{E}^7 \label{barile}
\end{eqnarray}
To construct in one stroke both the metric cone and its resolution,
namely the Calabi HyperK\"ahler metric $HK^{(2)}_{Calabi}$, we
introduce an additional coordinate that we name $\tau$ and we
introduce the following vielbein for an 8-dimensional manifold:
\begin{eqnarray}
  V^1 &=& A(\tau) \, \mathcal{E}^1\nonumber\\
  V^2 &=& A(\tau) \, \mathcal{E}^2 \nonumber\\
  V^3 &=& B(\tau)  \, \mathcal{E}^4 \nonumber\\
  V^4 &=& B(\tau)  \, \mathcal{E}^5 \nonumber\\
  V^5 &=& C(\tau) \, \mathcal{E}^6 \nonumber \\
  V^6 &=& C(\tau) \, \mathcal{E}^7 \nonumber\\
  V^7 &=& F(\tau) \, \mathcal{E}^3 \nonumber\\
  V^8 &=& \frac{\mathrm{d}\tau }{\sqrt{1-\frac{\ell ^4}{\tau
  ^4}}}\label{achtbein}
\end{eqnarray}
where $A(\tau),B(\tau),C(\tau),F(\tau)$ are functions of the
variable $\tau$  to be determined  and $\ell$ is a real parameter.
\par The choice (\ref{achtbein}) needs to be properly explained. We
have introduced a different scaling factor $A(\tau)$, $B(\tau)$,
$C(\tau)$, for each of the three doublets of Maurer Cartan forms
that are rotated one into the other by the generator $\mathbf{t}_3$,
as it is evident from the Maurer Cartan equations (\ref{barile}).
This choice respects the $\mathrm{U(1)}$ fibration of
$\mathrm{N^{0,1,0}}$ and it is mandatory. The fourth function
$F(\tau)$ multiplies the one-form $\mathcal{E}^3$ which is a
$\mathrm{U(1)}$ singlet. Hence it is a priori independent.  The
choice of the function of $\tau$ appearing in $V^8$ is not any
limitation of the ansatz, since any other function would amount to a
redefinition of the $\tau$ coordinate. It is just an educated guess
that simplifies the subsequent differential equations.
\par
Given the ansatz (\ref{achtbein}) we could start constructing the
spin connection $\Omega^{IJ}$ and the curvature $2$-form
$\mathfrak{R}^{IJ}$ for generic functions, yet, as the authors of
\cite{Cvetic:2001zb} do, imposing that the final manifold should be
HyperK\"ahler is much more restrictive and determines all the
undetermined functions.
\subsubsection{The three complex structures and the three
HyperK\"ahler forms} The advantage of working in the intrinsic
vielbein basis is that in this frame the three complex structures
$\mathbf{J}^x $ that must satisfy the algebra of quaternion
imaginary units:
\begin{equation}\label{ramificato}
    \mathbf{J}^x \cdot \mathbf{J}^y \, = \, - \, \delta^{xy} \,
    \mathbf{Id}_{8\times 8} \, + \, \epsilon^{xyz} \, \mathbf{J}^y
    \quad ; \quad x,y,z \, = \, 1,2,3
\end{equation}
are constant antisymmetric matrices. An explicit representation
representation of the algebra (\ref{ramificato}) that up to
$\mathrm{SO(8)}$ rotations is unique is provided by the following
matrices:
\begin{equation}\label{JJ1JJ2}
   \mathbf{J}^1 \, = \,  \left(
\begin{array}{cccccccc}
 0 & 0 & 0 & 0 & 0 & 0 & 0 & 1 \\
 0 & 0 & 0 & 0 & 0 & 0 & -1 & 0 \\
 0 & 0 & 0 & 0 & 0 & -1 & 0 & 0 \\
 0 & 0 & 0 & 0 & 1 & 0 & 0 & 0 \\
 0 & 0 & 0 & -1 & 0 & 0 & 0 & 0 \\
 0 & 0 & 1 & 0 & 0 & 0 & 0 & 0 \\
 0 & 1 & 0 & 0 & 0 & 0 & 0 & 0 \\
 -1 & 0 & 0 & 0 & 0 & 0 & 0 & 0 \\
\end{array}
\right) \quad ; \quad \mathbf{J}^2 \, = \, \left(
\begin{array}{cccccccc}
 0 & 0 & 0 & 0 & 0 & 0 & 1 & 0 \\
 0 & 0 & 0 & 0 & 0 & 0 & 0 & 1 \\
 0 & 0 & 0 & 0 & -1 & 0 & 0 & 0 \\
 0 & 0 & 0 & 0 & 0 & -1 & 0 & 0 \\
 0 & 0 & 1 & 0 & 0 & 0 & 0 & 0 \\
 0 & 0 & 0 & 1 & 0 & 0 & 0 & 0 \\
 -1 & 0 & 0 & 0 & 0 & 0 & 0 & 0 \\
 0 & -1 & 0 & 0 & 0 & 0 & 0 & 0 \\
\end{array}
\right)
\end{equation}
\begin{equation}\label{JJ3}
    \mathbf{J}^3 \, = \, \left(
\begin{array}{cccccccc}
 0 & -1 & 0 & 0 & 0 & 0 & 0 & 0 \\
 1 & 0 & 0 & 0 & 0 & 0 & 0 & 0 \\
 0 & 0 & 0 & -1 & 0 & 0 & 0 & 0 \\
 0 & 0 & 1 & 0 & 0 & 0 & 0 & 0 \\
 0 & 0 & 0 & 0 & 0 & 1 & 0 & 0 \\
 0 & 0 & 0 & 0 & -1 & 0 & 0 & 0 \\
 0 & 0 & 0 & 0 & 0 & 0 & 0 & 1 \\
 0 & 0 & 0 & 0 & 0 & 0 & -1 & 0 \\
\end{array}
\right)
\end{equation}
Correspondingly, by setting:
\begin{equation}\label{caleriformi}
    \mathbf{K}^x \, = \, \mathbf{J}^x_{IJ}\, V^I \,\wedge\, V^J
\end{equation}
we find the following three candidate HyperK\"ahler forms:
\begin{eqnarray}
  \mathbf{K}^1&=& 2 (V^1\wedge V^8-V^2\wedge V^7-V^3\wedge V^6+V^4\wedge V^5) \nonumber\\
  \mathbf{K}^2 &=& 2 (V^1\wedge V^7+V^2\wedge V^8-V^3\wedge V^5-V^4\wedge V^6) \nonumber\\
  \mathbf{K}^3  &=& -2 (V^1\wedge V^2+V^3\wedge V^4-V^5\wedge V^6-V^7\wedge
  V^8) \label{forminedisabbia}
\end{eqnarray}
In order to define a bona-fide HyperK\"ahler structure the above
triplet of $2$-forms must be closed. Imposing
\begin{equation}\label{rapitallo}
    \mathrm{d}\mathbf{K}^1\, = \,\mathrm{d}\mathbf{K}^2 \, = \,
    \mathrm{d}\mathbf{K}^3 \, = \, 0
\end{equation}
inserting the ansatz (\ref{achtbein}) and utilizing the Maurer-Cartan
equations (\ref{barile}) one obtains a collection of first
order differential and algebraic  constraints on the four functions
$A(\tau)$, $B(\tau)$, $C(\tau)$, $F(\tau)$ which has a unique,
easily retrievable solution:
\begin{equation}\label{solvente}
    A(\tau ) \, = \, \frac{\tau }{2},\quad B(\tau ) \, = \, \frac{\sqrt{\tau ^2+\ell ^2}}{2
   \sqrt{2}},\quad C(\tau ) \, = \, \frac{\sqrt{\tau ^2-\ell ^2}}{2 \sqrt{2}},\quad F(\tau ) \, = \,
   \frac{\sqrt{\tau ^4-\ell ^4}}{2 \tau }
\end{equation}
As one sees in the case $\ell \, = \, 0$ all the functions
degenerate in a coefficient times $\tau$. This means that the
corresponding metric line element is the metric cone over
$\mathrm{N^{0,1,0}}$. As we know such a manifold is singular. For
$\ell \neq 0$ we have instead the Calabi HyperK\"ahler metric which
is a smooth HyperK\"ahler manifold and corresponds to the resolution
of the conifold singularity. Obviously these are the same manifold
and the same metric as the manifold and the metric obtained in
section \ref{dividendus} by means of K\"ahler quotient. A precise
correspondence requires an identification between the coordinates
$w^a,z_a$ utilized there and the $7$ coordinates used for the
$\n010$ coset manifold plus the radial like coordinate $\tau$. We
did not dwell, at this level, on this cumbersome and boring
exercise. The precise identification of coordinates will also
provide the precise relation between the level parameter $\kappa$
and the resolution parameter $\ell$ appearing in the present
discussion.
\subsubsection{Spin connection and curvature} Having fixed all the
functions we can calculate the spin connection and the curvature of
the Calabi HyperK\"ahler manifold. From the torsion equation:
\begin{equation}\label{rasputella}
    \mathrm{d}V^I \, + \, \Omega^{IJ} \wedge V^J \, = \, 0
\end{equation}
we obtain a unique solution, as it is always the case, encoded in
the following one-form valued $8\times 8$ matrix:
{\fontsize{8.6}{8.2}
\begin{eqnarray}\label{larussa}
& \Omega^{IJ}\,  = \, & \nonumber\\
 &   \left(
\begin{array}{cccccccc}
 0 & \frac{\left(\ell ^4+\tau ^4\right) V_7}{\tau ^3 \sqrt{\tau ^4-\ell ^4}} &
   \frac{\sqrt{\frac{\ell ^2+\tau ^2}{\tau ^2-\ell ^2}} V_6}{\tau } &
   -\frac{\sqrt{\frac{\ell ^2+\tau ^2}{\tau ^2-\ell ^2}} V_5}{\tau } &
   \frac{\sqrt{1-\frac{2 \ell ^2}{\ell ^2+\tau ^2}} V_4}{\tau } &
   -\frac{\sqrt{1-\frac{2 \ell ^2}{\ell ^2+\tau ^2}} V_3}{\tau } &
   -\frac{\sqrt{\tau ^4-\ell ^4} V_2}{\tau ^3} & \frac{\sqrt{\tau ^4-\ell ^4}
   V_1}{\tau ^3} \\
 -\frac{\left(\ell ^4+\tau ^4\right) V_7}{\tau ^3 \sqrt{\tau ^4-\ell ^4}} & 0 &
   \frac{\sqrt{\frac{\ell ^2+\tau ^2}{\tau ^2-\ell ^2}} V_5}{\tau } &
   \frac{\sqrt{\frac{\ell ^2+\tau ^2}{\tau ^2-\ell ^2}} V_6}{\tau } &
   -\frac{\sqrt{1-\frac{2 \ell ^2}{\ell ^2+\tau ^2}} V_3}{\tau } &
   -\frac{\sqrt{1-\frac{2 \ell ^2}{\ell ^2+\tau ^2}} V_4}{\tau } &
   \frac{\sqrt{\tau ^4-\ell ^4} V_1}{\tau ^3} & \frac{\sqrt{\tau ^4-\ell ^4}
   V_2}{\tau ^3} \\
 -\frac{\sqrt{\frac{\ell ^2+\tau ^2}{\tau ^2-\ell ^2}} V_6}{\tau } &
   -\frac{\sqrt{\frac{\ell ^2+\tau ^2}{\tau ^2-\ell ^2}} V_5}{\tau } & 0 &
   \frac{V_7 \ell ^2}{\tau  \sqrt{\tau ^4-\ell ^4}}+\frac{\sqrt{3}
   \mathcal{E}_8}{2} & 0 & 0 & -\frac{\sqrt{\tau ^4-\ell ^4} V_4}{\tau ^3+\ell ^2
   \tau } & \frac{\sqrt{\tau ^4-\ell ^4} V_3}{\tau ^3+\ell ^2 \tau } \\
 \frac{\sqrt{\frac{\ell ^2+\tau ^2}{\tau ^2-\ell ^2}} V_5}{\tau } &
   -\frac{\sqrt{\frac{\ell ^2+\tau ^2}{\tau ^2-\ell ^2}} V_6}{\tau } & -\frac{V_7
   \ell ^2}{\tau  \sqrt{\tau ^4-\ell ^4}}-\frac{\sqrt{3} \mathcal{E}_8}{2} & 0 &
   0 & 0 & \frac{\sqrt{\tau ^4-\ell ^4} V_3}{\tau ^3+\ell ^2 \tau } &
   \frac{\sqrt{\tau ^4-\ell ^4} V_4}{\tau ^3+\ell ^2 \tau } \\
 -\frac{\sqrt{1-\frac{2 \ell ^2}{\ell ^2+\tau ^2}} V_4}{\tau } &
   \frac{\sqrt{1-\frac{2 \ell ^2}{\ell ^2+\tau ^2}} V_3}{\tau } & 0 & 0 & 0 &
   \frac{V_7 \ell ^2}{\tau  \sqrt{\tau ^4-\ell ^4}}+\frac{\sqrt{3}
   \mathcal{E}_8}{2} & \frac{\left(\ell ^2+\tau ^2\right) V_6}{\tau  \sqrt{\tau
   ^4-\ell ^4}} & \frac{\sqrt{1-\frac{\ell ^4}{\tau ^4}} \tau  V_5}{\tau ^2-\ell
   ^2} \\
 \frac{\sqrt{1-\frac{2 \ell ^2}{\ell ^2+\tau ^2}} V_3}{\tau } &
   \frac{\sqrt{1-\frac{2 \ell ^2}{\ell ^2+\tau ^2}} V_4}{\tau } & 0 & 0 &
   -\frac{V_7 \ell ^2}{\tau  \sqrt{\tau ^4-\ell ^4}}-\frac{\sqrt{3}
   \mathcal{E}_8}{2} & 0 & -\frac{\left(\ell ^2+\tau ^2\right) V_5}{\tau
   \sqrt{\tau ^4-\ell ^4}} & \frac{\sqrt{1-\frac{\ell ^4}{\tau ^4}} \tau
   V_6}{\tau ^2-\ell ^2} \\
 \frac{\sqrt{\tau ^4-\ell ^4} V_2}{\tau ^3} & -\frac{\sqrt{\tau ^4-\ell ^4}
   V_1}{\tau ^3} & \frac{\sqrt{\tau ^4-\ell ^4} V_4}{\tau ^3+\ell ^2 \tau } &
   -\frac{\sqrt{\tau ^4-\ell ^4} V_3}{\tau ^3+\ell ^2 \tau } & -\frac{\left(\ell
   ^2+\tau ^2\right) V_6}{\tau  \sqrt{\tau ^4-\ell ^4}} & \frac{\left(\ell
   ^2+\tau ^2\right) V_5}{\tau  \sqrt{\tau ^4-\ell ^4}} & 0 & \frac{\left(\ell
   ^4+\tau ^4\right) V_7}{\tau ^3 \sqrt{\tau ^4-\ell ^4}} \\
 -\frac{\sqrt{\tau ^4-\ell ^4} V_1}{\tau ^3} & -\frac{\sqrt{\tau ^4-\ell ^4}
   V_2}{\tau ^3} & -\frac{\sqrt{\tau ^4-\ell ^4} V_3}{\tau ^3+\ell ^2 \tau } &
   -\frac{\sqrt{\tau ^4-\ell ^4} V_4}{\tau ^3+\ell ^2 \tau } & -\frac{\left(\ell
   ^2+\tau ^2\right) V_5}{\tau  \sqrt{\tau ^4-\ell ^4}} & -\frac{\left(\ell
   ^2+\tau ^2\right) V_6}{\tau  \sqrt{\tau ^4-\ell ^4}} & -\frac{\left(\ell
   ^4+\tau ^4\right) V_7}{\tau ^3 \sqrt{\tau ^4-\ell ^4}} & 0 \\
\end{array}
\right) & \nonumber\\
\end{eqnarray}
} Next we calculate the curvature $2$-form: {\fontsize{9.6}{9.2}
\begin{equation}\label{lobotomista}
  \mathfrak{R}^{IJ} \, = \,   d\Omega^{IJ} \, + \, \Omega^{IK} \, \wedge \, \Omega^{KJ}
\end{equation}
and for it we get the following explicit rather simple form:
\begin{equation}\label{curvatura}
\begin{array}{rcl}
 \mathfrak{R}^{1,2} & = & \frac{2 \ell ^2 \left(\tau ^2 \left(V^3 \wedge V^4+V^5 \wedge V^6\right)+2
   \left(V^1 \wedge V^2+V^7 \wedge V^8\right) \ell ^2\right)}{\tau ^6} \\
 \mathfrak{R}^{1,3} & = & \frac{\left(V^1 \wedge V^3+V^2 \wedge V^4+V^5 \wedge V^7+V^6 \wedge V^8\right) \ell ^2}{\tau
   ^4} \\
 \mathfrak{R}^{1,4} & = & \frac{\left(-V^2 \wedge V^3+V^1 \wedge V^4+V^6 \wedge V^7-V^5 \wedge V^8\right) \ell ^2}{\tau
   ^4} \\
 \mathfrak{R}^{1,5} & = & \frac{\left(-V^1 \wedge V^5+V^2 \wedge V^6+V^3 \wedge V^7-V^4 \wedge V^8\right) \ell ^2}{\tau
   ^4} \\
 \mathfrak{R}^{1,6} & = & \frac{\left(-V^2 \wedge V^5-V^1 \wedge V^6+V^4 \wedge V^7+V^3 \wedge V^8\right) \ell ^2}{\tau
   ^4} \\
 \mathfrak{R}^{1,7} & = & \frac{2 \left(V^2 \wedge V^8-V^1 \wedge V^7\right) \ell ^4}{\tau ^6} \\
 \mathfrak{R}^{1,8} & = & -\frac{2 \left(V^2 \wedge V^7+V^1 \wedge V^8\right) \ell ^4}{\tau ^6} \\
 \mathfrak{R}^{2,3} & = & \frac{\left(V^2 \wedge V^3-V^1 \wedge V^4-V^6 \wedge V^7+V^5 \wedge V^8\right) \ell ^2}{\tau
   ^4} \\
 \mathfrak{R}^{2,4} & = & \frac{\left(V^1 \wedge V^3+V^2 \wedge V^4+V^5 \wedge V^7+V^6 \wedge V^8\right) \ell ^2}{\tau
   ^4} \\
 \mathfrak{R}^{2,5} & = & \frac{\left(-V^2 \wedge V^5-V^1 \wedge V^6+V^4 \wedge V^7+V^3 \wedge V^8\right) \ell ^2}{\tau
   ^4} \\
 \mathfrak{R}^{2,6} & = & \frac{\left(V^1 \wedge V^5-V^2 \wedge V^6-V^3 \wedge V^7+V^4 \wedge V^8\right) \ell ^2}{\tau
   ^4} \\
 \mathfrak{R}^{2,7} & = & -\frac{2 \left(V^2 \wedge V^7+V^1 \wedge V^8\right) \ell ^4}{\tau ^6} \\
 \mathfrak{R}^{2,8} & = & \frac{2 \left(V^1 \wedge V^7-V^2 \wedge V^8\right) \ell ^4}{\tau ^6} \\
 \mathfrak{R}^{3,4} & = & \frac{2 \left(2 \tau ^2 \left(V^3 \wedge V^4+V^5 \wedge V^6\right)+\left(V^1
   \wedge V^2+V^7 \wedge V^8\right) \ell ^2\right)}{\tau ^4} \\
 \mathfrak{R}^{3,5} & = & -\frac{2 \left(V^3 \wedge V^5-V^4 \wedge V^6\right)}{\tau ^2} \\
 \mathfrak{R}^{3,6} & = & -\frac{2 \left(V^4 \wedge V^5+V^3 \wedge V^6\right)}{\tau ^2} \\
 \mathfrak{R}^{3,7} & = & \frac{\left(V^1 \wedge V^5-V^2 \wedge V^6-V^3 \wedge V^7+V^4 \wedge V^8\right) \ell ^2}{\tau
   ^4} \\
 \mathfrak{R}^{3,8} & = & \frac{\left(V^2 \wedge V^5+V^1 \wedge V^6-V^4 \wedge V^7-V^3 \wedge V^8\right) \ell ^2}{\tau
   ^4} \\
 \mathfrak{R}^{4,5} & = & -\frac{2 \left(V^4 \wedge V^5+V^3 \wedge V^6\right)}{\tau ^2} \\
 \mathfrak{R}^{4,6} & = & \frac{2 \left(V^3 \wedge V^5-V^4 \wedge V^6\right)}{\tau ^2} \\
 \mathfrak{R}^{4,7} & = & \frac{\left(V^2 \wedge V^5+V^1 \wedge V^6-V^4 \wedge V^7-V^3 \wedge V^8\right) \ell ^2}{\tau
   ^4} \\
 \mathfrak{R}^{4,8} & = & \frac{\left(-V^1 \wedge V^5+V^2 \wedge V^6+V^3 \wedge V^7-V^4 \wedge V^8\right) \ell ^2}{\tau
   ^4} \\
 \mathfrak{R}^{5,6} & = & \frac{2 \left(2 \tau ^2 \left(V^3 \wedge V^4+V^5 \wedge V^6\right)+\left(V^1
   \wedge V^2+V^7 \wedge V^8\right) \ell ^2\right)}{\tau ^4} \\
 \mathfrak{R}^{5,7} & = & \frac{\left(V^1 \wedge V^3+V^2 \wedge V^4+V^5 \wedge V^7+V^6 \wedge V^8\right) \ell ^2}{\tau
   ^4} \\
 \mathfrak{R}^{5,8} & = & \frac{\left(V^2 \wedge V^3-V^1 \wedge V^4-V^6 \wedge V^7+V^5 \wedge V^8\right) \ell ^2}{\tau
   ^4} \\
 \mathfrak{R}^{6,7} & = & \frac{\left(-V^2 \wedge V^3+V^1 \wedge V^4+V^6 \wedge V^7-V^5 \wedge V^8\right) \ell ^2}{\tau
   ^4} \\
 \mathfrak{R}^{6,8} & = & \frac{\left(V^1 \wedge V^3+V^2 \wedge V^4+V^5 \wedge V^7+V^6 \wedge V^8\right) \ell ^2}{\tau
   ^4} \\
 \mathfrak{R}^{7,8} & = & \frac{2 \ell ^2 \left(\tau ^2 \left(V^3 \wedge V^4+V^5 \wedge V^6\right)+2
   \left(V^1 \wedge V^2+V^7 \wedge V^8\right) \ell ^2\right)}{\tau ^6} \\
\end{array}
\end{equation}
} From eq.(\ref{curvatura}) we easily extract the components of the
Riemann tensor:
\begin{equation}\label{Riemanntensor}
    \mathfrak{R}^{IJ} \, = \, \mathrm{Rie}^{IJ}_{\phantom{IJ}KL} \,
    V^K \wedge V^L
\end{equation}
and calculating the Ricci tensor we find that it duely vanishes:
\begin{equation}\label{Riccitensor}
    \mathrm{Ricci}^I_K \, \equiv
    \,\mathrm{Rie}^{IL}_{\phantom{IJ}KL} \, = \, 0
\end{equation}
The above result inserted into eq. (\ref{2.65}), together with eq.
(\ref{parascopio}) yields the explicit form of the $\mathrm{USp(4)}$
curvature two-form which is the following one:
\begin{equation}\label{baldus}
\begin{array}{rcl}
 \mathbb{R}^{1,1} & = & \frac{2 \ell ^4 \mathbf{e}^{1}\wedge
   \bar{\mathbf{e}}_{2}}{\tau ^6} \\
 \mathbb{R}^{1,2} & = & \frac{\ell ^2 \left(\tau ^2 (\mathbf{e}^{4}\wedge
   \bar{\mathbf{e}}_{4}-\mathbf{e}^{3}\wedge \bar{\mathbf{e}}_{3})+2 \ell ^2
   \mathbf{e}^{1}\wedge \bar{\mathbf{e}}_{1}-2 \ell ^2 \mathbf{e}^{2}\wedge
   \bar{\mathbf{e}}_{2}\right)}{\tau ^6} \\
 \mathbb{R}^{1,3} & = & -\frac{\ell ^2 (\mathbf{e}^{1}\wedge
   \bar{\mathbf{e}}_{4}+\mathbf{e}^{3}\wedge \bar{\mathbf{e}}_{2})}{\tau ^4}
   \\
 \mathbb{R}^{1,4} & = & \frac{\ell ^2 (\mathbf{e}^{4}\wedge
   \bar{\mathbf{e}}_{2}-\mathbf{e}^{1}\wedge \bar{\mathbf{e}}_{3})}{\tau ^4}
   \\
 \mathbb{R}^{2,2} & = & -\frac{2 \ell ^4 \mathbf{e}^{2}\wedge
   \bar{\mathbf{e}}_{1}}{\tau ^6} \\
 \mathbb{R}^{2,3} & = & \frac{\ell ^2 (\mathbf{e}^{2}\wedge
   \bar{\mathbf{e}}_{4}-\mathbf{e}^{3}\wedge \bar{\mathbf{e}}_{1})}{\tau ^4}
   \\
 \mathbb{R}^{2,4} & = & \frac{\ell ^2 (\mathbf{e}^{2}\wedge
   \bar{\mathbf{e}}_{3}+\mathbf{e}^{4}\wedge \bar{\mathbf{e}}_{1})}{\tau ^4}
   \\
 \mathbb{R}^{3,3} & = & \frac{2 \mathbf{e}^{3}\wedge \bar{\mathbf{e}}_{4}}{\tau
   ^2} \\
 \mathbb{R}^{3,4} & = & \frac{2 \tau ^2 (\mathbf{e}^{3}\wedge
   \bar{\mathbf{e}}_{3}-\mathbf{e}^{4}\wedge \bar{\mathbf{e}}_{4})+\ell ^2
   (-(\mathbf{e}^{1}\wedge \bar{\mathbf{e}}_{1}))+\ell ^2 \mathbf{e}^{2}\wedge
   \bar{\mathbf{e}}_{2}}{\tau ^4} \\
 \mathbb{R}^{4,4} & = & -\frac{2 \mathbf{e}^{4}\wedge \bar{\mathbf{e}}_{3}}{\tau
   ^2} \\
\end{array}
\end{equation}
\section{The self-dual closed $\Omega^{2,2}$-form on the Calabi
HyperK\"ahler manifold  $T^\star \mathbb{P}^2$ and the associated
deformed M2-Brane solution}\label{omma22} For the reasons specified
in the introduction  we want to find $\Omega^{2,2} \in
\bigwedge^{2,2}\left(T^\star HK^{(2)}_{Calabi}\right)$ such that
\begin{eqnarray}
  \star \Omega^{2,2} &=& \Omega^{2,2} \label{self} \\
  d \Omega^{2,2} &=& 0 \label{closed}
\end{eqnarray}
It is convenient to work in the coset frame where we can decompose
$\Omega^{2,2}$ along the real vielbein, $\left\{ V^I \right\}
I:1,...,8$. Using the complex structure $\mathbf{J}^3$ one can
always go back to the complex vielbein, $\left\{ \mathbf{e}^\alpha ,
\bar{\mathbf{e}}_{\beta} \right\}$, $\alpha,\beta=1,\dots,4$, which
are its eigenstates with eigenvalues $\{ \rm{ i , -i} \}$,
respectively. We find 21 $(2,2)$-self-dual independent basis
elements $\left\{ \mathfrak{S}^\alpha \right\}$. A generic solution
of (\ref{self}) is: $\Omega^{2,2} = \gamma_p \mathfrak{S}^p$. We
choose $\gamma_p = \gamma_p (\tau)$. Now we solve \footnote{This
step involves the torsionless equation $dV^I=-\Omega^I_J \wedge
V^J$. One gets some equations along $\mathscr{E}^8$ which is outside
the coset $\frac{\rm{SU}(3)}{\rm{U}(1)}$. These latter are 14
independent algebraic equations for $\{ \gamma_\alpha\}$.} the
differential equation (\ref{closed}). We get a 4 parameter solution.
We can use this solution to deform the M2-Brane $D=11$ Supergravity
background
\begin{eqnarray}
  ds_{11}^2 & = & H(\tau)^{-\frac{2}{3}} ds^2_{Mink_{1,2}} + H(\tau)^{\frac{1}{3}}
  ds^2_{HK^{(2)}_{Calabi}} \nonumber\\
  \mathbf{A}^{\left[ 3 \right]} &=& H(\tau)^{-1} {\rm Vol}_{Mink_{1,2}} \nonumber\\
  \mathbf{F}^{[ 4 ]} & = & d \mathbf{A}^{[3]} + \Omega^{2,2} \nonumber\\
  \Box H(\tau) & = & \star (\Omega^{2,2} \wedge \Omega^{2,2}) \label{brane}
\end{eqnarray}
To make this deformation consistent we impose $L^2(\ell,+
\infty)$-integrability\footnote{In this section we are searching for
a self-dual $(2,2)$-form. One could search for an antiself-dual
$(2,2)$-form $\mathfrak{w}$, the $L^2$-integrability condition will
imply $\mathfrak{w}=0$.} and reality for the source
$\star(\Omega^{2,2} \wedge \Omega^{2,2})$. Up to an overall constant
$\mathfrak{c}$ we find the following unique solution:
\begin{eqnarray}
  \Omega^{2,2} &=& -\frac{\mathfrak{c}}{2 \ell^2 \tau ^4 \left(\ell^2+\tau ^2\right)}\left( \mathbf{e}^1
  \wedge \mathbf{e}^2 \wedge \bar{\mathbf{e}}_1 \wedge \bar{\mathbf{e}}_2
  + \mathbf{e}^3 \wedge \mathbf{e}^4 \wedge \bar{\mathbf{e}}_3
  \wedge \bar{\mathbf{e}}_4 \right. \nonumber\\
   && \left. - \mathbf{e}^2 \wedge \mathbf{e}^4 \wedge \bar{\mathbf{e}}_2
   \wedge \bar{\mathbf{e}}_4 - \mathbf{e}^1 \wedge \mathbf{e}^3
   \wedge \bar{\mathbf{e}}_1 \wedge \bar{\mathbf{e}}_3\right)  \nonumber\\
         && +  \frac{\mathfrak{c}}{\tau ^2 \left(\ell^2+\tau ^2\right)^3}\left( \mathbf{e}^1
         \wedge \mathbf{e}^4 \wedge \bar{\mathbf{e}}_1 \wedge \bar{\mathbf{e}}_4
         + \mathbf{e}^1 \wedge \mathbf{e}^4 \wedge \bar{\mathbf{e}}_2
         \wedge \bar{\mathbf{e}}_3 \right. \nonumber\\
  && \left. + \mathbf{e}^2 \wedge \mathbf{e}^3 \wedge \bar{\mathbf{e}}_1
  \wedge \bar{\mathbf{e}}_4 + \mathbf{e}^2 \wedge \mathbf{e}^3
  \wedge \bar{\mathbf{e}}_2 \wedge \bar{\mathbf{e}}_3 \right) \nonumber\\
  \star(\Omega^{2,2} \wedge \Omega^{2,2}) &=& \frac{\mathfrak{c}^2 \left(\ell^8+6 \ell^6 \tau ^2
  +16 \ell^4 \tau ^4
  +6 \ell^2 \tau ^6+\tau ^8\right)}{\ell^4 \tau ^8 \left(\ell^2
  +\tau ^2\right)^6} \label{source}
\end{eqnarray}
Plugging (\ref{source}) in (\ref{brane}) we obtain a solution
involving only a new integration constant, namely the value of the
inhomogeneous harmonic function at infinity $H_\infty$:
\begin{equation}
  H(\tau)=\frac{\mathfrak{c}^2 \left(5 \tau ^6+40 \ell ^6+48 \tau ^2 \ell ^4
  +25 \tau ^4 \ell ^2\right)}{320 \tau ^2 \ell ^8 \left(\tau ^2+\ell ^2\right)^5}+H_\infty
\end{equation}
In this way, as already done in \cite{Cvetic:2001zb}, we have shown that there exists an exact M2-brane
solution of $D=11$ supergravity with the Calabi HyperK\"ahler
manfiold $HK^{(2)}_{Calabi}$ as transverse space and a self dual
flux of the $4$-form. For large values of $\tau$ with respect to the
resolution parameter $\ell$ the transverse space metric reduces to
the metric cone on the tri-sasakian manifold $\n010$.
\section{Conclusions}
\label{concluddo} In this paper, as we explained in the
introduction, we have generalized in a systematic way to curved
HyperK\"ahler manifolds the Gaiotto-Witten type of lagrangian for
$\mathcal{N}_3 \geq 4$ Chern-Simons gauge theories in $D=3$. The
enhancement conditions are fully geometrical and are encoded in the
weaker constraints (\ref{MMcon_triv}-\ref{MMcon_int}) to be
satisfied by the tri-holomorphic moment maps of the gauged isometries.
\par
In the perspective of the gauge/gravity correspondence, the
supersymmetric Chern Simons gauge theory is supposed to live on the
boundary of an asymptotic $\mathrm{AdS_4}$ manifold and a
challenging opportunity emerges since an infinite series of
HyperK\"ahler metrics satisfying the enhancement constraints are the
$HK^{n}_{Calabi}$ constructed on the total space of the cotangent
bundles $T^\star \mathbb{P}^{n}$ where the cases $n=1$ and $n=2$
respectively correspond to the Eguchi-Hanson gravitational instanton
and to the smooth resolution of the metric cone
$\mathcal{C}(\n010)$. The second example is the most interesting one
and provides the inspiration for further inquiries and developments
that we presently list:
\begin{description}
  \item[a)] The compactification of $D=11$ supergravity on
  $\mathrm{AdS_4}\times \n010$ is the unique one on a Sasakian
  homogeneous 7-manifold that yields $\mathcal{N}_4=3$ supersymmetry
  in $D=4$. In view of the discovered enhancement to
  $\mathcal{N}_3=4$ of the dual Chern-Simons theory on the
  $\mathrm{AdS_4}$-boundary when the flavor group $\mathrm{SU(3)}$
  is gauged, we would like to study the $\mathrm{AdS_4}\times \n010$
  vacuum in terms of an appropriate gauging of
  $\mathcal{N}_4=3,D=4$ supergravity.
  \item[b)] Utilizing the supergravity potential provided by the
  above mentioned gauging it would be interesting to find its moduli
  and deformations, looking for other extrema of the potential.
  \item[c)] As shown in this paper the Chern Simons theory on the
  $\mathrm{AdS_4}$ boundary where we gauge both the color group
  $\mathrm{U(N)}$ and the flavor group $\mathrm{SU(3)}$ is enhanced
  to $\mathcal{N}_3=4$ supersymmetry and admits a dual description
  in terms of a gauged-fixed supergroup Chern Simons theory with supergroup
  $\mathrm{SU(3|N)}$. Integrating out the color degrees of freedom
  we obtain a flavor Chern Simons gauge theory with gauge group
  $\mathrm{SU(3)}$ and target manifold
  $HK^{2}_{Calabi}$ which still preserves $\mathcal{N}_3=4$. It
  would be interesting to describe the same theory in the dual
  supergroup formulation.
  \item[d)] Since $HK^{2}_{Calabi}$ admits a self-dual (2,2)
  harmonic  form we can consider M2-brane solutions with self-dual
  internal fluxes. It would be interesting, in the framework of the gauge/gravity
  correspondence to retrieve the role of this flux in the
  Chern-Simons gauge theory on the boundary and to explore all the
  relations between the supergroup formulation, the $D=4$
  supergravity approach and the $D=11$ supergravity M2-brane
  solution.
\end{description}
We plan to investigate such multi-faceted questions in  new research
projects based on collaborations with Mario Trigiante, Daniele
Ruggeri and Laura Andrianapoli.
\section*{Acknowledgements} With pleasure we acknowledge, during the
development of the present research project very important and
clarifying discussions with our close friends and collaborators,
Laura Andrianopoli, Massimo Bianchi, Ugo Bruzzo, Dario Martelli and
Mario Trigiante.
\newpage
\appendix
\section{The example of the Eguchi-Hanson space}
\label{capponefinale} As mentioned in the introduction, the simplest example of curved
HyperK\"ahler manifold whose moment maps satisfy the constraints for supersymmetry enhancement
is the time honored Eguchi Hanson space \cite{eguccio}. \\
The EH space can be obtained as the HyperK\"ahel quotient  $\mathbb{C}^2 \times \mathbb{C}^{\star 2} \slash \, \slash_{HK}
U(1)$. We do not perform it explicitly since we would repeat the steps in \ref{dividendus}. Here, we briefly describe the EH geometry and we give the expression for the tri-holomorphic moment maps associated with the ${\rm SU}(2)$ action. The geometry of $T^\star \mathbb{P}^1$ (i. e. Eguchi-Hanson) is encoded in the following K\"ahler potential:
\begin{equation}
 \mathscr{K} = \sqrt{4R + \kappa^2} - \kappa log\left(\frac{\sqrt{4R + \kappa^2} +
       \kappa}{\sqrt{R}}\right)
\end{equation}
\begin{equation}
  R=|v|^2(1 + |u|^2)^2
\end{equation}
where $u$ is the coordinate on $\mathbb{P}^1$ and $v$ is the fibre coordinate. The metric is obtained as $g_{ij^\star}=\frac{\partial}{\partial z^i} \frac{\partial}{\partial z^{j^\star}} \mathscr{K}$, $z^i=(u,v)$. The Eguchi-Hanson space is an HyperK\"ahler space. The HyperK\"ahler form is the following:
\begin{eqnarray}
  \mathbf{K}^3 &=& {\rm i} g_{ij^\star} dz^i \wedge d\bar{z}^{j^\star} \nonumber\\
  \mathbf{K}^+ &=& 2(du \wedge dv) \nonumber\\
  \mathbf{K}^- &=& 2(d\bar{u} \wedge d\bar{v})
\end{eqnarray}
An element of the isometry group $\left( \begin{array}{cc} a & b \\ c & d \end{array} \right) \in
\rm{SU}(2) $ acts in the following way
\begin{equation}
  u \longrightarrow \frac{au+b}{cu+d} \,\,\, , \,\,\, v \longrightarrow v(cu+d)^2
\end{equation}
This isometry is generated by the following
holomorphic Killing vectors\footnote{$\sigma^\Lambda \,\,\, , \,\,\,
\Lambda=1,2,3$ are the standard Pauli matrices}
\begin{eqnarray}
  &k = k_\Lambda T^\Lambda \,\,\, , \,\,\, T^\Lambda = \frac{{\rm i}}{2}\sigma^\Lambda& \nonumber\\
  k_1=\frac{{\rm i}}{2}(u^2-1)\frac{\partial}{\partial u} + {\rm i}uv\frac{\partial}{\partial v}
  \,\,\, , \,\,\,
  &k_2=\frac{1}{2}(u^2+1)\frac{\partial}{\partial u} - uv \frac{\partial}{\partial v}& \,\,\, , \,\,\,
  k_3= {\rm i}u \frac{\partial}{\partial u} - {\rm i} v\frac{\partial}{\partial v}
\end{eqnarray}
Thanks to the HyperK\"ahler structure we can define the
tri-holomorphic moment maps
\begin{eqnarray}
  d\mathcal{P}^3_\Lambda &=& i_{k_\Lambda}\mathbf{K}^3 \nonumber\\
  d\mathcal{P}^+_\Lambda &=& i_{k_\Lambda}\mathbf{K}^+ \nonumber\\
   d\mathcal{P}^-_\Lambda &=& i_{k_\Lambda}\mathbf{K}^- \nonumber
\end{eqnarray}
These latter imply that $\mathscr{P}^3_\Lambda$ is defined modulo the real part of an holomorphic function while $\mathscr{P}^+_\Lambda$ is defined modulo a constant. Thanks to this freedom we can make the tri-holomorphic moment map equivariant, namely:
\begin{equation}
  \{\mathcal{P}^x_\Lambda,\mathcal{P}^x_\Sigma\}\equiv
i_{k_\Lambda}i_{k_\Sigma}\mathbf{K}^x=f_{\Lambda\Sigma}^\Gamma\mathcal{P}^x_\Gamma
\end{equation}
The equivariant real moment maps are
\begin{eqnarray}
  \mathscr{P}^3_1&=& -\frac{(u+\bar{u})\sqrt{4R+\kappa^2}}{2(1+|u|^2)} \nonumber\\
  \mathscr{P}^3_2&=& -{\rm i}\frac{(u-\bar{u})\sqrt{4R+\kappa^2}}{2(1+|u|^2)} \nonumber\\
  \mathscr{P}^3_3 &=& -\frac{(1-|u|^2)\sqrt{4R+\kappa^2}}{2(1+|u|^2)} \nonumber\\
\end{eqnarray}
The equivariant holomorphic moment maps are
\begin{eqnarray}
\mathscr{P}^+_1 &=& {\rm i}(1-u^2)v \nonumber\\
\mathscr{P}^+_2 &=& (1+u^2)v \nonumber\\
\mathscr{P}^+_3 &=& 2{\rm i} uv \nonumber
\end{eqnarray}
Now we choose $\mathfrak{m}_{\Lambda \Sigma} = \kappa_{\Lambda \Sigma}$ to be the Cartan-Killing metric of ${\rm SU}(2)$. We find
\begin{equation}
  \mathcal{P}^+ \cdot \mathcal{P}^+ = 0 \,\,\, , \,\,\, \mathcal{P}^+ \cdot \mathcal{P}^3 =0
\end{equation}
and
\begin{equation}
   \mathcal{P}^3 \cdot \mathcal{P}^3 = \frac{\kappa^2}{4}+R \,\,\, , \,\,\, \mathcal{P}^+ \cdot \mathcal{P}^- = 2R
 \end{equation}
 so that
\begin{equation}
 2\mathcal{P}^3 \cdot \mathcal{P}^3 - \mathcal{P}^+ \cdot \mathcal{P}^- = \frac{\kappa^2}{2}
\end{equation}
From these identities we see that the constraints (\ref{MMcon_triv}-\ref{MMcon_int}) hold true. \\
The Eguchi-Hanson space is the first element in the infinite series
of HyperK\"ahler manifolds $T^{\star}\mathbb{P}^{1+n}$. Now we
present the next case which is physically more interesting.
\section{Parameterizing the $\mathrm{N^{0,1,0}}$ coset
representative} In this appendix we study a suitable
parameterization of  the $\mathrm{N^{0,1,0}}$ coset which reflects
its double fibration structure.
\subsection{The double fibration and the coset representative of
the flag manifold} \label{sexbina} The first step in our
construction consists of establishing a good parameterization of the
coset $\mathrm{N^{0,1,0}}$ as described in equation
(\ref{cavernicolo}). To this effect we use a double fibration,
namely we regard the flag manifold $\mathfrak{m}^F_6$ as a
$\mathbb{P}^1$ fibration over $\mathbb{P}^2$:
\begin{equation}\label{lukardon}
    \mathfrak{m}^F_6 \, \stackrel{\pi}{\longrightarrow} \,
    \mathbb{P}^2 \quad ; \quad \forall p \in \mathbb{P}^2 \quad
    \pi^{-1}(p) \, \sim \mathbb{P}^1
\end{equation}
Regarding $\mathbb{P}^2$ as the standard coset manifold
$\mathrm{SU(3)/SU(2)\times U(1)}$, the usual complex coordinates
$u_{1,2}$ in which the $\mathrm{SU(3)}$ invariant K\"ahler metric on
$\mathbb{P}^2$ takes the familiar Fubini-Study form are encoded in
the following coset representative\footnote{In this definition, $u$ and $v$ must
not be confused with the flat $\mathbb{C}^3 \oplus \mathbb{C}^{\star 3}$ coordinates
related to the HyperK\"ahler quotient construction.}:
\begin{equation}\label{fubstudiaP2}
  \mathbb{L}_{\mathbb{P}^2} \, = \, \left(
\begin{array}{ccc}
 \frac{\frac{u_1 \bar{u}_1}{\sqrt{|\mathbf{u}|^2+1}}+u_2 \bar{u}_2}{|\mathbf{u}|^2} & \frac{u_1
   \left(\frac{1}{\sqrt{|\mathbf{u}|^2+1}}-1\right) \bar{u}_2}{|\mathbf{u}|^2} &
   \frac{u_1}{\sqrt{|\mathbf{u}|^2+1}} \\
 \frac{u_2 \left(\frac{1}{\sqrt{|\mathbf{u}|^2+1}}-1\right) \bar{u}_1}{|\mathbf{u}|^2} & \frac{\frac{u_2
   \bar{u}_2}{\sqrt{|\mathbf{u}|^2+1}}+u_1 \bar{u}_1}{|\mathbf{u}|^2} &
   \frac{u_2}{\sqrt{|\mathbf{u}|^2+1}} \\
 -\frac{\bar{u}_1}{\sqrt{|\mathbf{u}|^2+1}} & -\frac{\bar{u}_2}{\sqrt{|\mathbf{u}|^2+1}} &
   \frac{1}{\sqrt{|\mathbf{u}|^2+1}} \\
\end{array}
\right) \, \in \, \mathrm{SU(3)}\quad ; \quad |\mathbf{u}|^2 \equiv
|u_1|^2 +|u_2|^2
\end{equation}
Indeed, calculating the left-invariant $1$-form:
\begin{equation}\label{crucionP2}
    \Lambda_{\mathbb{P}^2} \, = \, \mathbb{L}_{\mathbb{P}^2}^\dagger \,
    \mathrm{d}\mathbb{L}_{\mathbb{P}^2}
\end{equation}
and defining the vierbein of the manifold $\mathbb{P}^2$
as\footnote{The formula below is justified because the four
generators of the subalgebra $\su(2)\oplus \uu(1)$ are
$\mathbf{t}_1,\mathbf{t}_2,\mathbf{t}_3,\mathbf{t}_8$, so that the
coset generators are
$\mathbf{t}_4,\mathbf{t}_5,\mathbf{t}_6,\mathbf{t}_7$.} :
\begin{equation}\label{vierbeinP2}
    \left\{E_{\mathbb{P}^2}^1,E_{\mathbb{P}^2}^2,E_{\mathbb{P}^2}^3,E_{\mathbb{P}^2}^4\right\}
    \, = \, -2 \left\{\mbox{Tr}\left(\mathbf{t}_4 \,\Lambda_{\mathbb{P}^2}\right),\mbox{Tr}\left(\mathbf{t}_5
    \,\Lambda_{\mathbb{P}^2}\right),
    \mbox{Tr}\left(\mathbf{t}_6 \,\Lambda_{\mathbb{P}^2}\right),\mbox{Tr}\left(\mathbf{t}_7
    \,\Lambda_{\mathbb{P}^2}\right)\right\}
\end{equation}
we obtain the standard Fubini-Study line element:
\begin{eqnarray}
  ds^2_{\mathbb{P}^2} &=& \sum_{I=1}^4 \,\left(E_{\mathbb{P}^2}^I\right)^2 \nonumber\\
   &=& \frac{\mathrm{d}u_1\,\mathrm{d}\bar{u}_1(1+u_2 \,\bar{u}_2)+\mathrm{d}u_2\,\mathrm{d}\bar{u}_2(1+u_1 \,\bar{u}_1)-u_1\,\bar{u}_2
   \mathrm{d}u_2 \mathrm{d}\bar{u}_1 -u_2\,\bar{u}_1
   \mathrm{d}u_1 \mathrm{d}\bar{u}_2}{(|\mathbf{u}|^2+1)^2}
\end{eqnarray}
Next we introduce the coset representative of $\mathrm{SU(2)/U(1)}$
immersed in $\mathrm{SU(3)}$. We set:
\begin{equation}\label{fubstudiaP1}
 \Lambda_{\mathbb{P}^1} \, = \,   \left(
\begin{array}{ccc}
 \sqrt{\frac{1}{|v|^2+1}} & \frac{v}{\sqrt{|v|^2+1}} & 0 \\
 -\frac{\bar{v}}{\sqrt{|v|^2+1}} & \sqrt{\frac{1}{|v|^2+1}} & 0 \\
 0 & 0 & 1 \\
\end{array}
\right)\, \in \, \mathrm{SU(3)}
\end{equation}
which, with the same logic as before, produces the standard
Fubini-Study metric on $\mathbb{P}^1$. In this case the zweibein of
$\mathbb{P}^1$ is obtained by tracing the left invariant one form:
\begin{equation}\label{crucionP2}
    \Lambda_{\mathbb{P}^1} \, = \, \mathbb{L}_{\mathbb{P}^1}^\dagger \,
    \mathrm{d}\mathbb{L}_{\mathbb{P}^1}
\end{equation}
with $\mathbf{t}_1$ and $\mathbf{t}_2$ that are the coset generators
inside the $\mathrm{SU(2)}$ subalgebra of $\mathrm{SU(3)}$ spanned
by the generators $\mathbf{t}_{1,2,3}$.
\par
In this way a convenient dense chart for the flag manifold
$\mathfrak{m}^F_6$ is provided by the three complex coordinates
$u_1,u_2,v$. Correspondingly we can define the coset representative
for $\mathfrak{m}^F_6$ as follows:
\begin{equation}\label{cartolinapostale}
    \mathbb{L}_{flag} \, = \, \mathbb{L}_{\mathbb{P}^2}\, \mathbb{L}_{\mathbb{P}^1}
\end{equation}
So doing the left invariant $1$-form of $\mathfrak{m}^F_6$ takes the
form:
\begin{equation}\label{brillato1}
    \Lambda_{flag} \, \equiv \, \mathbb{L}_{flag}^\dagger \,
    \mathrm{d}\mathbb{L}_{flag} \, = \, \Lambda_{\mathbb{P}^1} \,  +
    \,\mathbb{L}_{\mathbb{P}^1}^\dagger \, \Lambda_{\mathbb{P}^2} \mathbb{L}_{\mathbb{P}^1}
\end{equation}
which exposes the fibred structure (\ref{lukardon}) of the flag
manifold. The sechsbein of $\mathfrak{m}^F_6$ is provided
by\footnote{The generators of the coset manifold in this case are
$\mathbf{t}_1,\mathbf{t}_2,\mathbf{t}_4,\mathbf{t}_5,\mathbf{t}_6,\mathbf{t}_7$}:
\begin{eqnarray}\label{gramellone}
    &&\left\{E_{flag}^1,E_{flag}^2,E_{flag}^3,E_{flag}^4,E_{flag}^5,E_{flag}^6,\right\} \, = \, \nonumber\\
   && \, -2 \left\{\mbox{Tr}\left(\mathbf{t}_1
   \,\Lambda_{flag}\right),\mbox{Tr}\left(\mathbf{t}_2
    \,\Lambda_{flag}\right),
    \mbox{Tr}\left(\mathbf{t}_4
    \,\Lambda_{flag}\right),\mbox{Tr}\left(\mathbf{t}_5
    \,\Lambda_{flag}\right),\mbox{Tr}\left(\mathbf{t}_6
    \,\Lambda_{flag}\right),\mbox{Tr}\left(\mathbf{t}_7
    \,\Lambda_{flag}\right)
    \right\}
\end{eqnarray}
In the explicit calculation, if needed, of the coordinate dependence
of the vielbein $E_{flag}^I$, the structure of (\ref{brillato1}) of
the left-invariant one-form is very useful. Indeed naming:
\begin{equation}\label{piretti}
    \mathcal{E}_{\mathbf{n}}^\Sigma \, = \, - 2\, \mbox{Tr} \left( \mathbf{t}_\Sigma
    \,\Lambda_{\mathbf{n}}\right) \quad ; \quad
    \Lambda_{\mathbf{n}}\, \equiv \,
    \mathbb{L}^\dagger_{\mathbf{n}}\,\mathrm{d}\mathbb{L}_{\mathbf{n}}
\end{equation}
the components along the generators (\ref{strutturicosti}) of any
$\mathrm{SU(3)}$ left-invariant form  we see that under the subgroup
$\mathrm{SU(2)} \subset \mathrm{SU(3)}$ the generators and hence the
corresponding $1$-forms are organized in the following
representations:
\begin{equation}\label{crudura}
    \underbrace{\mathbf{t}_1,\mathbf{t}_2,\mathbf{t}_3}_{\text{triplet}\,
    =\,
    \text{adjoint}} \,\oplus
    \,\underbrace{\mathbf{t}_4,\mathbf{t}_5,\mathbf{t}_6,\mathbf{t}_7}_{\text{complex
    doublet} \, = \, \text{4-dim irrep} }\, \oplus \,
    \underbrace{\mathbf{t}_8}_{\text{singlet}}
\end{equation}
We name $x,y,\dots=1,2,3$ the indices of the triplet,
$\alpha,\beta,\dots=4,5,6,7$ the indices of the real quadruplet and
we keep $8$ for the singlet. According to this we conclude that:
\begin{eqnarray}
  \mathbb{L}_{\mathbb{P}^1} \, \mathbf{t}_x \mathbb{L}_{\mathbb{P}^1}^\dagger &=& \mathcal{S}_{xy}(v,\bar{v}) \, \mathbf{t}_y \nonumber\\
 \mathbb{L}_{\mathbb{P}^1} \, \mathbf{t}_\alpha \mathbb{L}_{\mathbb{P}^1}^\dagger &=& \mathcal{T}_{\alpha\beta}(v,\bar{v})
 \, \mathbf{t}_\beta \nonumber\\
 \mathbb{L}_{\mathbb{P}^1} \, \mathbf{t}_8
 \mathbb{L}_{\mathbb{P}^1}^\dagger &=&
 \mathbf{t}_8\label{saponettabasca}
\end{eqnarray}
where the $3\times3$ matrix $\mathcal{S}_{xy}(v,\bar{v})$ and the
$4\times 4$ one $\mathcal{T}_{\alpha\beta}(v,\bar{v})$ depend only
on the fibre coordinates $v,\bar{v}$. Looking now at the traces
appearing in equation (\ref{gramellone}), we see that the final form
of the sechsbein is as follows:
\begin{equation}
\begin{array}{rclcc}
  E^x_{flag} &=& \mathcal{E}_{\mathbb{P}^1}^x + \mathcal{S}^{xy}(v,\bar{v}) \, \mathcal{E}_{\mathbb{P}^2}^y  &; & x=1,2, \quad ;
  \quad y=1,2,3 \\
 E^\alpha_{flag}  &=& \mathcal{T}^{\alpha\beta}(v,\bar{v}) \, \mathcal{E}_{\mathbb{P}^2}^\beta &; & \alpha,\beta=4,5,6,7\\
 \end{array}
\end{equation}
By construction the $1$-forms $\mathcal{E}_{\mathbb{P}^1}^x$ depend
only on the coordinates $v,\bar{v}$, while
$\mathcal{E}_{\mathbb{P}^2}^x$ and
$\mathcal{E}_{\mathbb{P}^2}^\alpha$ depend only on the coordinates
$u_1,u_2,\bar{u}_1,\bar{u}_2$ of the $\mathbb{P}^2$ base manifold.
\subsection{The  coset representative of $\mathrm{N^{0,1,0}}$} The
next step of the construction consists of building  the coset
representative of the $7$-dimensional coset $\mathrm{N^{0,1,0}}$
regarded as a $\mathrm{U(1)}$ fibration over the flag manifold that
we studied in section \ref{sexbina}. The strategy is identical to
that used in the construction of the flag manifold vielbein. We
introduce the $\mathrm{U(1)}$ group element obtained by
exponentiating the generator $\mathbf{t}_3$:
\begin{equation}\label{u1}
    \mathbb{L}_{\mathrm{U(1)}}\, = \,\left(
\begin{array}{ccc}
 e^{\frac{i \psi }{2}} & 0 & 0 \\
 0 & e^{-\frac{i \psi }{2}} & 0 \\
 0 & 0 & 1 \\
\end{array}
\right)\, \in \, \mathrm{SU(3)}
\end{equation}
and we write the complete coset representative as follows:
\begin{equation}\label{carciofo}
    \mathbb{L}_{\mathrm{N^{010}}} \, = \,\mathbb{L}_{flag}\,\mathbb{L}_{\mathrm{U(1)}}
    \, = \, \mathbb{L}_{\mathbb{P}^2}\, \mathbb{L}_{\mathbb{P}^1}\,\mathbb{L}_{\mathrm{U(1)}}
\end{equation}
Introducing the complete left-invariant one form:
\begin{equation}\label{lefta}
    \Lambda_{\mathrm{N^{010}}} \, = \,
    \mathbb{L}_{\mathrm{N^{010}}}^\dagger \, \mathrm{d}\mathbb{L}_{\mathrm{N^{010}}}
\end{equation}
The vielbein of the $7$-manifold are given by:
\begin{eqnarray}\label{gramelloide}
    &&\left\{E_{\mathrm{N^{010}}}^1,E_{\mathrm{N^{010}}}^2,E_{\mathrm{N^{010}}}^3,E_{\mathrm{N^{010}}}^4,E_{\mathrm{N^{010}}}^5,
    E_{\mathrm{N^{010}}}^6,E_{\mathrm{N^{010}}}^7\right\} \, = \, \nonumber\\
   && \, -2 \left\{\mbox{Tr}\left(\mathbf{t}_1
   \,\Lambda_{\mathrm{N^{010}}}\right),\mbox{Tr}\left(\mathbf{t}_2
    \,\Lambda_{\mathrm{N^{010}}}\right),
    \mbox{Tr}\left(\mathbf{t}_4
    \,\Lambda_{\mathrm{N^{010}}}\right),\mbox{Tr}\left(\mathbf{t}_5
    \,\Lambda_{\mathrm{N^{010}}}\right),\mbox{Tr}\left(\mathbf{t}_6
    \,\Lambda_{\mathrm{N^{010}}}\right),\mbox{Tr}\left(\mathbf{t}_7
    \,\Lambda_{\mathrm{N^{010}}}\right),\mbox{Tr}\left(\mathbf{t}_3\Lambda_{\mathrm{N^{010}}}\right)
    \right\}\nonumber\\
\end{eqnarray}
and the doubled fibred-structure displayed in eq.(\ref{carciofo})
can be utilized to work out the explicit dependence of the
$7$-vielbein on the seven well-adapted coordinates $\psi, v,
u_1,u_2$ (one real and three complex) if this is needed. It suffices
to specialize to  $\mathrm{U_3(1)\subset \mathrm{SU(2)}}$ the
analysis performed in eq.s (\ref{saponettabasca}) for the full
subgroup $\mathrm{SU(2)}\subset \mathrm{SU(3)}$. The nested fibred
structure is also useful to work out the explicit transformations of
the coordinates $\mathbf{u},v,\psi$ of the manifold $\n010$ under
the isometry group $\mathrm{SU(3)}$. Indeed the compensator subgroup
$\mathrm{H}$ of each of the three factors in eq.(\ref{carciofo}) is
the $\mathrm{G}$ group of the next factor. Hence we expect the
following.
\par
Let:
\begin{equation}\label{romillusca}
    \mathfrak{g}=\left( \begin{array}{c||c}
                   A_{2\times 2} & B_{2\times 1} \\
                   \hline
                   \hline
                   C_{1\times 2} & D_{1\times 1}
                 \end{array}\right) \,  \in \, \mathrm{SU(3)}
\end{equation}
be a group element of $\mathrm{SU(3)}$. The transformation induced
on the $\mathbb{P}^2$ coordinates will be holomorphic and projective
linear fractional:
\begin{equation}\label{carinno1}
    \mathbf{u}^\prime \, = \, \left(A \,\mathbf{u} + B\right)\cdot
    \left(C \mathbf{u} + D\right)^{-1}
\end{equation}
that induced on the fibre coordinate will also be fractional linear,
but $\mathbf{u}$-dependent:
\begin{equation}\label{carinno1}
    v^\prime \, = \, \left(a(\mathfrak{g},\mathbf{u}) \,v + b(\mathfrak{g},d\mathbf{u})\right)\cdot
    \left(c(\mathfrak{g},\mathbf{u}) v + d(\mathfrak{g},\mathbf{u})\right)^{-1}
\end{equation}
where the coefficients $a,b,c,d$ appearing in the above formula are
those displayed by the compensator matrix in the subgroup
$\mathrm{SU(2) \times U_8(1)}$:
\begin{eqnarray}\label{primcompen}
    \mathfrak{H}_{comp|\mathbb{P}^2}\left(\mathfrak{g},\mathbf{u}\right)& =&
    \left(
    \begin{array}{cc||c}
    a(\mathfrak{g},\mathbf{u})& b(\mathfrak{g},\mathbf{u})& 0 \\
    c(\mathfrak{g},\mathbf{u}) & d(\mathfrak{g},\mathbf{u}) & 0 \\
    \hline
    \hline
    0 & 0 & 1 \\
    \end{array}
    \right)\cdot\underbrace{\left(
    \begin{array}{cc||c}
    \exp[\, i \, \mu(\mathfrak{g},\mathbf{u})] &0 &0 \\
    0 & \exp[\, i \, \mu(\mathfrak{g},\mathbf{u})] & 0 \\
    \hline
    \hline
    0 & 0 & \exp[-2 \, i \, \mu(\mathfrak{g},\mathbf{u})] \\
    \end{array}
    \right)}_{\mathfrak{H}_8\left(\mathfrak{g},\mathbf{u}\right)}\nonumber\\
\end{eqnarray}
As it happens in all coset manifolds and for any choice of the coset
representative, the compensator depends both on the point
$\mathbf{u}$ and on the choice of the group element $\mathfrak{g}$
acting as an isometry. At the next step we will have a compensator
depending both on the coordinates $\mathbf{u}$ and on the coordinate
$v$
\begin{equation}\label{seccompen}
    \mathfrak{H}_{comp|\mathbb{P}^1}\left(\mathfrak{g},\mathbf{u},v\right)\, =\,
    \left(
    \begin{array}{cc||c}
    \exp[\frac{i}{2} \,\lambda(\mathfrak{g},\mathbf{u},v)]& 0 & 0 \\
    0 & \exp[-\frac{i}{2} \,\lambda(\mathfrak{g},\mathbf{u},v)] & 0 \\
    \hline
    \hline
    0 & 0 & 1 \\
    \end{array}
    \right)\cdot \mathfrak{H}_8\left(\mathfrak{g},\mathbf{u}\right)
\end{equation}
and this determines the transformation of the coordinate $\psi$:
\begin{equation}\label{carismaticus}
    \psi^\prime \, = \, \psi+\lambda(\mathfrak{g},\mathbf{u},v)
\end{equation}
The procedure outlined above allows the construction of all Killing
vectors for the manifold $\mathrm{N^{0,1,0}}$ and eventually for the
resolution of its metric cone that is given by the Calabi
HyperK\"ahler manifold $HK^{(2)}_{Calabi}$. This is also the
starting point for the calculation of the moment maps of the
$\mathrm{SU(3)}$ isometries. For the actual construction of
$HK^{(2)}_{Calabi}$ we do not need any explicit parameterization of the
coset manifold. The Maurer-Cartan equations satisfied by the
left-invariant one-forms are completely sufficient.

\section{Calculation of moment maps in two cases}
In this section we calculate the moment maps in two cases relevant
to our discussion:
\begin{description}
  \item[a)] For the $\mathrm{SU(3)}$ isometry of the curved Calabi
  HyperK\"ahler manifold $HK^{(2)}_{Calabi}$
  \item[b)] For the linearly realized $\mathrm{SU(m)\times SU(n) \times
  U(1)}$ isometry of a flat HyperK\"ahler manifold with $2\times m
  \times n$ complex coordinates.
\end{description}
\subsection{The moment maps of the $\mathrm{SU(3)}$ isometries of the Calabi
metric on $T^*\mathbb{P}^2$} \label{su3momenti} In this section we calculate the moment
maps of the relevant isometry group $\mathrm{SU(3)}$ in the case of
the curved HyperK\"ahler manifold $T^*\mathbb{P}^2$ endowed with the
Calabi metric and we verify that they satisfy the weak constraint
(\ref{MMcon_triv}-\ref{MMcon_int}) necessary for the supersymmetry enhancement
discussed in the main text.
\subsubsection{Transformation of the $T^*\mathbb{P}^2$ complex
coordinates under the isometry group $\mathrm{SU(3)}$} We denote the
complex coordinates of $T^*\mathbb{P}^2$ as $q=(w^1,w^2\vert
z_1,z_2)$. The first pair provides a chart on the base
$\mathbb{P}^2$, while the second pair on the fibre. Then let us
write an element of the isometry group in a block form
\begin{equation}
    \mathfrak{g}=\left( \begin{array}{c||c}
                   A_{2\times 2} & B_{2\times 1} \\
                   \hline
                   \hline
                   C_{1\times 2} & D_{1\times 1}
                 \end{array}\right) \,  \in \, \mathrm{SU(3)}.
\end{equation}
 The transformation of the $\mathbb{P}^2$ coordinates will be holomorphic and
 projective linear fractional
\begin{equation}
\mathbf{w'}=(A\mathbf{w}+B)(C\mathbf{w}+D)^{-1}.
\end{equation}
This induces a transformation of the fibre coordinates given by the
inverse transformation of the differentials of the base coordinates.
Defining
\begin{align}
\delta=C\mathbf{w}+D \in \mathbb{C},&& M=(A\mathbf{w}+B)\otimes C
\in \text{Mat}_{2\times 2}
\end{align}
we obtain the transformation of the differentials of $\mathbf{w}$
\begin{equation}
d\mathbf{w'}=\frac{\delta A-M}{\delta^2}\equiv F^{-1} d\mathbf{w}.
\end{equation}
Thus the transformation of the fibre coordinates $\mathbf{z}$ takes
the form
\begin{equation}
\mathbf{z'}=F\mathbf{z}.
\end{equation}
Next, we verify that the above transformations implied by the
structure of a cotangent bundle leave invariant the K\"{a}hler
potential as derived by the HyperK\"{a}hler quotient construction
(see section \ref{dividendus}). This check supports the claim that
the manifold $HK_8$ emerging from the HyperK\"{a}hler quotient
construction can in fact be identified with $T^*\mathbb{P}^2$. The
K\"{a}hler potential reads
\begin{equation}
\mathcal{K}=F(R)+\kappa \log(1+\mathbf{w}^\dagger
\mathbf{w})\label{KPot}.
\end{equation}
It is not hard to observe that the second term is invariant up to a
real part of a holomorphic function, which can be absorbed by a
K\"{a}hler transformation
\begin{equation}
\log(1+\mathbf{w'}^\dagger \mathbf{w'})=\log(1+\mathbf{w}^\dagger
\mathbf{w})-
\underbrace{\Big(\log\delta+\log\bar{\delta}\Big)}_{\text{K\"{a}hler
transformation}}.
\end{equation}
Therefore it is enough to show that the function~\footnote{We are
thinking of $\mathbf{w}$ and $\mathbf{z}$ as column vectors.}
\begin{equation}\label{Rfnc}
R(q,\bar{q})=(1+\mathbf{w}^\dagger\mathbf{w})\left[\mathbf{z}^\dagger(\mathbf{1}_{2\times
2}
+\mathbf{\overline{w}}\mathbf{w}^T)\mathbf{z}\right]=(1+\|{\mathbf{w}}\|^2)(\|{\mathbf{z}}\|^2
+\lvert\mathbf{w}\cdot\mathbf{z}\rvert^2)
\end{equation}
is invariant. This can be achieved by computing finite
transformations of the coordinates with respect to all one-parameter
subgroups of $\mathrm{SU(3)}$
\begin{align}
\mathfrak{g}_\Lambda(t)=e^{t T_\Lambda} \;\;
\text{with}\;T_\Lambda=\frac{i}{2}\lambda_\Lambda \in \mathfrak{su}(3), \;\;
\Lambda=1,\ldots,8,
\end{align}
where $\lambda_\Lambda$ are the Gell-Mann matrices. We do not list
here the finite transformations of coordinates, rather write down
their infinitesimal form in terms of Killing vectors in the next
section. In any case, it can be explicitly verified that the
function $R$ is invariant under all finite transformations
associated with individual generators of the isometry group. This
concludes the proof that the K\"{a}hler potential of the
HyperK\"{a}hler quotient is invariant under the isometry group of
$T^*\mathbb{P}^2$.
\subsubsection{Killing vectors} The action of the isometry group
${\rm SU}(3)$ on the coordinates of $T^*\mathbb{P}^2$ yields at the
infinitesimal level the (holomorphic) Killing vectors
\begin{equation}
k_\Lambda^i=\left.\frac{d}{dt}\left(\mathfrak{g}_\Lambda(t) q^i\right)\right\vert_{t=0}.
\end{equation}
They are displayed in matrix form and the Killing vector
corresponding to the generator $T_\Lambda$ is stored in the
$\Lambda$-th row ~\footnote{For practical reasons we have shifted
the indices of $\mathbf{w}$ coordinates from top to the bottom.}
\begin{align}
k_\Lambda^i=\left(
\begin{array}{cccc}
 \frac{i w_2}{2} & \frac{i w_1}{2} & -\frac{1}{2} \left(i z_2\right) & -\frac{1}{2}
 \left(i z_1\right) \\
 \frac{w_2}{2} & -\frac{w_1}{2} & \frac{z_2}{2} & -\frac{z_1}{2} \\
 \frac{i w_1}{2} & -\frac{1}{2} \left(i w_2\right) & -\frac{1}{2} \left(i z_1\right) &
 \frac{i z_2}{2} \\
 -\frac{1}{2} i \left(w_1^2-1\right) & -\frac{1}{2} i w_1 w_2 & \frac{1}{2} i
 \left(2 w_1 z_1+w_2 z_2\right)
 & \frac{1}{2} i w_1 z_2 \\
 \frac{1}{2} \left(w_1^2+1\right) & \frac{w_1 w_2}{2} & -w_1 z_1-\frac{w_2 z_2}{2} &
 -\frac{1}{2} w_1 z_2 \\
 -\frac{1}{2} i w_1 w_2 & -\frac{1}{2} i \left(w_2^2-1\right) & \frac{1}{2} i w_2 z_1 & \frac{1}{2} i
 \left(w_1 z_1+2 w_2 z_2\right)
 \\
 \frac{w_1 w_2}{2} & \frac{1}{2} \left(w_2^2+1\right) & -\frac{1}{2} w_2 z_1 & -
 \frac{1}{2} w_1 z_1-w_2 z_2 \\
 \frac{1}{2} i \sqrt{3} w_1 &
 \frac{1}{2} i \sqrt{3} w_2 & -\frac{1}{2} i \sqrt{3} z_1 & -\frac{1}{2} i \sqrt{3} z_2 \\
 \label{KillVec}
\end{array}
\right).
\end{align}
These Killing vectors satisfy the $\mathfrak{su}(3)$ Lie algebra
\begin{equation}
\left[k_\Lambda,k_\Sigma\right]=-f_{\Lambda\Sigma}^\Gamma k_\Gamma
\end{equation}
with the structure constants defined as
\begin{equation}
f_{\Lambda\Sigma}^\Gamma=-2\text{Tr}\left(\left[T_\Lambda,T_\Sigma\right]T^\Gamma\right).
\end{equation}
\subsubsection{The $\mathrm{SU(3)}$ moment maps} The knowledge of
Killing vectors allows us to look for their potentials, i.e. the
associated moment maps $\mathcal{P}_\Lambda^x$. They are defined by the
formula
\begin{equation}
d\mathcal{P}_\Lambda^x=i_{k_\Lambda}\mathbf{K}^x,\;\; x=1,2,3.
\end{equation}
Here $\mathbf{K}^x$ is the triplet of K\"{a}hler forms. We pick $\mathbf{K}^3$ and
associate it with the real moment map $\mathcal{P}^3$. The remaining
two K\"{a}hler forms form (anti)-holomorphic combinations
$\mathbf{K}^\pm=\mathbf{K}^1\pm i \mathbf{K}^2$, which correspond to (anti)-holomorphic moment
maps $\mathcal{P}^\pm=\mathcal{P}^1 \pm i\mathcal{P}^2$
\begin{align}
d\mathcal{P}^3_\Lambda=i_{k_\Lambda}\mathbf{K}^3 \label{MM_real}\\
d\mathcal{P}^\pm_\Lambda=i_{k_\Lambda}\mathbf{K}^\pm \label{MM_cplx}.
\end{align}
The K\"{a}hler form $\mathbf{K}^3$ is associated with the metric and
therefore also with the K\"{a}hler potential as $\mathbf{K}^3=i\partial
\bar{\partial}\mathcal{K}$. Consequently, a general solution can be
constructed for the real moment map in \eqref{MM_real}. Indeed,
projecting it to $(1,0)$ and $(0,1)$ components leads to a system of
two equations
\begin{align}
\partial_i\mathcal{P}^3_\Lambda&=-ig_{ij^{\star}}\bar{k}^{j^{\star}}_\Lambda=-i\bar{k}^{j^{\star}}_\Lambda\partial_i\bar{\partial}_{j^{\star}}
\mathcal{K}=\partial_i\left(-i\bar{k}^{j^{\star}}_\Lambda\bar{\partial}_{j^{\star}}\mathcal{K}\right) \label{MM_real_10}\\
\bar{\partial}_{j^{\star}}\mathcal{P}^3_\Lambda&=ig_{ij^{\star}}k^i_\Lambda=ik^i_\Lambda\partial_i\bar{\partial}_{j^{\star}}
\mathcal{K}=\bar{\partial}_{j^{\star}}\left(ik^i_\Lambda\partial_i\mathcal{K}\right)
\label{MM_real_01}.
\end{align}
From \eqref{MM_real_10} follows
\begin{equation}
\mathcal{P}^3_\Lambda=-i\bar{k}^{j^{\star}}_\Lambda\bar{\partial}_{j^{\star}}\mathcal{K}+\bar{f}^1_\Lambda(\bar{q})\label{MM_10},
\end{equation}
while from \eqref{MM_real_01} one gets
\begin{equation}
\mathcal{P}^3_\Lambda=ik^j_\Lambda\partial_j\mathcal{K}+f^2_\Lambda(q)\label{MM_01}.
\end{equation}
Reality of $P^3$ implies $f^1_\Lambda(q)=f^2_\Lambda(q)\equiv h_\Lambda(q)$ and
one needs to take the symmetric combination to make it manifestly
real
\begin{equation}
\mathcal{P}^3
_\Lambda=\frac{i}{2}\left(k^j_\Lambda\partial_j\mathcal{K}-\bar{k}^{j^{\star}}_\Lambda\bar{\partial}_{j^{\star}}\mathcal{K}\right)
+\frac{1}{2}\left(h_\Lambda(q)+\bar{h}_\Lambda(\bar{q})\right)\label{MM_real_final}.
\end{equation}
Thus we just showed that there is freedom in shifting the real
moment map by a real part of a holomorphic function. It is important
as we will exploit this fact to impose equivariance on the moment
maps (with respect to the ${\rm SU}(3)$ action). Ultimately, only
the equivariant moment maps are supposed to fulfill constraints,
which allow for supersymmetry enhancement from $\mathcal{N}_3=3$ to
$\mathcal{N}_3=4$. On the other hand, the difference of the above
equations fixes the imaginary part of $h_\Lambda(q)$
\begin{equation}
h_\Lambda(q)-\bar{h}_\Lambda(\bar{q})=-i\left(k^j_\Lambda\partial_j\mathcal{K}+\bar{k}^{j^{\star}}_\Lambda\bar{\partial}_{j^{\star}}\mathcal{K}\right).
\end{equation}
In \eqref{MM_real_final} reality of $\mathcal{P}^3$ is manifest,
however it obscures the canonical relation for the moment maps as
potentials for the Killing vectors
\begin{equation}
k^i_\Lambda=-ig^{j^{\star}i}\bar{\partial}_{j^{\star}}\mathcal{P}^3_\Lambda\label{Kill_pot}.
\end{equation}
The apparent problem gets resolved once we apply the equation that
fixes $\text{Im}\left(h_\Lambda\right)$. In fact substituting either for
$h_\Lambda(q)$ or $\bar{h}_\Lambda(\bar{q})$ brings us back to \eqref{MM_10}
or \eqref{MM_01}, respectively. This operation of course does not
break reality, just makes it less manifest. In other words, the
expression for $\text{Im}\left(h_\Lambda\right)$ allows us to transfer
between two equivalent forms for $\mathcal{P}^3$ -- one that is
manifestly real and the other one that clearly displays the
canonical relation \eqref{Kill_pot} between the Killing vector and
the associated moment map.

In order to get an explicit expression for $\mathcal{P}^3$, we plug
in the Killing vectors \eqref{KillVec} and the K\"{a}hler potential
given in \eqref{KPot} into \eqref{MM_real_final}. It is best to keep
the function $F(R)$ implicit during the computation. Even so, the
final formulae for the real moment map are not as neat as for the
rest of the objects. We list the result component-wise for each
generator of $\mathfrak{su}(3)$
\begin{align}\label{P3result}
\mathcal{P}^3_1&=-\frac{\kappa  \left(w_2 \overline{w}_1+w_1
\overline{w}_2\right)}{2 \left(\|{\mathbf{w}}\|^2+1\right)}-
\frac{1}{2} \Bigg[\overline{z}_1 \bigg(z_1 \big(|w_1|^2+1\big) \big(w_2 \overline{w}_1+w_1 \overline{w}_2\big)  \notag \\
&+z_2 \left(w_2^2
\overline{w}_1^2+|w_1|^2|w_2|^2-(1+\|{\mathbf{w}}\|^2)\right)\bigg)
+\overline{z}_2
\bigg(z_2 \left(w_2 \overline{w}_1+w_1 \overline{w}_2\right) \left(|w_2|^2+1\right)  \notag \\
&+z_1 \left(w_1^2 \overline{w}_2^2+|w_1|^2|w_2|^2-(1+\|{\mathbf{w}}\|^2)\right)\bigg)\Bigg] F'(R)  \notag \\
\mathcal{P}^3_2&=\frac{i \kappa  \left(w_2 \overline{w}_1-w_1
\overline{w}_2\right)}{2 (\|{\mathbf{w}}\|^2+1)}+\frac{i}{2}
\Bigg[\overline{z}_1 \bigg(z_1(|w_1|^2+1)  \left(w_2 \overline{w}_1-w_1 \overline{w}_2\right)  \notag \\
&+z_2 \left(w_2^2
\overline{w}_1^2-|w_1|^2|w_2|^2+(1+\|{\mathbf{w}}\|^2)\right)\bigg)
+\overline{z}_2 \bigg(z_2
\left(w_2 \overline{w}_1-w_1 \overline{w}_2\right)\left(|w_2|^2+1\right)  \notag \\
&-z_1\left(w_1^2 \overline{w}_2^2-|w_1|^2|w_2|^2+(1+\|{\mathbf{w}}\|^2)\right)\bigg)\Bigg]F'(R) \notag \\
\mathcal{P}^3_3&=-\frac{\kappa  (|w_1|^2-|w_2|^2)}{2
(\|{\mathbf{w}}\|^2+1)}+\frac{1}{2} \Bigg[z_1
\bigg(\overline{z}_2w_1
\overline{w}_2  (|w_2|^2-|w_1|^2)+\overline{z}_1\Big(|w_1|^2 \left(|w_2|^2 -|w_1|^2\right)  \notag \\
&+2 |w_2|^2 +1\Big)+z_2 \bigg(\overline{z}_1\overline{w}_1w_2\left(|w_1|^2-|w_2|^2\right) \notag \\
&+\overline{z}_2\Big(|w_2|^2 \left(|w_2|^2 -|w_1|^2\right)-2 |w_1|^2 -1\Big)\bigg)\Bigg]F'(R) \notag \\
\mathcal{P}^3_4&=\frac{\kappa  (\|{\mathbf{w}}\|^2-1)
\left(\overline{w}_1+w_1\right)}{4 (\|{\mathbf{w}}\|^2+1)}-
\frac{1}{2} \Bigg[\overline{z}_1 \bigg(z_1
\left(\overline{w}_1+w_1\right) \left(1+
|w_1|^2+1+\|{\mathbf{w}}\|^2\right)
\notag \\
&-w_2 z_2
\left(\overline{w}_1(w_1+\overline{w}_1)+\|{\mathbf{w}}\|^2+1\right)\bigg)+\overline{z}_2
\bigg(z_1 \overline{w}_2 \left(w_1(w_1+\overline{w}_1)+\|{\mathbf{w}}\|^2+1\right) \notag \\
&+z_2 \left(\overline{w}_1+w_1\right)(1+|w_2|^2)\bigg)\Bigg]F'(R) \notag \\
\end{align}
\begin{align}
 \mathcal{P}^3_5&=\frac{i \kappa  (\|{\mathbf{w}}\|^2-1)
\left(w_1-\overline{w}_1\right)}{4 (\|{\mathbf{w}}\|^2+1)}
+\frac{i}{2} \Bigg[ \overline{z}_2 \bigg(z_1 \overline{w}_2
\big(w_1(\overline{w}_1-w_1)+\|{\mathbf{w}}\|^2+1)\big)
\notag \\
&-z_2 \left(w_1-\overline{w}_1\right)(1+|w_2|^2)\bigg)-
\overline{z}_1 \bigg(z_1 \left(w_1-\overline{w}_1\right) (1+
|w_1|^2+1
+\|{\mathbf{w}}\|^2) \notag \\
&+w_2 z_2 \left(\overline{w}_1(w_1-\overline{w}_1)+1+\|{\mathbf{w}}\|^2\right)\bigg)\Bigg]F'(R) \notag \\
\mathcal{P}^3_6&=\frac{\kappa  (\|{\mathbf{w}}\|^2-1)
\left(\overline{w}_2+w_2\right)}{4
(\|{\mathbf{w}}\|^2+1)}-\frac{1}{2}
\Bigg[\overline{z}_1 \bigg(z_2 \overline{w}_1 \left(w_2(w_2+\overline{w}_2)+1+\|{\mathbf{w}}\|^2\right) \notag \\
&+z_1
\left(\overline{w}_2+w_2\right)(1+|w_1|^2)\bigg)+\overline{z}_2
\bigg(z_2 \left(\overline{w}_2+w_2\right) (1+ |w_2|^2+1+
\|{\mathbf{w}}\|^2) \notag \\
&+w_1 z_1 \left(\overline{w}_2(w_2+\overline{w}_2)+1+\|{\mathbf{w}}\|^2\right)\bigg)\Bigg]F'(R) \notag \\
\mathcal{P}^3_7&=\frac{i \kappa  (\|{\mathbf{w}}\|^2-1)
\left(w_2-\overline{w}_2\right)}{4 (\|{\mathbf{w}}\|^2+1)}-
\frac{i}{2} \Bigg[z_2\bigg( \overline{w}_1 \overline{z}_1
\Big(w_2(w_2-\overline{w}_2)-\left(1+\|{\mathbf{w}}\|^2\right)\Big)
\notag \\
&+\overline{z}_2\left(w_2-\overline{w}_2\right)
(1+|w_2|^2+1+\|{\mathbf{w}}\|^2)\bigg)+z_1
\bigg(w_1  \overline{z}_2 \left(\overline{w}_2(w_2-\overline{w}_2)+1+\|{\mathbf{w}}\|^2\right) \notag \\
&+\overline{z}_1 \left(w_2-\overline{w}_2\right) (1+|w_1|^2) \bigg)\Bigg] F'(R) \notag \\
\mathcal{P}^3_8&=-\frac{\sqrt{3} \kappa  \|{\mathbf{w}}\|^2}{2
(\|{\mathbf{w}}\|^2+1)}- \frac{\sqrt{3}}{2}
\Bigg[\|{\mathbf{w}}\|^2|\mathbf{w}\cdot\mathbf{z}|^2-\|{\mathbf{z}}\|^2\Bigg]F'(R)
.
\end{align}
In the expressions above we still keep the freedom of adding
$\text{Re}(h_\Lambda)$.

In the next step we wish to solve for the holomorphic moment map. At
this point we have to make an ansatz for the holomorphic K\"{a}hler
form. Let us propose that it has the canonical form (based on our
experience with $T^*\mathbb{P}^1$, i.e. the Eguchi--Hanson space)
\begin{equation} \label{KFhol}
\mathbf{K}^+=2\left(dw^1\wedge dz_1+dw^2\wedge dz_2\right).
\end{equation}
In the following we verify that this assumption is indeed correct.
To do so, we recall the relation between the triplet of K\"{a}hler
forms, complex structures and the metric
\begin{equation}
\mathbf{J}^x=\mathbf{K}^xg^{-1} \label{CplxKahler}.
\end{equation}
With the definitions $\mathbf{J}^\pm=\mathbf{J}^1\pm i\mathbf{J}^2$, the quaternionic algebra
$\mathbf{J}^x\mathbf{J}^y=-\delta^{xy}\mathbf{1} +\epsilon^{xy}_{\phantom{xy}z}\mathbf{J}^z$
translates to
\begin{align}
\left[\mathbf{J}^+,\mathbf{J}^3\right]&=2i\mathbf{J}^+  \label{Qalgp3}\\
\left[\mathbf{J}^-,\mathbf{J}^3\right]&=-2i\mathbf{J}^- \label{Qalgm3}\\
\left[\mathbf{J}^+,\mathbf{J}^-\right]&=-4i\mathbf{J}^3 \label{Qalgpm}
\end{align}
Writing \eqref{CplxKahler} in matrix notation leads to
\begin{equation} \label{CplxKahlerMat}
\left(\begin{array}{c|c} i \mathbf{1} & \mathbf{J}^+ \\ \hline \mathbf{J}^- & -i
\mathbf{1}
\end{array}\right)
\equiv \left(\begin{array}{c|c} \mathbf{K}^+ & ig \\ \hline -i\overline{g} &
\mathbf{K}^-
\end{array}\right)
\left(\begin{array}{c|c} 0 & \overline{g}^{-1} \\ \hline g^{-1} & 0
\end{array}\right)
=\left(\begin{array}{c|c} i \mathbf{1} & \mathbf{K}^+\overline{g}^{-1} \\
\hline \mathbf{K}^-g^{-1} & -i \mathbf{1}
\end{array}\right).
\end{equation}
It is immediate to check that relations \eqref{Qalgp3},
\eqref{Qalgm3} are trivially satisfied. A true restriction is
provided by equation \eqref{Qalgpm}. Utilizing the matrix notation
\eqref{CplxKahlerMat}, it turns it into the following constraints on
$\mathbf{K}^\pm$
\begin{align}
\mathbf{K}^+\overline{g}^{-1 }\mathbf{K}^-g^{-1}&=4\mathbf{1}_{4\times 4} \\
\mathbf{K}^-g^{-1}\mathbf{K}^+\overline{g}^{-1 }&=4\mathbf{1}_{4\times 4}.
\end{align}
We showed that these relations are satisfied (it is eaiser to check
the inverse of them to avoid inverting the metric), which proves
that our ansatz for the holomorphic K\"{a}hler form in \eqref{KFhol}
is consistent with the quaternionic algebra of the triplet of
complex structures and standard formulae of HyperK\"{a}hler
geometry.
\par
Once we are sure that we have the correct holomorphic K\"{a}hler
form, we can compute its associated holomorphic moment map
$\mathcal{P}^+$. To get it, one has to solve a very simple (in our
case) system of first order partial differential equations given in
\eqref{MM_cplx}. The solution is straightforward and takes the form
\begin{align} \label{Ppresult}
\mathcal{P^+}=
\begin{pmatrix}
i(w^1z_2+w^2z_1)\\(-w^1z_2+w^2z_1)\\i(w^1z_1-w^2z_2)\\i(-w^1(w^1z_1+w^2z_2)+z_1)\\
(w^1(w^1z_1+w^2z_2)+z_1)
\\i(-w^2(w^1z_1+w^2z_2)+z_2)\\(w^2(w^1z_1+w^2z_2)+z_2)\\i\sqrt{3}(w^1z_1+w^2z_2).
\end{pmatrix}
\end{align}
At this stage we have to impose equivariance on the moment maps
\begin{equation}
\{\mathcal{P}^x_\Lambda,\mathcal{P}^x_\Sigma\}\equiv
i_{k_\Lambda}i_{k_\Sigma}\mathbf{K}^x=f_{\Lambda\Sigma}^\Gamma\mathcal{P}^x_\Gamma,
\end{equation}
which fixes the freedom of shifts by $\text{Re}(h_\Lambda)$. Equivariance
requires
\begin{equation}
h_\Lambda(q)=-\frac{\kappa}{2}\left(0,0,0,w_1,i w_1,w_2,i
w_2,-\frac{4}{\sqrt{3}}\right).
\end{equation}
\subsubsection{Verification of the supersymmetry enhancing conditions
on moment maps} Having settled equivariance of the moment maps, we
are finally in a position for checking the supersymmetry enhancing
constraints on the moment maps. These constraints were spelled out
in eq.s(\ref{MMcon_triv},\ref{MMcon_int}). In this particular case
we choose $\mathfrak{m}_{\Lambda \Sigma} = \kappa_{\Lambda \Sigma }$ to be the Cartan-Killing metric
of $\rm{SU}(3)$.
\par
Substituting our explicit expressions for the moment maps summarized
in \eqref{P3result} and \eqref{Ppresult} to the above equations
gives
\begin{equation}
 \left(\mathcal{P}^+\cdot\mathcal{P}^+\right)=\left(\mathcal{P}^-\cdot\mathcal{P}^-\right)=
 \left(\mathcal{P}^+\cdot\mathcal{P}^3\right)=\left(\mathcal{P}^-\cdot\mathcal{P}^3\right)=0,
\end{equation}
while the argument of the most interesting constraint
\eqref{MMcon_int} reduces to
\begin{equation}
 2\mathcal{P}^3\cdot \mathcal{P}^3-\mathcal{P}^+\cdot\mathcal{P}^-=\frac{2\kappa^2}{3}.
\end{equation}
\par
It is actually interesting to show the individual pieces from which
the constraint is built. They depend in a very simple way on the
function $R$ defined in \eqref{Rfnc}
\begin{align}
\mathcal{P}^3\cdot \mathcal{P}^3&=\frac{\kappa^2}{3}+\,R \\
\mathcal{P}^+\cdot\mathcal{P}^-&=2 \,R.
\end{align}
It is plausible that such structure for the various scalar products
of moment maps holds true for the whole series of HyperK\"{a}hler
Calabi manifolds $T^*\mathbb{P}^n$ (the function $R$ generalizes in
a straightforward way for the whole series).
\par
Since the argument of the constraint \eqref{MMcon_int} depends only
on the resolution parameter $\kappa$ (related to the scale of the
metric $\ell$), all constraints on the moment maps are satisfied,
which implies supersymmetry enhancement from $\mathcal{N}_3=3$ to
$\mathcal{N}_3=4$. Thus the conclusion of this analysis is that
super Chern--Simons theory with target space $T^*\mathbb{P}^2$ and
gauge group $\rm{SU}(3)$ acting non-linearly on the target space has
actually $\mathcal{N}_3=4$ supersymmetry.
\subsection{The moment maps of $\mathfrak{su}(n)\oplus
\mathfrak{su}(m)\oplus \mathfrak{u}(1)$ acting on a flat
HyperK\"ahler manifold}\label{piattume}  Let us start by specifying
the gauge Lie algebra $\mathfrak{g}=\mathfrak{su}(n)\oplus
\mathfrak{su}(m)\oplus \mathfrak{u}(1)$\footnote{In this discussion
we are assuming $m \neq n$.} and the representations of the
hypermultiplet scalars that provide coordinates of the flat (Hyper
K\"ahler) target space $\mathbb{C}^{nm}\oplus\mathbb{C}^{nm}$
\begin{align}
    \label{eq:target_reps}
\mathbf{u}\in (n,\overline{m}), \quad  \mathbf{v}\in
(m,\overline{n}).
\end{align}
To be completely explicit we write $\mathbf{u}$, $\mathbf{v}$ in
matrix form
\begin{align}
&\mathbf{u}\equiv u^i_{\phantom{i}\widehat{k}}  \in \textrm{Mat(n$\times$m)},
\quad \mathbf{v}\equiv v^{\widehat{k}}_{\phantom{k}i} \in \textrm{Mat(m$\times$n)},
&& i=1,\ldots,n \notag \\
& &&\widehat{k}=1,\ldots,m
\end{align}
and treat them as independent, i.e. not hermitian conjugate. In
these coordinates the triplet of canonical HyperK\"ahler forms
reads
\begin{align}
\mathbf{K}^+&=\textrm{Tr}(d\mathbf{u}\wedge d\mathbf{v}) \\
\mathbf{K}^-&=-\textrm{Tr}(d\mathbf{v}^{\dagger}\wedge d\mathbf{u}^\dagger) \\
\mathbf{K}^3&=\frac{i}{2}\textrm{Tr}(d\mathbf{u}\wedge
d\mathbf{u}^\dagger-d\mathbf{v}^\dagger\wedge d\mathbf{v}).
\end{align}
The action of the gauge group on $\mathbf{u}$ is by $A\in
\mathrm{SU(n)}$ on the left and the dual of $B\in \mathrm{SU(m)}$ on
the right and similarly for $\mathbf{v}$
\begin{align}\label{eq:gauge_action}
\mathbf{u}\mapsto\mathbf{A}\mathbf{u}\left(\mathbf{B}^{-1}\right)^T=
\mathbf{A}\mathbf{u}\overline{\mathbf{B}},
\quad
\mathbf{v}\mapsto\mathbf{B}\mathbf{v}\left(\mathbf{A}^{-1}\right)^T=
\mathbf{B}\mathbf{v}\overline{\mathbf{A}},
\end{align}
while the $\mathfrak{u}(1)$ acts as
\begin{align}
    \mathbf{u}\mapsto e^{i\phi}\mathbf{u},\quad \mathbf{v}\mapsto e^{-i\phi}\mathbf{v}.
\end{align}
Writing $\mathbf{A}$ and $\mathbf{B}$ in infinitesimal form\footnote{Here the adjoint index
$\Lambda$ splits in $(a,\widehat{a}, \bullet)$ for $(\mathfrak{su}(n)$,$\mathfrak{su}(m)$,
$\mathfrak{u}(1))$ respectively.}
\begin{align}
\mathbf{A}^{(a)}&=\exp^{it\mathbf{T}^a},\;\; \mathbf{T}^a:\textrm{ generator of }  \mathfrak{su}(n) \\
\mathbf{B}^{(\widehat{a})}&=\exp^{it\mathbf{T}^{\widehat{a}}},\;\;
\mathbf{T}^{\widehat{a}}:\textrm{ generator of } \mathfrak{su}(m)
\end{align}
and defining the Killing vectors as generators of the gauge group
action in~\eqref{eq:gauge_action}
\begin{align}
    k^a_{\mathfrak{su}(n)}=\textrm{Tr}\left[ \left( \frac{d}{dt}\left.
    \mathbf{u}_{\textrm{new}}^{(a)}\right\vert_{t=0}\right)\frac{\partial}{\partial\mathbf{u}}
    + \left( \frac{d}{dt}\left.\mathbf{v}_{\textrm{new}}^{(a)}\right\vert_{t=0}\right)
    \frac{\partial}{\partial\mathbf{v}} + c.c.\right],
\end{align}
leads to
\begin{align}\label{eq:KV_n}
k^a_{\mathfrak{su}(n)}=i\textrm{Tr}\left(
\frac{\partial}{\partial\mathbf{u}}\mathbf{T}^a\mathbf{u}-\mathbf{u}^\dagger
\overline{\mathbf{T}}^a\frac{\partial}{\partial\mathbf{u}^\dagger}
-\mathbf{v}\overline{\mathbf{T}}^a\frac{\partial}{\partial\mathbf{v}}
+\frac{\partial}{\partial\mathbf{v}^\dagger}\mathbf{T}^a\mathbf{v}^\dagger\right).
\end{align}
A word by word derivation holds true also for $\mathfrak{su}(m)$ and
$\mathfrak{u}(1)$. The final formulae for the corresponding Killing
vectors are
\begin{align}
    k^{\widehat{a}}_{\mathfrak{su}(m)}&=i\textrm{Tr}\left(
    -\mathbf{u}\overline{\mathbf{T}}^{\widehat{a}}\frac{\partial}{\partial\mathbf{u}}
    +\frac{\partial}{\partial\mathbf{u}^\dagger}\mathbf{T}^{\widehat{a}}\mathbf{u}^\dagger
    +\frac{\partial}{\partial\mathbf{v}}\mathbf{T}^{\widehat{a}}\mathbf{v}
    -\mathbf{v}^\dagger\overline{\mathbf{T}}^{\widehat{a}}\frac{\partial}{\partial\mathbf{v}^\dagger}
    \right) \label{eq:KV_m} \\
k_{\mathfrak{u}(1)}&=i\textrm{Tr}\left(
\mathbf{u}\frac{\partial}{\partial\mathbf{u}}
-\mathbf{u}^\dagger\frac{\partial}{\partial\mathbf{u}^\dagger}
-\mathbf{v}\frac{\partial}{\partial\mathbf{v}}
+\mathbf{v}^\dagger\frac{\partial}{\partial\mathbf{v}^\dagger}
\right) \label{eq:KV_1}.
\end{align}
In the next step one computes the moment maps defined as
\begin{align}
    d\mathscr{P}^{+a}_{\mathfrak{su}(n)}&=i_{k^a}\mathbf{K}^+ \\
    d\mathscr{P}^{-a}_{\mathfrak{su}(n)}&=i_{k^a}\mathbf{K}^- \\
    d\mathscr{P}^{3a}_{\mathfrak{su}(n)}&=i_{k^a}\mathbf{K}^3
\end{align}
for $\mathfrak{su}(n)$ and equivalently also for $\mathfrak{su}(m)$
and $\mathfrak{u}(1)$. The resulting expressions are respectively
\begin{align}
    \mathscr{P}^{+a}_{\mathfrak{su}(n)}&=i \textrm{Tr}\left(\mathbf{v}\mathbf{T}^a\mathbf{u} \right)\\
    \mathscr{P}^{-a}_{\mathfrak{su}(n)}&=-i\textrm{Tr}\left(
    \mathbf{u}^\dagger\mathbf{T}^a\mathbf{v}^\dagger \right)\\
    \mathscr{P}^{3a}_{\mathfrak{su}(n)}&=-\frac{1}{2}\textrm{Tr}\left(
    \mathbf{u}^\dagger\mathbf{T}^a\mathbf{u}-\mathbf{v}\mathbf{T}^a\mathbf{v}^\dagger \right)
\end{align}
for $\mathfrak{su}(n)$,
\begin{align}
    \mathscr{P}^{+\widehat{a}}_{\mathfrak{su}(m)}&=-i \textrm{Tr}\left(\mathbf{u}
    \mathbf{T}^{\widehat{a}}\mathbf{v} \right)\\
    \mathscr{P}^{-\widehat{a}}_{\mathfrak{su}(m)}&=i\textrm{Tr}\left(
    \mathbf{v}^\dagger\mathbf{T}^{\widehat{a}}\mathbf{u}^\dagger \right)\\
    \mathscr{P}^{3\widehat{a}}_{\mathfrak{su}(m)}&=-\frac{1}{2}\textrm{Tr}\left(
    -\mathbf{u}\mathbf{T}^{\widehat{a}}\mathbf{u}^\dagger
    +\mathbf{v}^\dagger\mathbf{T}^{\widehat{a}}\mathbf{v}\right)\\
\end{align}
for $\mathfrak{su}(m)$ and finally
\begin{align}
\mathscr{P}^{+}_{\mathfrak{u}(1)}   =& i\textrm{Tr}\left( \mathbf{u}\mathbf{v} \right) \\
    \mathscr{P}^{-}_{\mathfrak{u}(1)}   =& -i\textrm{Tr}\left( \mathbf{v}^\dagger\mathbf{u}^\dagger
 \right) \\
    \mathscr{P}^{3}_{\mathfrak{u}(1)}=  & -\frac{1}{2}\textrm{Tr}\left(\mathbf{u}\mathbf{u}^\dagger
    -\mathbf{v}^\dagger\mathbf{v} \right)
\end{align}
for $\mathfrak{u}(1)$.

Next, we would like to verify that the moment map constraints
\begin{align}
\begin{split}\label{eq:MM_con1} &\partial_i\left(\mathscr{P}^+\cdot\mathscr{P}^+\right)
=\partial_{\bar{j}}\left(\mathscr{P}^-\cdot\mathscr{P}^-\right)=0 \\
    &\partial_i\left(\mathscr{P}^+\cdot\mathscr{P}^3\right)=
    \partial_{\bar{j}}\left(\mathscr{P}^-\cdot\mathscr{P}^3\right)=0 \\
    &\partial_{\bar{j}}\left(\mathscr{P}^+\cdot\mathscr{P}^3\right)=
    \partial_i\left(\mathscr{P}^-\cdot\mathscr{P}^3\right)=0
\end{split}\\
&\partial_i\partial_{\bar{j}}\left(2\mathscr{P}^3\cdot
\mathscr{P}^3-\mathscr{P}^+\cdot\mathscr{P}^-\right)=0
\label{eq:MM_con2},
\end{align}
which imply supersymmetry enhancement are satisfied. This requires
in particular finding the correct quadratic form (denoted by a
$\cdot$ in the formulae above) on the gauge Lie algebra
$\mathfrak{su}(n)\oplus\mathfrak{su}(m)\oplus\mathfrak{u}(1)$. In
fact, we will see that for a flat target space a stronger version of
the constraints holds true, such that the products of the moment
maps in parenthesis vanish by themselves. We will explicitly check
the most involved constraint in the last line. The rest of the
constraints can be easily verified to vanish as well. Parameterizing
the quadratic form on
$\mathfrak{su}(n)\oplus\mathfrak{su}(m)\oplus\mathfrak{u}(1)$ by a
relative sign between the Killing form on $\mathfrak{su}(n)$ and $\mathfrak{su}(m)$ and
by a constant $c$ for the $\mathfrak{u}(1)$ factor one gets for
instance
\begin{align}
\mathscr{P}^3\cdot \mathscr{P}^3= \mathbf{\kappa}_{\mathfrak{su}(n)}\left(
\mathscr{P}^3_{\mathfrak{su}(n)},\mathscr{P}^3_{\mathfrak{su}(n)}
\right)   \pm  \mathbf{\kappa}_{\mathfrak{su}(m)}\left(
\mathscr{P}^3_{\mathfrak{su}(m)},\mathscr{P}^3_{\mathfrak{su}(m)}
\right) + c
\mathscr{P}^3_{\mathfrak{u}(1)}\mathscr{P}^3_{\mathfrak{u}}(1).
\end{align}
Employing the completeness relation for $\mathfrak{su}(n)$ (which
implicitly fixes the normalization of the generators)
\begin{align}
\left( \mathbf{T}^a \right)^j_{\phantom{j}i} \left( \mathbf{T}_a
\right)^p_{\phantom{p}q}=\frac{1}{2} \left(
\delta^j_{\phantom{j}q}\delta^p_{\phantom{p}i} -
\frac{1}{n}\delta^j_{\phantom{j}i}\delta^p_{\phantom{p}q} \right)
\end{align}
we arrive at
\begin{align}
2\mathscr{P}^3\cdot \mathscr{P}^3&=
\Bigg[ \frac{1}{4}\left(\textrm{Tr}(\mathbf{u}^\dagger\mathbf{u}\mathbf{u}^\dagger\mathbf{u})
-\frac{1}{n}\left( \textrm{Tr}(\mathbf{u}^\dagger\mathbf{u}) \right)^2 \right)
-\frac{1}{2}\left( \textrm{Tr}(\mathbf{u}^\dagger\mathbf{v}^\dagger\mathbf{v}\mathbf{u})
-\frac{1}{n}\textrm{Tr}(\mathbf{u}^\dagger\mathbf{u})\textrm{Tr}(\mathbf{v}^\dagger\mathbf{v}) \right)
\notag \\
&+\frac{1}{4}\left( \textrm{Tr}(\mathbf{v}\mathbf{v}^\dagger\mathbf{v}\mathbf{v}^\dagger)
-\frac{1}{n}\left( \textrm{Tr}(\mathbf{v}\mathbf{v}^\dagger) \right)^2 \right) \Bigg] \notag \\
&\pm \Bigg[ \frac{1}{4}\left( \textrm{Tr}(\mathbf{u}\mathbf{u}^\dagger\mathbf{u}\mathbf{u}^\dagger)
-\frac{1}{m}\left( \textrm{Tr}(\mathbf{u}\mathbf{u}^\dagger) \right)^2 \right)
-\frac{1}{2}\left( \textrm{Tr}(\mathbf{u}\mathbf{v}\mathbf{v}^\dagger\mathbf{u}^\dagger)
-\frac{1}{m}\textrm{Tr}(\mathbf{u}\mathbf{u}^\dagger)\textrm{Tr}(\mathbf{v}^\dagger\mathbf{v}) \right)
\notag \\
 &+\frac{1}{4}\left(
\textrm{Tr}(\mathbf{v}^\dagger\mathbf{v}\mathbf{v}^\dagger\mathbf{v})
-\frac{1}{m}\left( \textrm{Tr}(\mathbf{v}^\dagger\mathbf{v}) \right)^2 \right)\Bigg] \notag \\
& +c\Bigg[\frac{1}{2}\left( \textrm{Tr}(\mathbf{u}
{\mathbf{u}}^\dagger) \right)^2
-\textrm{Tr}(\mathbf{u}\mathbf{u}^\dagger)\textrm{Tr}(\mathbf{v}\mathbf{v}^\dagger)
+\frac{1}{2}\left( \textrm{Tr}(\mathbf{v}\mathbf{v}^\dagger)\right)^2
\Bigg]
\end{align}
and
\begin{align}
\mathscr{P}^+\cdot\mathscr{P}^-&=\frac{1}{2}\left(
\textrm{Tr}(\mathbf{v}\mathbf{v}^\dagger\mathbf{u}^\dagger\mathbf{u})
-\frac{1}{n}\textrm{Tr}(\mathbf{v}\mathbf{u})\textrm{Tr}(\mathbf{u}^\dagger\mathbf{v}^\dagger) \right)
\notag \\
&\pm\left(
\textrm{Tr}(\mathbf{u}\mathbf{u}^\dagger\mathbf{v}^\dagger\mathbf{v})
-\frac{1}{m}\textrm{Tr}(\mathbf{u}\mathbf{v})\textrm{Tr}(\mathbf{v}^\dagger\mathbf{u}^\dagger)
\right)+c\textrm{Tr}(\mathbf{u}\mathbf{v})\textrm{Tr}(\mathbf{v}^\dagger\mathbf{u}^\dagger).
\end{align}
Subtracting the two expressions above and imposing the result to
vanish fixes the relative minus sign between the Killing forms for
$\mathfrak{su}(n)$ and $\mathfrak{su}(m)$ and the constant $c$ as
\begin{align}
c=\frac{m-n}{2mn}.
\end{align}
Therefore we can conclude that with the choice of
$\mathfrak{m}$ for $\mathfrak{su}(n)\oplus\mathfrak{su}(m)\oplus\mathfrak{u}(1)$\footnote{We can also conclude that if $m = n$ the gauge Lie algebra would be $\su(n) \oplus \su(n)$. The choice of $\mathfrak{m}$ would be the same but the $\mathfrak{u}(1)$ piece would be excluded. }
\begin{align}
\pm \left(
\mathbf{1}_{\mathfrak{su}(n)},-\mathbf{1}_{\mathfrak{su}(m)},\frac{m-n}{2mn}
\right)
\end{align}
the constraints~\eqref{eq:MM_con1} and~\eqref{eq:MM_con2} are
satisfied and thus supersymmetry is enhanced from $\mathscr{N}=3$ to $\mathscr{N}=4$.
\section{Gamma matrix conventions and R-symmetry}
\label{rsimmaconvo} In view of what we have discussed in the
introduction we provide here a careful consideration of the
$R$--symmetry enhancement that occurs when we dimensionally reduce
an $\mathcal{N}_4=2$ gauge theory from $D=4$ down to $D=3$
\subsection{The enhancement of R--symmetry}
Prior to dimensional reduction $D=4\to D=3$ the $R$--symmetry of an
$\mathcal{N}_4$-extended gauge theory in four-dimensions is
$\mathrm{U(\mathcal{N}_4)}=\mathrm{U(1)} \times
\mathrm{SU(\mathcal{N}_4)}$, whose infinitesimal action on the
$\mathcal{N}_4$  Majorana supercharges is the following:
\begin{equation}
  \delta Q_A =\left [ A_{AB} + i S_{AB} \, \gamma_5 \right] \, Q_B
  \quad ; \quad A,B=1,\dots, \mathcal{N}_4
\label{uNact}
\end{equation}
where $A_{AB} =-A_{BA}$ is an antisymmetric matrix and
$S_{AB}=S_{BA}$ is a symmetric one. Taking the chiral projection of
the Majorana spinor:
\begin{eqnarray}
  \mathcal{Q}_A &=& \ft{1}{2} \left( 1 +\gamma_5\right) Q^A
  \nonumber\\
\mathcal{Q}^A &=& \ft{1}{2} \left( 1 -\gamma_5\right) Q^A
\label{chiralpro}
\end{eqnarray}
we obtain the standard complex action of $u(\mathcal{N}_4)$:
\begin{equation}
  \delta\mathcal{Q}_A =\left [ A_{AB} + i S_{AB} \right] \,=
  \mathcal{Q}_B \equiv U_A^{\phantom{A}B} \, \mathcal{Q}_B
\label{mathcalQ}
\end{equation}
and the complex conjugate transformation for $\mathcal{Q}^B$. The
same transformations apply to the other spinors (gauginos,
hyperinos, etc) and bosons (the HyperK\"ahlerian vielbein $U^{\alpha
A}$) with the same $R$--symmetry index.
\par
After dimensional reduction the $R$--symmetry of the
three--dimensional theory is enhanced from
$\mathrm{U(\mathcal{N}_4)}$ to $\mathrm{SO(2\mathcal{N}_4)}$. This
is essentially due to the splitting of each four--component Majorana
spinor into a doublet of two--component Majorana spinors. It is
important to follow the details of this enhancement mechanism since
it is at the level of this symmetry that one finds the new dynamical
patterns possible in $D=3$ and not available in $D=4$ in particular
the breaking of $\mathcal{N}_3=4$ supersymmetry down to
$\mathcal{N}_3=3$ via the introduction of a Chern Simons term. The
re-enhancement mechanism from $\mathcal{N}_3\,=\,3  \to \,
\mathcal{N}_3 \, = \, 4$, that constitutes the main issue of the
present paper, is nothing else but the restoration of the original
$\mathrm{SO(4)}$ arising in dimensional reduction.
\par
To begin with let us recall the embedding of
$\mathrm{U(\mathcal{N}_4)}$ into $\mathrm{SO(2\mathcal{N}_4)}$ and
the structure of the coset space
$\mathrm{SO(2\mathcal{N}_4)}/\mathrm{U(\mathcal{N}_4)}$. Let us work
at the Lie algebra level and set:
\begin{equation}
\begin{array}{rclcrcl}
\mathbb{G} & \equiv & \so(2\mathcal{N}_4) &; & \mathbb{H} & \equiv & \uu(\mathcal{N}_4)\\
\mathbb{G} & = & \mathbb{H}\oplus \mathbb{K} & ; & \left[
\mathbb{H},\mathbb{H}\right] &= & \mathbb{H} \quad ,\quad \left[
\mathbb{H},\mathbb{K}\right]=\mathbb{K}\quad ,\quad \left[
\mathbb{K},\mathbb{K}\right]=\mathbb{H}  \\
\end{array}
\label{hkcoset}
\end{equation}
A generic antisymmetric $2\mathcal{N}_4 \times 2\mathcal{N}_4$
matrix can be decomposed as follows:
\begin{equation}
 \mathbf{M} \in \so(2\mathcal{N}_4) \quad \Rightarrow \quad  \mathbf{M} \, = \, \underbrace{
 \left(  \begin{array}{c|c}
  A & -S \\
  \hline
  S & A \\
\end{array} \right) }_{h\in\mathbb{H}=\uu(\mathcal{N}_4)} \oplus \underbrace{
 \left(  \begin{array}{c|c}
  B & C \\
  \hline
  C & -B \
  \end{array} \right) }_{k\in\mathbb{K} } \nonumber\\
\label{Hkdecom}
\end{equation}
where:
\begin{equation}
  A=-A^T \quad , \quad B=-B^T \quad , \quad C=-C^T
\label{asymABC}
\end{equation}
are antisymmetric $\mathcal{N}_4\times \mathcal{N}_4$ matrices and
\begin{equation}
  S=S^T
\label{symmS}
\end{equation}
is instead symmetric. The first matrix on the left-hand side of
(\ref{Hkdecom} belongs to $\uu(\mathcal{N}_4)$ subalgebra, while the
second matrix belongs to the orthogonal subspace $ \mathbb{K}$ whose
dimension is $\mathcal{N}_4(\mathcal{N}_4-1)$. We will see how this
decomposition is relevant to the enhancement of $R$--symmetry after
dimensional reduction.
\par
To grasp this phenomenon in a clean way we need to choose a well
adapted gamma matrix basis.
\subsection{The gamma matrix basis} In three dimensions we follow
the conventions of \cite{Fabbri:1999ay} and we set:
\begin{equation}
  \begin{array}{ccccccccccc}
    \gamma^0 & = & \sigma^2 & = &\left( \begin{array}{cc}
      0 & -i \\
      i & 0 \\
    \end{array}\right)  & ; & \gamma^1 & =&-i \, \sigma^3 &= &
     \left( \begin{array}{cc}
       -i & 0 \\
       0 & i \\
\end{array}\right)  \\
\null &  \null & \null & \null & \null & \null & \null & \null &
\null &
        \null & \null \\
    \gamma^2 & = &-i \,  \sigma^1 & = &\left( \begin{array}{cc}
      0 & -i \\
      -i & 0 \\
    \end{array}\right)  & ; & C_{[3]} & = & -i\, \sigma^2 &=&
     \left( \begin{array}{cc}
       0 & -1 \\
       1 & 0 \\
\end{array}\right) \
  \end{array}
\label{3dgamma}
\end{equation}
Then if we explicitly write the Majorana condition for a spinor
$\theta$ we obtain:
\begin{equation}
  \theta= \theta^{c} \equiv  C_{[3]} \, {\bar \theta} = i \, \theta^*
\label{maiocond}
\end{equation}
so that we can write:
\begin{equation}
  \theta= \exp [i\pi/4] \, \left( \begin{array}{c}
    \theta_1 \\
    \theta_2 \
  \end{array}\right) \quad \mbox{with} \quad \theta_{1,2}^* =
  \theta_{1,2} \quad \mbox{real}
\label{expmaio}
\end{equation}
To make nice contact between  four and three dimensions which is
instrumental in order to derive the $D=3$ supersymmetry
transformations of hypermultiplet fields from their $D=4$ susy
transformations we choose the following basis of $D=4$ gamma
matrices:
\begin{equation}
  \begin{array}{ccccc}
    \gamma^0_{[4]} & = & \left( \begin{array}{c|c}
      \gamma^0 & 0 \\
      \hline
      0 & -\gamma^0 \\
    \end{array}\right)  & = & \sigma^2 \otimes \sigma^3 \\
 \gamma^1_{[4]} & = & \left( \begin{array}{c|c}
      \gamma^1 & 0 \\
      \hline
      0 & -\gamma^1 \\
    \end{array}\right)  & = & -i \, \sigma^3 \otimes \sigma^3
  \\
\gamma^2_{[4]} & = & \left( \begin{array}{c|c}
      \gamma^2 & 0 \\
      \hline
      0 & -\gamma^2 \\
    \end{array}\right)  & = & -i \, \sigma^1 \otimes \sigma^3
   \\
\gamma^3_{[4]} & = & \left( \begin{array}{c|c}
      0     &- {\bf 1} \\
      \hline
      {\bf 1} & 0   \\
    \end{array}\right)  & = & -i \, {\bf 1} \otimes \sigma^2
   \\
\gamma^5_{[4]} & = & \left( \begin{array}{c|c}
      0     &- {\bf 1} \\
      \hline
      -{\bf 1} & 0   \\
    \end{array}\right)  & = & -  \, {\bf 1} \otimes \sigma^1
  \\
  C_{[4]} & = & \left( \begin{array}{c|c}
       C_{[3]}    &0\\
      \hline
      0 &  C_{[3]}   \\
    \end{array}\right)  & = & -i \, \sigma^2 \otimes {\bf 1} \
   \end{array}
\label{4gamma}
\end{equation}
We can now verify the decomposition of a four dimensional Majorana
spinor under dimensional reduction. We set:
\begin{equation}
  \psi=\left( \begin{array}{c}
    \vartheta_1 \\
    \vartheta_2 \
  \end{array}\right)
\label{psi=due}
\end{equation}
and we obtain:
\begin{equation}
\psi=\left( \begin{array}{c}
    \vartheta_1 \\
    \vartheta_2 \
  \end{array}\right)=  C_{[4]} \, {\bar \psi}^T = \left( \begin{array}{c}
    i \, \vartheta^*_1 \\
    -i \, \vartheta^*_2 \
  \end{array}\right)
\label{psixichi}
\end{equation}
so that we can conclude:
\begin{equation}
  \vartheta_1=\xi_1 \quad ; \quad \vartheta_2= -i \xi_2 , \quad \mbox{where} \quad
  \xi_{1,2} = \mbox{Majorana spinors in $D=3$}
\label{2majora}
\end{equation}
\subsection{Dimensional reduction of the supersymmetry algebra}
Let us now consider the dimensional reduction of the
$\mathcal{N}_4$--extended supersymmetry algebra that we write in its
dual form utilizing Maurer Cartan equations:
\begin{equation}
    dV^{\hat a}   = \, \ft{i}{2}\, {\bar \Psi}_A \, \wedge\,
    \gamma^{\hat a}_{[4]} \, \Psi_A \quad ; \quad \hat a, \hat
    b=0,1,2,3
\label{torsion4}
\end{equation}
where $V^{\hat a}$ is the vielbein $1$--form of rigid superspace and
$\Psi_A$ ($A=1,\dots, \mathcal{N}_4$ ) is the  gravitino 1--form,
namely an $\mathcal{N}_4$-tuplet of Majorana spinor fermionic
one-forms. Using the above defined gamma matrix basis we immediately
find:
\begin{eqnarray}
  dV^{a} &=&\, \ft{i}{2}\, \left[{\bar \xi}^{1}_A \,\wedge \,
    \gamma^{  a}_{[3]} \, \xi^1_A + {\bar \xi}^{2}_A \,\wedge\,
    \gamma^{  a}_{[3]} \, \xi^2_A \right] \quad ; \quad a=0,1,2\label{3dsusy}\\
  dV^{3} &=& \,\ft{1}{2}\,\left[{\bar \xi}^{1}_A \, \wedge
     \, \xi^2_A + {\bar \xi}^{2}_A \, \wedge
      \, \xi^1_A \right] \label{3dcentral}
\end{eqnarray}
The  supersymmetry algebra (\ref{torsion4}) is invariant against the
$\uu(\mathcal{N}_4)$ transformations (\ref{uNact}) where $Q_A$ is
replaced by $\Psi_A$. The same is obviously true of eq.s
(\ref{3dsusy},\ref{3dcentral}) which are just a transcription of the
same algebra. However if we delete  eq.(\ref{3dcentral}), then eq.
(\ref{3dsusy}) which is the $2\mathcal{N}_4$--extended supersymmetry
algebra in $D=3$ is invariant against $\so(2\mathcal{N}_4)$
transformations: it suffices to consider $\left( \xi^1_A,
\xi^2_A\right) $ as a column $2\mathcal{N}_4$ vector in the defining
representation of $\so(2\mathcal{N}_4)$. Disregarding
eq.(\ref{3dcentral}) has a clearcut physical meaning. Indeed $V^3$
is the 1--form dual to the translation generator in the 3-rd
direction, namely $P_3$. Hence, in the dual language, disregarding
eq.(\ref{3dcentral}) means that we set $P_3=0$. This is just the
very idea of dimensional reduction: we restrict our attention to
field configurations that have zero momentum in the third direction,
namely that are independent from $x^3$. On the $P_3=0$ slice we have
an enhancement of $R$--symmetry which is promoted from
$\uu(\mathcal{N}_4)$ to $\so(2\mathcal{N}_4)$.
\subsection{The relevant case $\mathcal{N}_4=2$}
Let us now consider the chiral projections of the Majorana gravitino
one-forms $\Psi_A$ pertaining to the four-dimensional theory. We
have:
\begin{eqnarray}
  \psi_A &=& \ft{1}{2}\, \left( 1+\gamma^5_{[4]} \right ) \Psi_A =\left( \begin{array}{c}
     \chi_A \\
    -\chi_A \
  \end{array}\right)=\left( \begin{array}{c}
      \ft{1}{2}\left( \xi^1_A + i \,\xi^2_A\right) \\
    -\ft{1}{2}\left( \xi^1_A +i \, \xi^2_A\right) \
  \end{array}\right)\nonumber\\
  \psi^A &=& \ft{1}{2}\, \left( 1-\gamma^5_{[4]} \right ) \Psi_A =\left( \begin{array}{c}
     \chi^A \\
    \chi^A \
  \end{array}\right)=\left( \begin{array}{c}
      \ft{1}{2}\left( \xi^1_A - i \,\xi^2_A\right) \\
    \ft{1}{2}\left( \xi^1_A -i \, \xi^2_A\right) \
  \end{array}\right)
\label{trigormir}
\end{eqnarray}
where $\chi_A$ is a generic 2--component spinor in $D=3$ (no
Majorana condition) and $\chi^A$ is just its conjugate:
\begin{equation}
  \chi_A^c = C_{[3]} {\bar \chi}_A^T \equiv \chi^A
\label{chiachino}
\end{equation}
The $R$--symmetry transformations on the $\chi_A$ gravitino 1--forms

are easily derived by comparing with equations (\ref{Hkdecom}). We
find:
\begin{eqnarray}
  \delta  \, \chi_A &=& U_{A}^{\phantom{A}B} \chi_B + \mathcal{A}_{AB}
  \, \chi^B \label{dechi}\\
\delta  \, \chi^A &=&  U^{A}_{\phantom{A}B} \chi^B +
\bar{\mathcal{A}}^{AB}
  \, \chi_B \label{dechicong}
\label{UandA}
\end{eqnarray}
where:
\begin{eqnarray}
  U_{A}^{\phantom{A}B}&=& A_{AB} + i \, S_{AB} \, \in \, \mathbb{H} =
  \uu(\mathcal{N}_4)\label{nocong} \\
\mathcal{A}_{AB}&=& B_{AB} + i \, C_{AB} \, \in \, \mathbb{K} =
  \so(2\mathcal{N}_4)/\uu(\mathcal{N}_4)\label{cong}
\label{mixcong}
\end{eqnarray}
Eq.s (\ref{dechi},\ref{dechicong}) are the holomorphic transcription
of eq.s (\ref{Hkdecom}), namely the decomposition of the adjoint of
$\so(2\mathcal{N}_4)$ with respect to $\uu(\mathcal{N}_4)$:
\begin{equation}
  \mbox{adj} \, \so(2\mathcal{N}_4) = \mbox{adj} \, \uu(\mathcal{N}_4) \oplus  {\wedge^2
  \mbox{fundamental}} \oplus  {\wedge^2
  \mbox{anti-fundamental}}
\label{adjoino}
\end{equation}
On the other hand recalling eq.(\ref{trigormir}) we see that the
transformation under $R$-symmetry of the doublet of spinors
$\{\xi^1_A \, , \, \xi^2_A \}$ is the following one:
\begin{equation}\label{crostatica}
    \delta_R \left ( \begin{array}{c} \xi^1_A \\ \hline
    \xi^2_A \\ \end{array} \right) \, = \, \mathbf{M} \,\left ( \begin{array}{c} \xi^1_A \\ \hline
    \xi^2_A \\ \end{array} \right) \, = \, \left(\begin{array}{c|c} A+B &
    C-S \\ \hline
    C+S & A-B  \end{array} \right)\, \left ( \begin{array}{c} \xi^1_A \\ \hline
    \xi^2_A \\ \end{array} \right)
\end{equation}
where the $4\times 4$ antisymmetric matrix $\mathbf{M}$ is that
defined in eq. (\ref{Hkdecom}). It follows that the doublet of
spinors $\{\xi^1_A \, , \, \xi^2_A \}$ transforms in the fundamental
defining representation of  $\so(2\mathcal{N}_4)$. The subalgebra
$\uu(\mathcal{N}_4)$ inherited from higher dimensions is that which
does not mix the complex supercharges with their conjugates. The
enhancement of $R$--symmmetry produced by the dimensional reduction
is given by the antisymmetric representation $ {\wedge^2
\mbox{fundamental}}$ that mixes conjugate supercharges.
\par
  There are two important observations:
\begin{enumerate}
  \item When $\mathcal{N}_4=1$ there is no $R$--symmetry enhancement. This is the
  only case where the $\wedge^2 \mbox{fundamental}$ vanishes and we
  have $\mathrm{U(1)}\sim \mathrm{SO(2)}$ both in four and three dimensions.
  \item When $\mathcal{N}_4=2$ the enhanced $R$-symmetry algebra is:
\begin{equation}
  \so(4)\sim \su(2)_L \oplus \su(2)_R
\label{so4}
\end{equation}
which, because of the accidental isomorphism splits into two simple
subalgebras. The supercharges, which before reduction were in the
fundamental $2$ of $\uu(2)$ are, after reduction in the fundamental
${\bf 4}$ of $\so(4)$. With respect to the decomposition (\ref{so4})
one has $4\sim (2,2)$. With respect to the diagonal subalgebra
$\su(2)_{diag}= \mbox{diag} \left( \su(2)_L \oplus \su(2)_R \right
)$ we have: ${\bf 4} \to {\bf 3} \oplus {\bf 1}$, so that we can
decompose the supercharges into a triplet plus a singlet and
consider new terms in the lagrangian that violate the fourth
supercharge preserving the other three. This is the way to construct
$\mathcal{N}_3=3$ theories in three dimensions. \item The
enhancement of specially constructed Chern Simons theories from
$\mathcal{N}_3\,=\,3$ to $\mathcal{N}_3\,=\,4$ is associated with a
full reinstallment of the natural $\so(4)$ produced by a
hypothetical dimensional reduction from $D=4$.
\end{enumerate}
Let us now focus on   $\mathcal{N}_4=2$   and reconsider the
specific form of the decomposition (\ref{Hkdecom}) in this case.  A
complete basis for the antisymmetric $4\times 4$ matrices, namely
for the $\mathfrak{so}(4)$ Lie algebra is provided by the  't Hooft  matrices.
These are
  $4\times 4$ real antisymmetric, (anti)self--dual
  matrices which satisfy the following relations:
\begin{equation}
     \left|  \begin{array}{rcl}
\stackrel{\pm ~}{J^{ab}_x} & = &
\pm \, \epsilon^{abcd} \,\stackrel{\pm ~}{J^{ab}_x}  \\
\stackrel{\pm ~}{J_x} \, \stackrel{\pm ~}{J_y}& = &
-\delta_{xy}-\epsilon^{xyz} \,
\stackrel{\pm ~}{J_z} \\
\left [ \stackrel{+ ~}{J_x} \, , \,  \stackrel{- ~}{J_x}\right ] & =
& 0 \
\end{array} \right \} \quad
\left(  x,y,..=1,2,3 \quad;\quad a,b,c,..=1,2,3,4\right)
\label{thofto}
\end{equation}
The explicit form for the 't Hooft matrices is the following:
\begin{equation}
  \begin{array}{cccccc}
    \stackrel{+ ~}{J_1} & = & \left( \begin{array}{cccc} 0 & 0 & 0 & -1 \\ 0 & 0 & -1 & 0
    \\
0 & 1 & 0 & 0 \\
      1 & 0 & 0 & 0 \\  \end{array}  \right) & \stackrel{- ~}{J_1} & = &
      \left( \begin{array}{cccc} 0 & 0 & 0 & 1 \\ 0 & 0 & -1 & 0 \\
        0 & 1 & 0 & 0 \\ -1 & 0 & 0 & 0 \\  \end{array}  \right) \\
       \null  & \null  & \null  & \null  & \null  & \null  \\
    \stackrel{+ ~}{J_2}  &  = &  \left( \begin{array}{cccc} 0 & -1 & 0 & 0 \\ 1 & 0 & 0 & 0
    \\
0 & 0 & 0 & -1 \\
      0 & 0 & 1 & 0 \\  \end{array}   \right) & \stackrel{- ~}{J_2} & = &
      \left( \begin{array}{cccc} 0 & 1 & 0 & 0 \\
        -1 & 0 & 0 & 0 \\ 0 & 0 & 0 & -1 \\
        0 & 0 & 1 & 0 \\  \end{array}   \right)  \\
 \null & \null & \null & \null & \null & \null \\
    \stackrel{+ ~}{J_2} & = & \left( \begin{array}{cccc} 0 & 0 & 1 & 0 \\ 0 & 0 & 0 & -1
    \\
-1 & 0 & 0 & 0 \\
        0 & 1 & 0 & 0 \\ \end{array} \right) & \stackrel{- ~}{J_3} & = &
        \left( \begin{array}{cccc} 0 & 0 & -1 & 0 \\
        0 & 0 & 0 & -1 \\ 1 & 0 & 0 & 0 \\
      0 & 1 & 0 & 0 \\ \end{array} \right) \
  \end{array}
\label{jpm}
\end{equation}
Comparing eq.(\ref{Hkdecom}) with eq.(\ref{jpm}) one sees that the
generators of the $\uu(2) \subset \so(4)$ subalgebra already present
in four dimensions are:
\begin{equation}
  \uu(2) = \mbox{span} \left[ \stackrel{+ ~}{J_1}, \stackrel{+ ~}{J_2}
  , \stackrel{+ ~}{J_3} , \stackrel{- ~}{J_3} \right ]
\label{u2inso4}
\end{equation}
Indeed $\stackrel{- ~}{J_2}$ contains the trace part of the
symmetric $2 \times 2$ matrix $S_{AB}$. On the other hand the
$\wedge^2{{\rm fundamental}}$ that enhances $R$-symmetry in three
dimensions is provided by:
\begin{equation}
  \wedge^2{{\rm fundamental}}=\mbox{span} \left[ \stackrel{- ~}{J_1}, \stackrel{- ~}{J_2}
   \right ]
\label{wedge2}
\end{equation}
If we consider the explicit form of the diagonal $\su(2)_{diag}$
generators we find:
\begin{eqnarray}
J^1_{diag}=\ft{1}{2} \left( \stackrel{+ ~}{J_1}+ \stackrel{-
~}{J_1}\right)
  = \left( \begin{array}{cccc}
    0 & 0 & 0 & 0 \\
    0 & 0 & -1 & 0 \\
    0 & 1 & 0 & 0 \\
    0 & 0 & 0 & 0 \
  \end{array}\right) \nonumber\\
  J^2_{diag}=\ft{1}{2} \left( \stackrel{+ ~}{J_2}+ \stackrel{- ~}{J_2}\right)
  = \left( \begin{array}{cccc}
    0 & 0 & 0 & 0 \\
    0 & 0 & 0 & 0 \\
    0 & 0 & 0 & -1 \\
    0 & 0 & 1 & 0 \
  \end{array}\right) \nonumber\\
   J^3_{diag}=\ft{1}{2} \left( \stackrel{+ ~}{J_3}+ \stackrel{- ~}{J_3}\right)
  = \left(\begin{array}{cccc}
    0 & 0   & 0 & 0 \\
    0 & 0 & 0  & -1 \\
    0 & 0  & 0 & 0 \\
    0 & 1 & 0 & 0 \
  \end{array}\right) \nonumber\\
\label{su2diag}
\end{eqnarray}
As one sees these are the rotation generators on the three space
spanned by the axes (2-3-4). The first direction being left
invariant. Hence the triplet of unbroken supersymmetries of an
$\mathcal{N}_3=3$ theory are given by:
\begin{eqnarray}
1st & = & \xi^2_1 =  - i \, \left( \chi_1 - \chi^1 \right)  \nonumber\\
2nd & = &  \xi^1_2= \left( \chi_2 + \chi^2\right) \nonumber\\
3rd &=& \xi^2_2 = -i \, \left ( \chi_2 - \chi^2\right)
\label{chi2chi2}
\end{eqnarray}
The above conclusion applies to the supersymmetry parameters in the
same way. Utilizing the above described gamma matrix basis and
naming $\kappa_A$, $\kappa^A$ the chiral supersymmetry parameters in
$D=4$ we can write:
\begin{equation}\label{carriolarotta}
    \eta_A \, = \, \left(
                         \begin{array}{c}
                           \ft 12 (\varepsilon_A^{1} + i \,\varepsilon_A^{2}) \\
                           -\ft 12 (\varepsilon_A^{1} + i \,\varepsilon_A^{2}) \\
                         \end{array}
                       \right) \quad ; \quad \eta^A \, = \, \left(
                         \begin{array}{c}
                           \ft 12 (\varepsilon_A^{1} - i \,\varepsilon_A^{2}) \\
                           \ft 12 (\varepsilon_A^{1} - i \,\varepsilon_A^{2}) \\
                         \end{array}
                       \right)
\end{equation}
where $\varepsilon^{X}_A$ ($X=1,2$, $A=1,2$) are $2\times 2 \, = \,
4$ anticommuting $D=3$ Majorana spinors   that constitute the
supersymmetry parameters of an $\mathcal{N}_3\, = \,4 $
supersymmetry algebra in the three-dimensional theory. The generic
$\mathcal{N}_3=3$ Chern Simons theory admits only
$\varepsilon^{2}_A$, $\varepsilon_1^2$. When there is enhancement,
the missing parameter $\varepsilon_1^1$ is reinstalled.

\end{document}